\documentclass[12pt]{article}
\usepackage{url} 
\usepackage{hyperref}
\hypersetup{citecolor=blue,
    colorlinks=true,
    linkcolor=blue,
    filecolor=magenta,      
    urlcolor=blue,
    linktocpage=true
}
\usepackage{array}
\usepackage{soul}
\usepackage{booktabs}
\usepackage{multirow}
\usepackage{float}
\usepackage{dsfont}
\usepackage{color}
\usepackage{eurosym}
\usepackage{subcaption}
\usepackage{framed}
\usepackage{mathtools}
\usepackage[utf8]{inputenc}
\usepackage{algorithm}
\usepackage{algpseudocode}

\usepackage[linguistics]{forest}
\usepackage{adjustbox}
\usepackage{caption}
\usepackage{natbib}
\usepackage{enumerate}
\usepackage[mathscr]{euscript}
\usepackage{graphicx}
\usepackage{multicol}
\usepackage{amsmath, amssymb, amsthm, amsfonts}
\usepackage{latexsym}
\usepackage{psfrag,epsfig,epsf} 

\newcommand{\blind}{1}

\newtheorem{prop}{Proposition}
\newtheorem{theorem}{Theorem}
\newtheorem{lemma}{Lemma}
\newtheorem{corollary}{Corollary}
\newtheorem{assumption}{Assumption}

\newtheorem{remark}{Remark}

\newcommand{\Expect}[1]{\mathbb{E}\left[{#1}\right]}
\newcommand{\Expects}[2]{\mathbb{E}_{{#1}}\left[{#2}\right]}
\newcommand{\Var}[1]{\mathsf{Var}\left[{#1}\right]}

\newcommand{\phat}{\widehat{p}}
\newcommand{\thetahat}{\widehat{\theta}}
\newcommand{\thetatil}{\widetilde{\theta}}
\newcommand{\md}{\mathsf{d}}
\newcommand{\cX}{\mathcal{X}}
\newcommand{\IE}{\mathbb{I}_{\mathsf{E}}}
\newcommand{\ESJD}{\mathsf{ESJD}}
\newcommand{\Normal}{\mathsf{N}}
\newcommand{\Poisson}{\mathsf{Pois}}

\newcommand{\Multinomial}{\mathsf{Multinomial}}
\newcommand{\Uniform}{\mathsf{Uniform}}
\newcommand{\calA}{\mathcal{A}}
\newcommand{\Ubar}{\overline{U}}
\newcommand{\uhat}{\widehat{u}}
\newcommand{\vhat}{\widehat{v}}
\newcommand{\zhat}{\widehat{z}}

\newcommand{\hell}{\mathfrak{h}}

\newcommand{\CSadd}[1]{\textcolor{black}{{#1}}}

\begin{document}

\def\spacingset#1{\renewcommand{\baselinestretch}%
{#1}\small\normalsize} \spacingset{1}


\if1\blind
{
  \title{\bf Metropolis--Hastings with Scalable Subsampling}
  \author{Estev\~ao Prado, Christopher Nemeth\thanks{
     The author was supported by EPSRC grants EP/V022636/1 and EP/Y028783/1.}\hspace{.2cm} and Chris Sherlock \\
    School of Mathematical Sciences, Lancaster University, UK}
  \maketitle
  \vspace{-1cm}
} \fi

\if0\blind
{
  \bigskip
  \bigskip
  \bigskip
  \begin{center}
    {\LARGE\bf Metropolis--Hastings with Scalable Subsampling}
\end{center}
  \medskip
} \fi


\begin{abstract}
The Metropolis--Hastings (MH) algorithm is one of the most widely used Markov Chain Monte Carlo schemes for generating samples from Bayesian posterior distributions. The algorithm is asymptotically exact, flexible and easy to implement. However, in the context of Bayesian inference for large datasets,  evaluating the likelihood on the full data for thousands of iterations until convergence can be prohibitively expensive. This paper introduces a new subsample MH algorithm that satisfies detailed balance with respect to the target posterior and utilises control variates to enable exact, efficient Bayesian inference on datasets with large numbers of observations. Through theoretical results, simulation experiments and real-world applications on certain generalised linear models, we demonstrate that our method requires substantially smaller subsamples and is computationally more efficient than the standard MH algorithm and other exact subsample MH algorithms.

\end{abstract}

\noindent%
\vspace{-.5cm}

\spacingset{1.9} 

\section{Introduction}

In the Bayesian framework, the posterior distribution of the model parameters contains all information needed for inference.
The Metropolis--Hastings (MH) algorithm \citep[e.g.,][]{brooks2011handbook} is a core Markov Chain Monte Carlo (MCMC)  method commonly used in Bayesian statistics to sample from posterior distributions. At each iteration, it proposes at a new parameter value and then decides whether or not to accept it. The acceptance decision requires the full (unnormalised) posterior density evaluated at the proposed parameter, and this, in turn, requires the evaluation of every single term in the likelihood. In this big-data era, where datasets consist of millions or even billions of data points, each full likelihood evaluation is computationally expensive and the cost of repeating this at every iteration, as required by the standard MH algorithm, becomes prohibitive.

Optimisation-based algorithms, such as variational and Laplace approximations 
are still feasible, but are inexact. In this context, a new class of MCMC methods has gained attention: \textit{scalable MCMC} algorithms; see \cite{fearnhead2024scalable} for a recent review. To alleviate the computational burden, some works propose partitioning the data and then, in parallel, performing posterior sampling on each partition across multiple CPUs. This `divide-and-conquer' approach approximates the final posterior by combining the subposteriors from each subsample. One of the challenges with this approach is to efficiently yet accurately combine the subposteriors in cases where the target distribution is non-Gaussian; see, for example, \cite{neiswanger2013asymptotically, nemeth2018merging}.

Alternatively, we can reduce the computational cost by running a single  MCMC chain but using a random subsample of the data at each iteration \citep[e.g.,][]{bardenet2017markov, quiroz2019speeding, llorente2024survey}. 
 However, just as with divide-and-conquer approaches, the distribution targeted by the algorithm is typically only an approximation to the true posterior. For example, \cite{quiroz2019speeding} propose a Metropolis-based MCMC algorithm that builds upon the pseudo-marginal approach \citep{andrieu2009pseudo}.   The MH acceptance ratio is replaced with an approximation based on subsamples. The error in the targeted posterior is reduced through the use of control variates \citep[e.g.,][]{ripley2009stochastic} and an approximate bias-correction term. 

There are also subsampling MCMC methods that are constructed in such a way that the invariant distribution of the resulting Markov chain is the \emph{true posterior distribution}. All of these methods require that the full likelihood is a product of terms (typically, one term for each data point) and require assumptions on the individual terms in the product. 
For instance, \cite{maclaurin2014firefly} assume that for each observation there exists a strictly positive lower bound on the log-likelihood that accumulates tractably when summed over many observations; it then augments the parameter space using Bernoulli latent variables. \cite{cornish2019scalable} and \cite{zhang2020asymptotically} are less restrictive and instead replace the Metropolis--Hastings acceptance probability with an alternative that uses a small number of subsamples and, hence, has a high variance. \cite{cornish2019scalable} uses control variates to reduce this variance, whilst \cite{zhang2020asymptotically} incorporates an additional parameter to  mitigate the variance's deleterious effect on the variable acceptance rate.

We propose a new, exact MH algorithm that makes use of control variates and subsampling  to produce a Markov chain that satisfies detailed balance with respect to the true posterior. We establish general bounds on the discrepancy between the log-likelihood difference between current and proposed parameter values and the log-likelihood difference suggested by the control variates; these bounds are typically much tighter than those in previous works, especially when the dimension, $d$, of the parameter vector is moderate or large. \CSadd{We derive specific, tight bounds for four regression models including logistic regression}. More broadly, we establish that our algorithm is optimal within a particular class of data-subsampling algorithms. All of this leads to a substantial speed-up in the posterior sampling. We also conduct an asymptotic analysis comparing our proposed bounds with previous approaches, demonstrating theoretically and empirically that the computational cost of using our bounds is at least a factor of $d^{1/2}$ better than previous bounds.

Section \ref{MH_efficient_subsampling_sec}  introduces our algorithm and proves that it targets the correct posterior. Section \ref{section:related_works} discusses existing subsample-based MH algorithms. Section \ref{section:theoretical_results} presents theoretical results on the exactness and optimality of our algorithm and guidance on its tuning; proofs are deferred to Appendix \ref{appendix_proofs}. Sections \ref{section:simulation} and \ref{section:real_datasets} present simulation experiments and real-world applications using the logistic, probit and Poisson regression models, where our proposed method is compared against the equivalent Metropolis--Hastings algorithm and other exact subsample-based MH algorithms. 
Section \ref{section:discussion} concludes with a discussion. Additional simulation experiments, asymptotic analyses comparing bounds and results related to the acceptance rate of the proposed algorithms are presented in Appendices \ref{section:tuning_guidelines} to 
\ref{sec.alternative.acc.tuna}.

\section{Metropolis--Hastings with Scalable Subsampling}
\label{MH_efficient_subsampling_sec}

We are interested in sampling from a posterior distribution with density $\pi(\theta| y)$, where $\theta \in \cX \subseteq \mathbb{R}^{d}$ denotes the parameter vector and $y = (y_1, \ldots, y_n)$ are the observed data. Assuming a prior $p(\theta)$, and data that are conditionally independent with log likelihood terms $\ell_i(\theta):=\log p(y_i| \theta)$, the posterior density is
\begin{equation}
\pi(\theta|y) \propto p(\theta) \prod_{i=1}^{n} p(y_i | \theta)\equiv p(\theta) p(y_{1:n}|\theta)
=
p(\theta)
\exp\left[\sum_{i=1}^n \ell_i(\theta)\right]. \nonumber
\end{equation}
Given a current value, $\theta$, and a proposal density, $q(\theta'|\theta)$, the Metropolis--Hastings acceptance probability is $1\wedge [q(\theta|\theta')\pi(\theta'|y)/\{q(\theta'|\theta)\pi(\theta| y)\}]$. The contribution of the computationally expensive likelihood to this ratio is
\[
\exp\left[\sum_{i=1}^n\{\ell_i(\theta')-\ell_i(\theta)\}\right].
\]
To allow us to subsample terms from this sum whilst controlling the variance of the subsequent estimator, we create a deterministic \emph{control variate} for each difference, $\ell_i(\theta')-\ell_i(\theta)$. 
Let $\thetahat\in \cX$ be (ideally, but for strict correctness of the algorithm, not necessarily) close to the posterior mode. We choose control variates to be the approximations to $\ell_i(\theta')-\ell_i(\theta)$ that arise from first- and second-order Taylor expansion of $\ell_i$ about $\thetahat$:
\begin{align}
\label{eqn.CVone}
r_i^{(1)}(\theta,\theta';\thetahat)
&:=(\theta'-\theta)^\top g_i(\thetahat),\\
r_i^{(2)}(\theta,\theta';\thetahat)
&:=
(\theta'-\theta)^\top g_i(\thetahat)
+
(\theta'-\theta)^\top H_i(\thetahat)\left\{\frac{1}{2}(\theta'+\theta)-\thetahat\right\},
\label{eqn.CVtwo}
\end{align}
where $g_i(\theta)=\nabla \ell_i(\theta)$ and $H_i(\theta)=\nabla \nabla ^\top \ell_i(\theta)$ is the Hessian. Whichever control variate, $k\in \{1,2\}$, is used, our algorithm requires the existence of bounds $M^{(k)}(\theta,\theta')\ge 0$ and $c^{(k)}_1,\dots, c^{(k)}_n>0$ such that $M^{(k)}(\theta,\theta')=M^{(k)}(\theta',\theta)$ for all $\theta,\theta'\in \cX$ and
\begin{equation}
\label{eqn.basicBound}
 |\ell_i(\theta^\prime)-\ell_i(\theta)-r_i^{(k)}(\theta,\theta'; \thetahat)| \le c^{(k)}_i M^{(k)}(\theta,\theta').
\end{equation}
In the following, for notational convenience, we drop the superscript $(k)$ from $r^{(k)}_i(\theta,\theta'; \thetahat)$, $c^{(k)}_i$ and $M^{(k)}(\theta,\theta')$ and use the notation $r_i(\theta,\theta'; \thetahat)$, $c_i$, $M(\theta,\theta')$, respectively. Such bounds for logistic regression and robust linear regression models when $k=1, 2$ are given in \cite{cornish2019scalable}. In Section \ref{sec.bounds}, we provide much-improved bounds for logistic, probit\CSadd{, robust linear} and a form of Poisson regression. We define the difference between the control variate and the truth to be
\begin{equation}
  \label{eqn.define.Delta}
\Delta_i:=r_i(\theta,\theta'; \thetahat)-\{\ell_i(\theta')-\ell_i(\theta)\},
\end{equation}
 and we introduce the functions $\phi_i$ and $\phi_i'$ for $i=1,\ldots,n$ which are chosen to satisfy the following properties for all $\theta,\theta' \in \cX$ and $i=1,\dots,n$:
\begin{equation}
\label{eqn.three.conditions}
    \text{F1:} \ 0 \leq \phi_i \leq c_i M(\theta, \theta^\prime); \quad 
    \text{F2:} \ 0 \leq \phi'_i \leq c_i M(\theta, \theta^\prime); \quad \text{F3:} \ \phi_i-\phi_i'=\Delta_i.
\end{equation}
F1--F3 ensure that these are bounded, non-negative functions. 
In this paper, we choose
\begin{align}
  \label{eqn.define.phi}
  \phi_i &=\gamma \max[0,\Delta_i]+(1-\gamma)\left\{c_i M(\theta,\theta')+\min[0,\Delta_i]\right\}~~~
\mbox{for some}~ \gamma\in[0,1];
\end{align}
\CSadd{Corollary \ref{cor.efficiency.ess} in Section \ref{sec.optimal.phi} shows that $\gamma=0$ always leads to the most efficient algorithm so, in practice, we choose this.}
Interchanging $\theta \leftrightarrow \theta'$, we also define:
\begin{align}
  \label{eqn.define.phip}
\phi'_i&=\gamma \max[0,\Delta'_i]+(1-\gamma)\left\{c_i M(\theta',\theta)+\min[0,\Delta'_i]\right\},
\end{align}
where $\Delta_i'=r_i(\theta',\theta;\thetahat)-\{\ell_i(\theta)-\ell_i(\theta')\}=-\Delta_i$, since $r_i(\theta',\theta;\thetahat)=-r_i(\theta,\theta';\thetahat)$ for both $r^{(1)}$ and $r^{(2)}$. 
From \eqref{eqn.basicBound} and \eqref{eqn.define.Delta}, both the multiplicand of $\gamma$ and that of $1-\gamma$ in \eqref{eqn.define.phi} are between $0$ and $c_i M(\theta,\theta')$, proving F1, and, by an analogous argument, as $M(\theta,\theta')=M(\theta',\theta)$, F2. 
F3 holds because $\max[0,\Delta_i]+\min[0,\Delta_i]=0+\Delta_i=\Delta_i$.

\CSadd{F1 permits simulation of the number of times each datum will be used this iteration:}
\begin{equation}
\label{sample_s_i}
S_i \stackrel{iid}{\sim} \Poisson\left(\phi_i\right),~~~i=1,\dots,n.
\end{equation}
\CSadd{Typically, $\phi_i<<1$ so $S_i$ is  likely to be $0$. The acceptance probability for the algorithm is
\begin{equation}
\alpha(\theta,\theta';s,\thetahat)=   1 \wedge \frac{q(\theta^\prime|\theta)p(\theta^\prime)}{q(\theta | \theta^\prime)p(\theta)} 
    \exp\left[\sum_{i=1}^n r_i(\theta,\theta';\thetahat)\right]
    \prod_{i=1}^{n} \left\{ \frac{\phi_i^\prime}{\phi_i} \right\}^{s_{i}}. 
    \label{tuna_ratioNODA}
\end{equation}}
\CSadd{From F3,
\begin{equation}
\label{eqn.ratioOfSprobs}
\frac{p(s_i|\theta',\theta)}{p(s_i|\theta,\theta')}
=
\frac{\exp(-\phi_i'){\phi_i'}^{s_i}/s_i!}
{\exp(-\phi_i){\phi_i}^{s_i}/s_i!}
=
\exp(\Delta_i)\frac{{\phi_i'}^{s_i}}{{\phi_i}^{s_i}}.
\end{equation}}
\CSadd{The product of all $n$ such ratios contains $\exp(\sum_{i=1}^n\Delta_i)$, which contains the key term $p(y|\theta')/p(y|\theta)$. Indeed, the expectation over $S:=(S_1,\dots,S_n$) of the acceptance ratio is exactly the standard MH ratio (see Appendix \ref{sec.proofRatUnb}), so our algorithm is a particular case of \emph{randomised Metropolis--Hastings} \cite[]{NicFoxWatt2012}. A direct proof that the invariant distribution is $\pi(\theta|y)$ is also straightforward, as we now show.} 
\CSadd{
\begin{prop}
\label{propDB}
    A MH-based Markov chain that proposes via $q(\theta'|\theta)$ and \eqref{sample_s_i} and accepts $\theta'$ with a probability of \eqref{tuna_ratioNODA} satisfies detailed balance with respect to $\pi(\theta|y)\propto p(\theta) \exp[\sum_{i=1}^n \ell_i(\theta)]$.
\end{prop}
}
\begin{proof}
   Write $p(s|\theta,\theta'):=\prod_{i=1}^n p(s_i|\theta,\theta')$. \CSadd{Detailed balance holds with respect to $\pi(\theta|y)$ if }
\begin{equation}
\label{eqn.generalDB}
\pi(\theta|y)q(\theta'|\theta)p(s|\theta,\theta')\alpha(\theta,\theta';s,\thetahat)
=
\pi(\theta'|y)q(\theta|\theta')p(s|\theta',\theta)\alpha(\theta',\theta;s,\thetahat).
\end{equation}
This clearly holds when the acceptance probability is 
\begin{align}
\nonumber
    \alpha(\theta, \theta^\prime;s,\thetahat) & = 1 \wedge \frac{\pi(\theta^\prime|y) q(\theta|\theta^\prime) p(s | \theta^\prime, \theta )}{\pi(\theta|y) q(\theta^\prime|\theta) p(s | \theta,\theta^\prime )} \\
    \nonumber
    & = 1 \wedge \frac{q(\theta^\prime|\theta)p(\theta^\prime)}{q(\theta | \theta^\prime)p(\theta)} 
    \exp\left[\sum_{i=1}^n \ell_i(\theta')-\ell_i(\theta)\right]\exp\left[\sum_{i=1}^n \Delta_i\right]
    \prod_{i=1}^{n} \left\{ \frac{\phi_i^\prime}{\phi_i} \right\}^{s_{i}},
    \end{align}
    \CSadd{where the second line follows from \eqref{eqn.ratioOfSprobs} and because $\pi(\theta|y)\propto p(\theta)\exp(\sum_{i=1}^n \ell_i(\theta))$. This simplifies to \eqref{tuna_ratioNODA} on substituting for $\Delta_i$ from \eqref{eqn.define.Delta}.}
\end{proof}

Directly sampling independent Poisson variables $S_1,\dots,S_n$ would be an $\mathcal{O}(n)$ operation that is performed at every Metropolis-Hastings iteration, and therefore not practical. To avoid such costly simulations, we use Poisson thinning \citep{lewis1979simulation, cornish2019scalable, zhang2020asymptotically}. Defining $C:=\sum_{i=1}^nc_i$, we proceed as follows:
\begin{enumerate}
\itemsep 0ex
\item Sample $B \sim \Poisson(CM(\theta,\theta'))$.
\item Conditional on $B$, sample $(T_1,\dots,T_B)\sim \Multinomial(B;c_1/C,\dots,c_n/C)$. $\{T_j\}_{j=1}^B$ contains the indices (possibly repeated) of data that could be used this iteration.
\item For each $j\in 1,\dots,B$, include $T_j$ in $\mathcal{I}$ with a probability $\phi_{T_j}/\{c_{T_j} M(\theta,\theta')\}$. For each $i=1,\dots,n$, $S_i$ is the number of times the index $i$ occurs in $\mathcal{I}$. 
\end{enumerate}
Marginalising Steps 1 and 2 over $B$, the number of times each index $i$ appears in the set $\{T_j\}_{j=1}^B$ has a $\Poisson(c_i M(\theta,\theta'))$ distribution, independent of the frequencies of all other indices. Step 3 then thins down to $\Poisson(\phi_i)$. Calculating $C$ has a single $\mathcal{O}(n)$ cost. Similarly, after \CSadd{the $\mathcal{O}(n)$ set-up cost of Walker's alias method \cite[]{KronmalPeterson1979}}, Step 2 has a per-iteration cost of $\mathcal{O}(B)$ since it assigns an index to each of the $B$ values of $T_j$ that are non-zero.  Lastly, Step 3 also has a per-iteration cost of $\mathcal{O}(B)$.  \CSadd{The calculations in Appendix \ref{app.ourBounds} show that $B=\Theta(d^{1/2})$ when first-order control variates are used and $\Theta(d/\sqrt{n})$ with second-order control variates: the per-iteration cost does not grow with $n$.} The final set-up cost of the algorithm is in finding a suitable $\thetahat$, which is also $\mathcal{O}(n)$.

\subsection{The MHSS algorithm}

\CSadd{Here we present the Metropolis--Hastings with Scalable Subsampling (MH-SS) algorithm. In the case of regression models (our focus), each of the $n$ data points, $y_i$ ($i=1,\dots,n$), has an associated covariate vector, $x_i$ and a log-likelihood function $\ell_i(\theta)$, implicitly a function of $x_i$ through $\eta=x_i^\top \theta$, and of $y_i$. We abbreviate ``first-order control variates" to CV1 and ``second-order control variates" to CV2. Algorithm \ref{mh_ss_algorithm_set_up}  describes the one-off set up, then Algorithm \ref{mh_ss_algorithm} gives the iterative MCMC procedure. }

\begin{algorithm}[H]
\linespread{1.05}\selectfont
\caption{MH-SS: Set Up for Regression Models}
\CSadd{
	\begin{algorithmic}[1]
    \State {For the particular type of regression model, look up the appropriate form for $K_1(y)$ (if CV1) or $L_1(y)$ (if CV2) from Section 4.2.2, or, if this is a new model type, derive it.}
    \State{Set $c_i=\|x_i\|^2 K_1(y_i)$ (if CV1) or $c_i=\|x_i\|^3 L_1(y_i)/2$ (if CV2), $i=1,\dots,n$.}
    \State{Set $C=\sum_{i=1}^n c_i$.}
    \State{Set up Walker's alias method for sampling indices $1,\dots,n$ with replacement and with probabilities proportional to $c_1,\dots,c_n$.}
    \State{Find $\thetahat$, the (approximate) posterior mode.}
    \State{For $i=1,\dots,n$, find $g_i=\left.\nabla \ell_i(\theta)\right|_{\thetahat}$; if CV2, also find $H_i=\left.\nabla \nabla^\top \ell_i(\theta)\right|_{\thetahat}$.}
    \State{Set $g=\sum_{i=1}^n g_i$; if CV2, also set $H=\sum_{i=1}^n H_i$.}   
	\end{algorithmic} 
    }
\label{mh_ss_algorithm_set_up}
\end{algorithm}

\CSadd{The particular forms for the $c_i$ (Steps 1 and 2 of the set up) and $M(\theta,\theta')$ (Step 7 of the MCMC procedure) are derived in Section \ref{section:theoretical_results}; see Corollary \ref{cor.DotBounds}, in particular. Theorem \ref{theorem.remainders} provides the basis for these results, and alternative, looser bound for models that do not rely on a linear predictor.}

Algorithm \ref{mh_ss_algorithm} uses a delayed-acceptance (DA) formulation \citep{christen2005markov}. This uses the approximation from the Taylor expansion about $\thetahat$ to pre-screen each proposal. If the proposal is rejected at the pre-screening stage, the algorithm proceeds to the next iteration,  reducing the computational cost still further, \CSadd{while maintaining the exactness of the algorithm}. Here $\alpha(\theta,\theta';s)$ in \eqref{tuna_ratioNODA} is replaced with $\alpha_1(\theta,\theta';\thetahat)\alpha_{2,MHSS}(\theta,\theta';s,\thetahat)$, where
\begin{align}
\label{eqn.StageOneAcc}
\alpha_1\left(\theta, \theta^{\prime};\thetahat\right) 
&= 
1 \wedge \frac{p(\theta')q(\theta|\theta')}{p(\theta)q(\theta'|\theta)}\exp\left[\sum_{i=1}^n r_i(\theta,\theta';\thetahat)\right],\\
\label{eqn.StageTwoAcc}
\alpha_{2,MHSS} \left(\theta, \theta^{\prime};s,\thetahat\right) 
&= 
1\wedge \prod_{i=1}^n\left(\frac{\phi_i'}{\phi_i}\right)^{s_i}
\equiv 1 \wedge \exp \left[ \sum_{i \in \mathcal{I}} s_{i}  \left\{\log \left( \phi_i^\prime \right) - \log \left(\phi_i \right) \right\} \right].
\end{align}
\CSadd{The detailed detailed balance condition \eqref{eqn.generalDB} is still satisfied as we have simply split the product in \eqref{tuna_ratioNODA} over two terms, so the delayed-acceptance algorithm also targets $\pi(\theta|y)$.}

\begin{algorithm}[H]
\linespread{1.05}\selectfont
	\caption{Metropolis--Hastings with Scalable Subsampling (MH-SS)} 
	\begin{algorithmic}[1]
	\State \textbf{Initialise}: $\theta$. 
		\For {($t=1$ to $T$)}
                \State Sample $\theta^\prime \sim q(.|\theta)$ and $u \sim \Uniform\left(0,1\right)$.
                \State \CSadd{Calculate $\sum_{i=1}^n r_i(\theta,\theta';\thetahat)$ using \eqref{eqn.CVone} and $g$ (if CV1) or \eqref{eqn.CVtwo}, $g$ and $H$ (if CV2).}
                    \State Compute the DA-MH Stage 1 acceptance probability $\alpha_1\left(\theta, \theta^{\prime};\thetahat\right)$ in \eqref{eqn.StageOneAcc}. 
                    \If {$u\le \alpha_1\left(\theta, \theta^{\prime};\thetahat\right)$}
                    \State{\CSadd{Set $M(\theta,\theta')$ according to Corollary \ref{cor.DotBounds}.}}
                                    \If {$C M(\theta, \theta^\prime) \ge n$} \Comment{Metropolis--Hastings step on the full data}
                    \State Compute the DA Metropolis--Hastings Stage 2 acceptance probability:
                        \qquad\qquad 
                        \[
                        \alpha_{2,MH}(\theta,\theta';\thetahat)=1\wedge \frac{p(y_{1:n}|\theta')}{p(y_{1:n}|\theta)}\exp\left[-\sum_{i=1}^n r_i(\theta,\theta';\thetahat)\right].
                        \]
                    \State Sample $u \sim \Uniform\left(0,1\right)$: if $u\le \alpha_{2, MH}\left(\theta, \theta^{\prime};\thetahat\right)$, set $\theta = \theta^{\prime}$.
                    \Else \Comment{Metropolis--Hastings step on subsamples}
                        \State Sample $B \sim \mbox{Poisson}(C M(\theta, \theta^\prime))$
                        \State $\mathcal{I} \leftarrow \emptyset$.
                        \For {($b=1$ to $B$)} \Comment{Subsample creation step}
                            \State Sample $i_b$ such that $P(i_b = i) = c_i/C$ for $i = 1, \ldots, n$.
                            \State{\CSadd{Obtain $r_{i_b}$ via \eqref{eqn.CVone} (CV1) or \eqref{eqn.CVtwo} (CV2), then  $\phi_{i_b}$  via  \eqref{eqn.define.Delta} \&  \eqref{eqn.define.phi} with $\gamma=0$.}}
                            \State Add $i_b$ to $\mathcal{I}$ with probability $\frac{\phi_{i_b}}{c_{i_b} M(\theta, \theta^\prime)}$.
                        \EndFor
                        \State{\CSadd{For $i\in \mathcal{I}$, obtain $\phi'_i$ via \eqref{eqn.three.conditions}, F3.}}
                        \State Compute the DA MH-SS Stage 2 acceptance probability in \eqref{eqn.StageTwoAcc}.     
                        \State Sample $u \sim \Uniform\left(0,1\right)$: if $u\le \alpha_{2, MHSS}\left(\theta, \theta^{\prime};\thetahat\right)$, set $\theta = \theta^\prime$.
                    \EndIf
                \EndIf
        \EndFor
		\State \textbf{Output}: Samples from the posterior distribution $\pi(\theta)$. 
	\end{algorithmic} 
\label{mh_ss_algorithm}
\end{algorithm}

 The CPU cost of evaluating $\alpha_1(\theta, \theta^\prime)$ is negligible, as the relevant summaries of the  $r_i(\theta,\theta';\thetahat)$ are precomputed \CSadd{at Step 7 of Algorithm \ref{mh_ss_algorithm_set_up}}. This brings down the computational cost per iteration in evaluating the control variates to $\mathcal{O}(d)$ for first-order and $\mathcal{O}(d^{2})$ for second-order. When the pre-screening step has been passed successfully and $\mathbb{E}(B)=CM(\theta,\theta') \ge n$, then subsampling is not used, and the remainder of a delayed-acceptance Metropolis--Hastings step is performed on the full data. Otherwise, on passing the pre-screening, the proposal $\theta^\prime$ is accepted with probability $\alpha_{2,MHSS}(\theta, \theta^\prime;s)$ in \eqref{eqn.StageTwoAcc}, which is calculated taking into account at most $B$ observations, those with non-zero $s_i$.

We focus on the random-walk proposal with a scaling parameter of $\lambda>0$: 
\begin{equation}
\label{eqn.gen.RWM.prop}
\theta'\sim \mathsf{N}(\theta,\lambda^2 V/d),
\end{equation}
where $d=\mathsf{dim}(\theta)$ and $V$ is a $d\times d$ positive-definite matrix (usually) approximating the posterior variance.
Appendix \ref{app.general.proposals} discusses alternatives.

\section{Existing \CSadd{exact} subsampling MH algorithms}
\label{section:related_works}
\cite{maclaurin2014firefly} propose the Firefly Monte Carlo (FlyMC) algorithm, which augments the parameter space by introducing latent Bernoulli random variables. The success probabilities are defined so that the $\theta$ marginal of the invariant distribution of the Markov chain is $\pi$. Crucially, the evaluation of the joint posterior only requires likelihood terms corresponding to the non-zero Bernoulli variables. The biggest limitation of this algorithm is the requirement of a global exponential-family lower bound that is needed on each likelihood term $B_i(\theta) \le \ell_i(\theta)$. A secondary consideration is the trade-off between the MCMC mixing of the Bernoulli variables and the computational cost of updating them.

\subsection{Scalable Metropolis--Hastings}
\label{subsec.SMH}

Similar to our proposed algorithm, scalable MH \citep[SMH;][]{cornish2019scalable} also assumes that the log-likelihood difference is bounded as in \eqref{eqn.basicBound}; however, the delayed-acceptance idea is applied repeatedly to give an overall acceptance probability of
\begin{equation}
\label{eqn.SMHacc}
    \alpha_1(\theta,\theta')\prod_{i=1}^n 1\wedge \exp\left(\Delta_i\right).
\end{equation}
We explain the remainder of the algorithm using a formulation that is equivalent to that in \cite{cornish2019scalable}, but using quantities more familiar from MH-SS.
$B$ and $T_1,\dots,T_B$ are (effectively) simulated via the same mechanism as in MH-SS so that $N_i$, the number of repeats of the index $i$ in the set $\{T_j\}_{j=1}^B$ has a $\mathsf{Pois}(c_iM(\theta,\theta'))$ distribution. Then, for each $i$, $A_i\sim\mathsf{Binom}(N_i,p_i)$ is created, where $p_i=-\log[1\wedge \exp(\Delta_i)]/[c_iM(\theta,\theta')]$. Since $A_i\sim \mathsf{Pois}(-\log[1\wedge \exp(\Delta_i)])$, $\mathbb{P}(\sum_{i=1}^n A_i=0)$ is exactly the product term in \eqref{eqn.SMHacc}. As $(1\wedge a)(1\wedge b)\le 1\wedge ab$, the acceptance probability \eqref{eqn.SMHacc} is always lower than ours. Experiments with first-order control variates later in this section show that to avoid the acceptance rate dropping unacceptably low \CSadd{as $n$ increases}, the scaling parameter in the proposal $q(\theta'|\theta)$ must be reduced to well below the optimal value for the equivalent MH algorithm. Our bounds $c_iM(\theta,\theta')$ are also typically tighter than those in \cite{cornish2019scalable}, especially in moderate to high dimensions.

\subsection{TunaMH}
\label{subsec.TunaMH}

The TunaMH algorithm of \cite{zhang2020asymptotically} again uses \eqref{eqn.basicBound}, but without control variates. It creates quantities $\phi_i$ and $\phi_i'$, but with $\phi_i=[\ell_i(\theta)-\ell_i(\theta')+c_iM(\theta,\theta')]/2$, a special case of \eqref{eqn.define.phi}, with $\gamma=\frac12$. Section \ref{sec.optimal.phi} shows that the optimal choice of $\gamma$ is always $0$\CSadd{, the value used in MH-SS}. 

In TunaMH, where there are no control variates, $c_iM(\theta,\theta')$ is much larger (because the quantity it is bounding is much larger); however, it is also, typically, a much tighter bound on $\phi_i$ and $\phi_i'$. As a consequence, with $\phi$ and $\phi'$ varying across the range from $0$ to $c_iM(\theta,\theta')$, the acceptance ratio in \eqref{sample_s_i} has a high variance and is often very low. TunaMH adds an extra term, $\chi$, to each Poisson expectation, $\phi_i$. This lower bounds both the numerator and denominator of every term in the product in \eqref{sample_s_i} and, if $\chi$ is sufficiently large compared with each $\phi_i$ and $\phi_i'$, it stabilises the acceptance rate. The \emph{tuna}bility of $\chi$ gives the algorithm its name. However, this additional expectation increases $\Expect{B}$ and hence the computational cost, leading to an apparent trade-off between the acceptance rate and the subsample size. 

In fact, it is a three-way trade-off. Decreasing the scaling parameter, $\lambda$, for the jumps in the MH proposal, $q(\theta'|\theta)$, decreases $\|\theta'-\theta\|$ and hence $\ell_i(\theta')-\ell_i(\theta)$ and, thus, $M(\theta,\theta')$. In this way, each $\phi_i$ and $\phi_i'$ becomes smaller compared with $\chi$; the smaller $\lambda$ helps stabilise the acceptance probability. In reducing $\|\theta'-\theta\|$, it also reduces the computational cost via a smaller $\Expect{B}$ (the size of the final subsample) through smaller $\phi_i$. The advice in \cite{zhang2020asymptotically} is to add only a small amount to each expectation so that no more than one extra observation is used on average in each subsample, and to aim for an empirical acceptance rate of around 60\%. As a consequence, the Metropolis--Hastings scaling $\lambda$ is reduced substantially compared with a standard random walk Metropolis algorithm, the Markov chain mixes more slowly and although the algorithm is more efficient than MH with \emph{the same} $\lambda$ but no subsampling, it brings no substantial improvement over an \emph{optimally scaled} MH algorithm, as we show in Figure \ref{fig:issues_SMH_Tuna}. \CSadd{Appendix \ref{sec.alternative.acc.tuna}, shows the same phenomenon when an acceptance rate of $0.23$ is targeted instead.}


\subsection{Numerical comparison}
\label{subsec.NumCompbase}

\begin{figure}[htbp]
\begin{center}
   \begin{subfigure}{0.32\textwidth}
        \caption{}
          \includegraphics[scale=0.25]{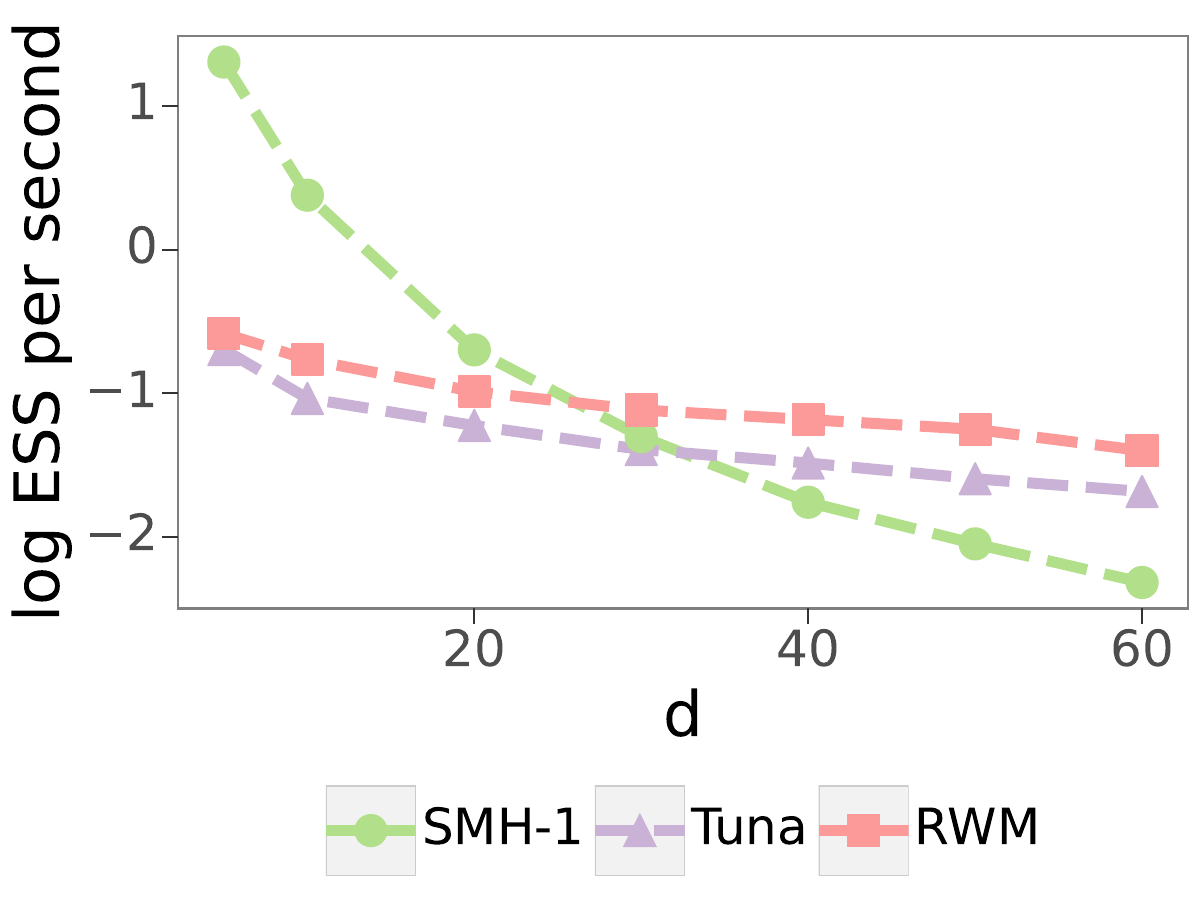}
   \end{subfigure}
   \begin{subfigure}{0.32\textwidth} 
        \caption{}
          \includegraphics[scale=0.25]{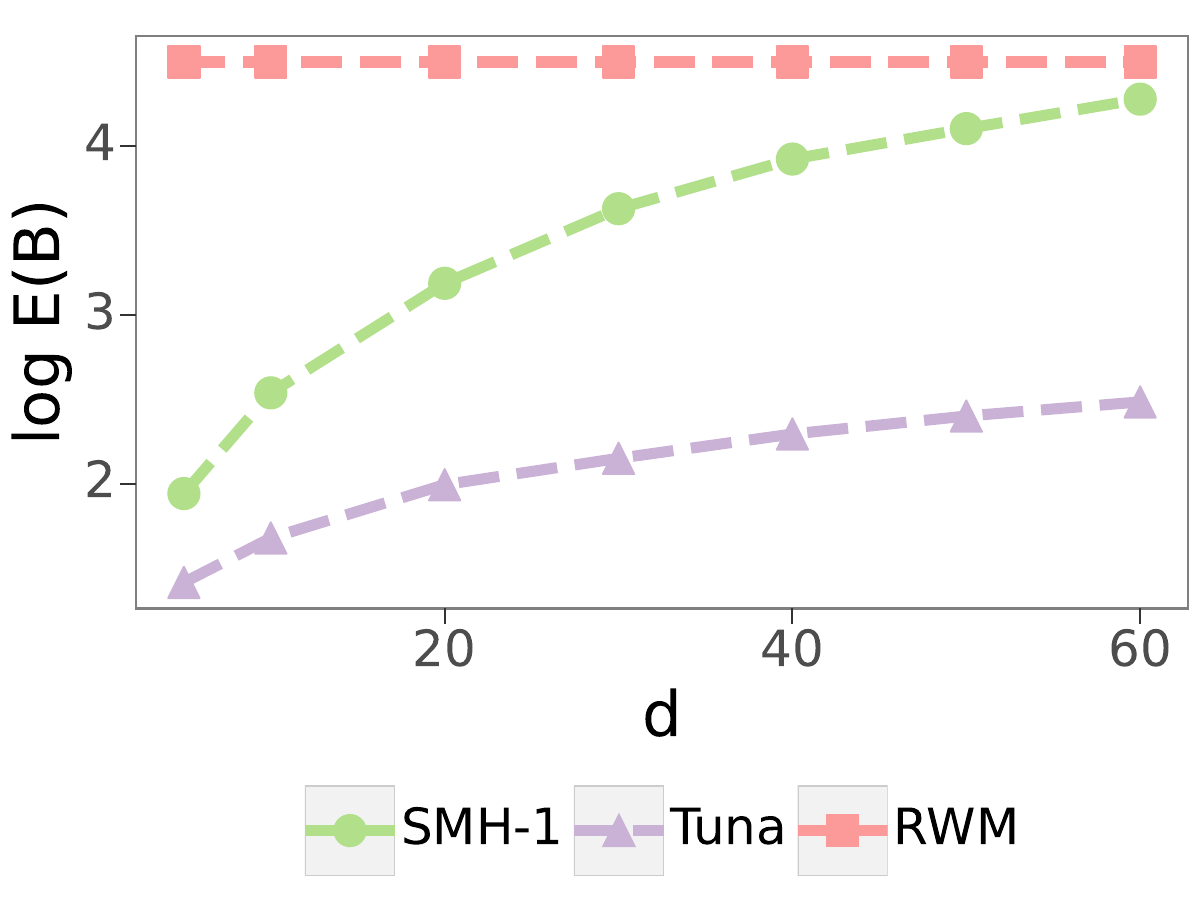}
   \end{subfigure}
   \begin{subfigure}{0.32\textwidth} 
        \caption{}
          \includegraphics[scale=0.25]{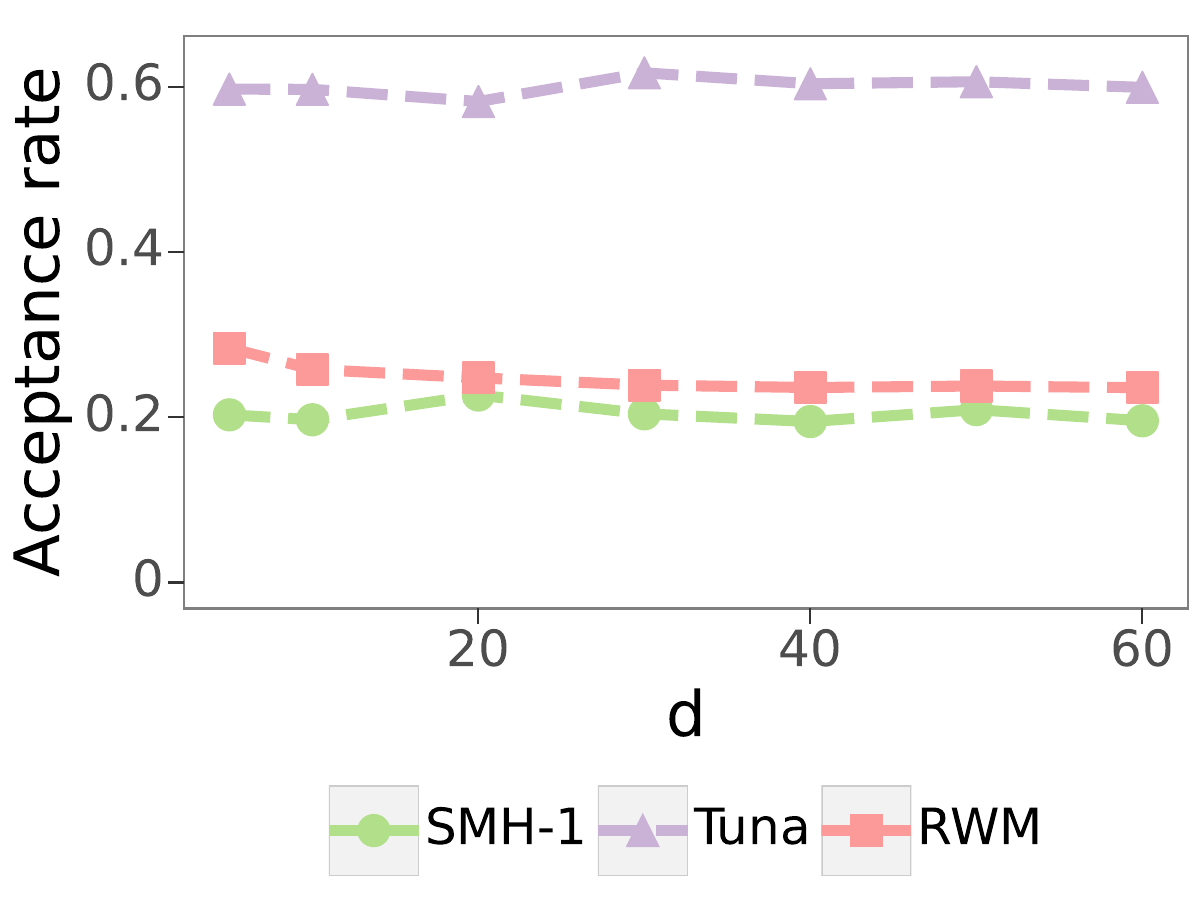}
   \end{subfigure}
\vspace{-0.3cm}
\caption{Acceptance rates and ESS per second for SMH-1, Tuna and RWM. The results are based on synthetic datasets generated from a logistic regression model with $n = 31,622$ observations. The y-axis of panels (a) and (b) are presented in the logarithm base 10.}
\label{fig:issues_SMH_Tuna}
\end{center}
\end{figure}

We compare MH, TunaMH and SMH with first-order control variates (referred to as SMH-1), as dimension $d$ increases for a logistic regression model with $n = 30,000$ observations. All algorithms use $q(\theta^\prime | \theta)$ from \eqref{eqn.gen.RWM.prop}, where $V = -[H(\hat{\theta})]^{-1}$ is the negative inverse Hessian at an estimate, $\widehat{\theta}$, of the mode. Following the literature on optimal scaling for this particular case of MH, the random-walk Metropolis algorithm \citep[RWM,][]{gelman1997weak, roberts2001optimal, sherlock2009optimal}, we set $\lambda=2.38$. This choice leads, in the limit as $d \to \infty$, to an acceptance rate of approximately $23.4\%$, as shown in panel (c) of Figure \ref{fig:issues_SMH_Tuna}.  In sharp contrast, for Tuna, $\lambda$ is set from $0.03$ ($d = 5$) to $0.1$ ($d = 60$) to maintain an acceptance rate of around $60\%$, as recommended in \cite{zhang2020asymptotically}; see Table \ref{tab:values_chi_simulation_related_works} in Appendix \ref{appendix:experimental_details} for all values. As recommended in \cite{zhang2020asymptotically}, we set the Tuna $\chi$ so that the expected number of additional data points was typically just below $1$.

Figure \ref{fig:issues_SMH_Tuna} also shows the effective sample size (ESS) per second and  average batch sizes.  Panel (b) shows that Tuna uses a fraction of the observations, making its per-iteration cost substantially lower than that of the RWM. However, Panel (a) demonstrates that TunaMH is no more efficient than the RWM, especially as $d$ increases; the much smaller $\lambda$, needed to keep the acceptance rate reasonable, substantially reduces the per iteration mixing.  

The bounds on the log-likelihood differences used in the SMH algorithm are quite loose (see Section \ref{section:simulation}), even for moderate $d$, especially when using first-order control variates. In practice, looser bounds imply that larger subsamples need to be evaluated in the MH ratio, leading to increased computational time and lower efficiency. Panel (a) of Figure \ref{fig:issues_SMH_Tuna} shows that SMH-1's ESS per second deteriorates more quickly with dimension than that of Tuna and the RWM. The explanation for this is two-fold. Firstly, Panel (b) shows the steady increase in SMH's average batch size as the number of variables in the linear predictor of the logistic regression model increases. The greater the batch size, the slower the algorithm. 
Secondly, to keep a constant acceptance rate at which SMH-1 is most efficient as dimension $d$ changes, it is necessary to reduce the scaling parameter in the proposal distribution. For the results presented in Figure \ref{fig:issues_SMH_Tuna}, $\lambda$ varied from $2$ ($d = 5$) to $0.65$ ($d = 60$).

In Sections \ref{section:simulation} and \ref{section:real_datasets}, we empirically show through simulation experiments and applications to real-world datasets that our proposed method, MH-SS, is substantially more efficient than Tuna and the RWM and that it requires substantially fewer observations than SMH for both first- and second-order control variates. The improvements over TunaMH are a direct consequence of the the use of control variates with a good bound on $M(\theta,\theta')$ and of the optimal choice of $\gamma$. The improvement over SMH is due to the control variate bound being tighter and scaling better with dimension than that in SMH and because the MH-SS acceptance probability is a product of two terms, rather than of many terms, all bounded above by $1$. Our tighter bounds on the log-likelihood differences are introduced in Section \ref{sec.bounds}. As we will show in Section \ref{sec.optimal.phi}, the optimal choice of $\gamma$ in \eqref{eqn.define.phi} leads to larger acceptance rates, which helps maximise the computational efficiency of our algorithms. In addition, in Appendix \ref{section:asymptotic_analysis_bounds} (see also Section \ref{sec.compCost}), through an asymptotic analysis, we compare our proposed bounds with those of SMH and show that our bounds are at least a factor of $d^{1/2}$ tighter. Finally, Appendix \ref{section:tuning_guidelines} studies both the scaling and acceptance rate that maximise the efficiency for our proposed methods in the case of a RWM proposal. In contrast to the standard RWM algorithm, our results suggest that the optimal acceptance rate for the MH-SS algorithm is approximately $45\%$.

\section{Theoretical results}
\label{section:theoretical_results}
We now present theoretical results to support our proposed method. First, we demonstrate that the acceptance rate of MH-SS is maximised when $\gamma$ is set to zero in \eqref{eqn.define.phi}. Then, we present general bounds for the log-likelihood differences for first- and second-order control variates. The proofs of all theoretical results in this section are deferred to Appendix \ref{appendix_proofs}.

\subsection{Optimality with respect to \texorpdfstring{$\phi$}{phi}}
\label{sec.optimal.phi}
\begin{figure}[H]
\begin{center}
     \includegraphics[scale=0.6]{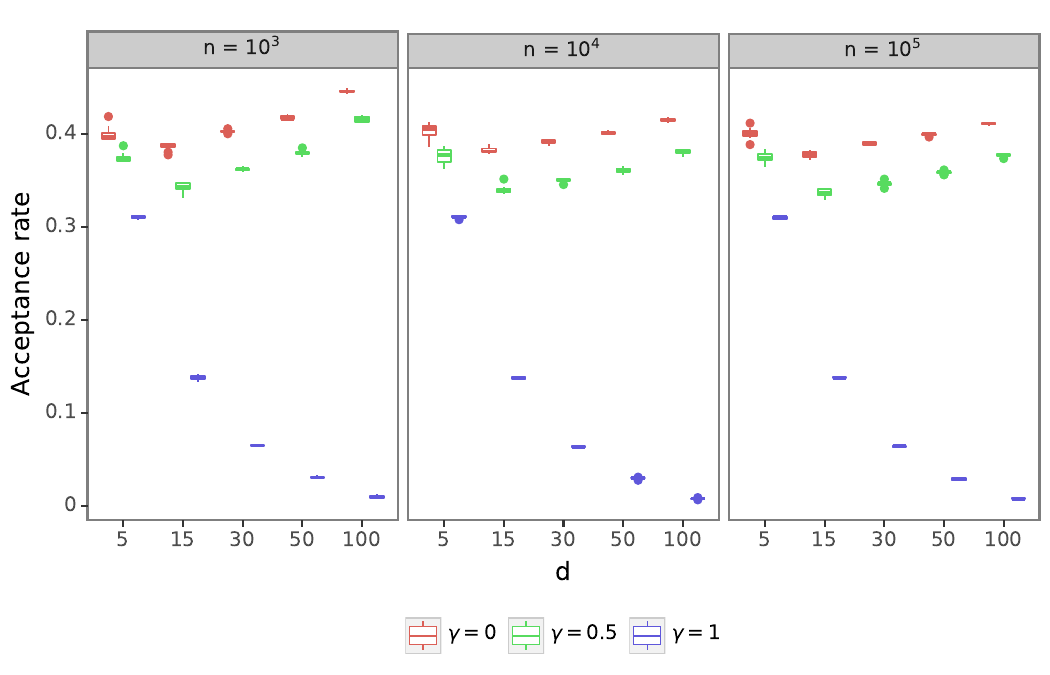}
     \captionsetup{justification=centering}
     \label{interaction_plot_b}
\end{center}
\vspace{-1cm}
\caption{Acceptance rates of MH-SS with first-order control variates.}
\label{fig_compare_phi_functions}
\end{figure}

\begin{theorem}\label{thrm.optimal.gamma}
Let $\alpha_2^{(\gamma)}(\theta,\theta')$ be the acceptance probability $\alpha_{2, MHSS}(\theta,\theta')$ from \eqref{eqn.StageTwoAcc} when $\phi_i$ and $\phi'_i$ are defined as in \eqref{eqn.define.phi} and \eqref{eqn.define.phip}. Then for any fixed $\theta$ and $\theta'$, $\alpha_2^{(\gamma)}(\theta,\theta')$ is a decreasing function of $\gamma$, so is maximised when $\gamma=0$.
\end{theorem}
Theorem \ref{thrm.optimal.gamma} is proved in Appendix \ref{proof.thrm.optimal.gamma}.  Since $\alpha_1(\theta,\theta')$ in \eqref{eqn.StageOneAcc} does not depend on $\gamma$, this implies that the overall acceptance probability is maximised when $\gamma=0$. Furthermore, the number of samples used in the calculation depends only on the bounds $c_iM(\theta,\theta')$ and so the computational cost is not affected by $\gamma$. Since the proposal, $q(\theta'|\theta)$, is the same, whatever the value of $\gamma$, \cite{peskun1973optimum} and \cite{tierney1998note} lead to the following result:
\begin{corollary} \label{cor.efficiency.ess}
The efficiency of Algorithm \ref{mh_ss_algorithm} in terms of effective samples per second is maximised when $\gamma=0$. 
\end{corollary}
\vspace{-0.5cm}
To illustrate the practical implications of Theorem \ref{thrm.optimal.gamma}, we perform simulation experiments using our method with first-order control variates on a logistic regression model. For each combination of $n = (10^3, 10^4, 10^5)$ and $d = (5, 10, 30, 50, 100)$, we set $\gamma$ to $0$, $0.5$ and $1$. Setting $\gamma = 0.5$ corresponds to employing the $\phi$ function in \eqref{eqn.define.phi} used by \cite{zhang2020asymptotically} (albeit with control variates and without the additional Tuna term), thereby illustrating the acceptance rate of our method if we were to use this sub-optimal choice of $\gamma$. Figure \ref{fig_compare_phi_functions} presents acceptance rates of our proposed method with first-order control variates obtained from 10 Monte Carlo simulations. Unsurprisingly, the acceptance rates of MH-SS are maximised when $\gamma = 0$, regardless of the values of $n$ and $d$, which is consistent with Theorem \ref{thrm.optimal.gamma}. Conversely, $\gamma = 0.5$ consistently yields lower acceptance rates compared to $\gamma = 0$, while $\gamma = 1$ exhibits the poorest results. Since the algorithms differ only in terms of the value of $\gamma$, Corollary \ref{cor.efficiency.ess} states that the one with the highest acceptance rate inevitably exhibits the largest effective sample size. We also investigated the effect of $\gamma$ on acceptance rates of our method with second-order control variates. Whilst the same ordering with $\gamma$ was evident, the differences were less pronounced, and we omit them for the sake of conciseness.

\subsection{Bounds on the remainder terms}
\label{sec.bounds}
Our method assumes \eqref{eqn.basicBound}. Such bounds are provided in \cite{cornish2019scalable}; however, these are loose as dimension $d$ increases. In this section, we provide tighter generic bounds and then refine these further in the case of regression models.

\subsubsection{General bounds on the remainders}
We first prove general bounds which will be useful in all of our examples in Sections \ref{section:simulation} and \ref{section:real_datasets} and many others besides. These general bounds then lead to specific forms in each example; see the end of Subsection \ref{sec.bounds} for cases of logistic regression, probit regression\CSadd{, robust regression} and a form of Poisson regression. \CSadd{To simplify notation, and because Theorem \ref{theorem.remainders} as stated is more general than the subsampling situation for which we employ it, for most of this section, we drop the subscript $i$ in $r_i^{(k)}$, and in $\ell_i$ and its derivatives.} Theorem \ref{theorem.remainders} is proved in Appendix \ref{proof.theorem.remainders}. 

\begin{theorem}
  \label{theorem.remainders}
  Let $\theta,\theta',\thetahat\in \cX \subseteq\mathbb{R}^d$ and let $\mathcal{A}$ be the triangle with vertices at $\theta,\theta',\thetahat$. Let $\ell:\mathbb{R}^d\to \mathbb{R}$ and let $g(\theta)=\nabla \ell(\theta)$ and $H(\theta)=\nabla \nabla^\top \ell(\theta)$. Let $\|v\|$ indicate the $L_2$ norm of $v$.
  Set
  \begin{align}
    r^{(1)}(\theta,\theta';\thetahat)
    &:=(\theta'-\theta)^\top g(\thetahat),\\
    r^{(2)}(\theta,\theta';\thetahat)
    &:=
(\theta'-\theta)^\top g(\thetahat)
+
(\theta'-\theta)^\top H(\thetahat)\left\{\frac{1}{2}(\theta'+\theta)-\thetahat\right\}.
  \end{align}
  \begin{itemize}
  \item
  Suppose that there exists $x\in \mathbb{R}^d$ and $K_1\ge 0$ such that for any $\theta_1\in \mathcal{A}$ and $u,v \in \mathbb{R}^d$, 
  \begin{equation}
    \label{eqn.bounded.Hessian}
|u^\top H(\theta_1) v| \le |x^\top u|~|x^\top v| ~K_1.
  \end{equation}
  Then
  \begin{equation}
    \label{eqn.gen.rem.one}
    \left|
    \ell(\theta')-\ell(\theta)
    -r^{(1)}(\theta,\theta';\thetahat)
    \right|
    \le
    |x^\top (\theta'-\theta)| \max\left(|x^\top (\theta-\thetahat)|,|x^\top (\theta'-\thetahat)|\right)K_1.
  \end{equation}
  If, instead, for any $\theta_1\in \mathcal{A}$ and $u,v \in \mathbb{R}^d$, there exists $K_2\ge 0$ such that
$|u^\top H(\theta_1) v| \le \|u\|~\|v\|~K_2$,
then
\[
    \left|
    \ell(\theta')-\ell(\theta)
    -r^{(1)}(\theta,\theta';\thetahat)
    \right|
    \le
    \|\theta'-\theta\| \max\left(\|\theta-\thetahat\|,\|\theta'-\thetahat\|\right)K_2.
\]
\item
  Alternatively, suppose that there exists $x\in \mathbb{R}^d$ and $L_1\ge 0$ such that for any $\theta_1,\theta_2\in \mathcal{A}$ and $u,v \in \mathbb{R}^d$
  \begin{equation}
    \label{eqn.genLip}
    \left|
u^\top \left\{H(\theta_2)-H(\theta_1)\right\} v
\right|
\le |x^\top u| ~|x^\top v| ~ |x^\top(\theta_2-\theta_1)|~L_1.
  \end{equation}
  Then
  \begin{align}
    \left|
    \ell(\theta')-\ell(\theta)
    -r^{(2)}(\theta,\theta';\thetahat)
    \right|
    &\le
\frac{1}{12}    |x^\top (\theta'-\theta)|^3~L_1\nonumber +
    \frac{1}{2}|x^\top(\theta'-\theta)|~|x^\top(\theta-\thetahat)|^2~L_1    \nonumber\\
&+
    \frac{1}{2}|x^\top(\theta'-\theta)|~|x^\top(\theta'-\thetahat)|^2~L_1.\label{eqn.gen.rem.two}
  \end{align}
  If, instead, for any $\theta_1,\theta_2\in\calA$ and $u,v\in \mathbb{R}^d$, there exists $L_2\ge 0$ such that\\
 $   \left|
u^\top \left\{H(\theta_2)-H(\theta_1)\right\} v
\right|
\le \|u\| ~\|v\| ~ \|\theta_2-\theta_1\|~L_2$,
then
  \begin{align*}
    \left|
    \ell(\theta')-\ell(\theta)
    -r^{(2)}(\theta,\theta';\thetahat)
    \right|
    &\le
    \frac{1}{12}\|\theta'-\theta\|^3~L_2
    +
    \frac{1}{2}\|\theta'-\theta\|~\|\theta-\thetahat\|^2~L_2 \\
   &+
    \frac{1}{2}\|\theta'-\theta\|~\|\theta'-\thetahat\|^2~L_2.
  \end{align*}
  \end{itemize}
  \end{theorem}

\begin{remark}
Let $\kappa$ be the magnitude of the largest eigenvalue of a symmetric matrix $W$, where $W=H(\theta_1)$ \CSadd{(for first-order covariates)} or $W=H(\theta_2)-H(\theta_1)$ \CSadd{(for second-order covariates)}. In general, $a^\top W b\le \kappa \|a\|~ \|b\|$, which motivates the two unnumbered conditions\CSadd{ that start with "If, instead", and lead to $K_2$ and $L_2$ as the largest eigenvalues of the respective $W$ matrices defined earlier in this remark}. 
\end{remark}

\begin{remark}
In the special case of a single observation from a regression model, we obtain a more subtle bound by taking $x$ to be a covariate vector and using the forms in \eqref{eqn.gen.rem.one} and \eqref{eqn.gen.rem.two}\CSadd{, for which we require the bounds $K_1$ and $L_1$. In Section \ref{sec:regModelBounds} we show that these are bounds on the absolute value of the second ($K_1$) and third ($L_1$) derivatives of the log-likelihood with respect to the linear predictor,  and derive $K_1$ and $L_1$ for specific regression models.}
\end{remark}

\subsubsection{Regression models: Further improvement on bounds}
\label{sec:regModelBounds}
We now specialise to the case where each observation is a realisation from a regression model with a covariate vector, $x$, regression coefficients, $\beta$, and observation, $y$, such that the likelihood for the observation is $\ell(\beta)=\mathfrak{h}(\eta;y)$, where $\eta=x^\top \beta$. 

In the following, we abbreviate $\mathfrak{h}(\eta;y)$ to $\mathfrak{h}(\eta)$. Firstly,
\[
g(\beta)=
\nabla_{\beta}\ell = (\nabla_{\beta}~ \eta)~ \mathfrak{h}'(\eta)=x\mathfrak{h}(\eta)
~~~\mbox{and}~~~
H(\beta)=\nabla_{\beta} \nabla_{\beta}^\top \ell
=
xx^\top \mathfrak{h}''(\eta).
\]
Thus, $|\CSadd{u}^\top H(\beta)\CSadd{v}| =| x^\top \CSadd{u}|~ |x^\top \CSadd{v}|~|\mathfrak{h}''(\eta)|$; also, with $\eta_1=x^\top \beta_1$ and $\eta_2=x^\top \beta_2$,
\begin{align*}
|\CSadd{u}^\top \{H(\beta_2)-H(\beta_1)\}\CSadd{v}|
&=
|\CSadd{u}^\top x \{\mathfrak{h}''(\eta_2)-\mathfrak{h}''(\eta_1)\} x^\top \CSadd{v}|
=
| x^\top \CSadd{u}|~ |x^\top \CSadd{v}|~|\eta_2-\eta_1| ~|\mathfrak{h}'''(\eta_*)|\\
&=
| x^\top \CSadd{u}|~ |x^\top \CSadd{v}|~|x^\top(\beta_2-\beta_1)| ~|\mathfrak{h}'''(\eta_*)|,
\end{align*}                                                                    
for some $\eta_*\in [\eta_1,\eta_2]$. \CSadd{Thus}, we can equate  \CSadd{$K_1$ and $L_1$ in }\eqref{eqn.bounded.Hessian} and \eqref{eqn.genLip} \CSadd{with uniform bounds on}  $|\mathfrak{h}''(\eta;y)|$ and $|\mathfrak{h}'''(\eta;y)|$\CSadd{, respectively}.

Before doing so, however, we tighten our general bounds further in the case of a regression model\CSadd{; \emph{i.e.}, in both \eqref{eqn.gen.rem.one} and \eqref{eqn.gen.rem.two}, we can improve over the obvious bound of $|x^\top u|~|x^\top v|\le \|x\|^2\|u\|\|v\|$.} The improvement\CSadd{, which may be of interest in its own right,} helps because in high-dimensions,``most'' vectors are almost orthogonal so the inner product $\omega$ that we will define is often close to $0$. Lemma \ref{lemma.dotprod} is proved in Appendix \ref{proof.lemma.dotprod}. 

\begin{lemma}\label{lemma.dotprod} Let $u,v,x\in\mathbb{R}^d$ and let $\omega:=u^\top v/(\|u\| \|v\|)$, then for $k>0$, $|x^\top u|^k|x^\top v|\le D_k(\omega) \|u\|^k \|v\| \|x\|^{k+1}$, where
  \[
  D_k(\omega)=
  \frac{1}{a_k(\omega)(k+1)^{(k+1)/2}}
  \left\{
k+|\omega|a_k(\omega)
  \right\}^{(k+1)/2},
  \]
  and
  $a_k(\omega)=\sqrt{k+\frac{1}{4}(k-1)^2\omega^2}-\frac{1}{2}(k-1)|\omega|$.
\end{lemma}
\begin{remark}
    We make three points. Firstly, when $k=1$, we have $a_1=1$, so $D_1(\omega)=\{1 + |\omega|\}/2$, giving a typical improvement of nearly a factor of $2$ over the obvious application of the Cauchy-Schwarz inequality. Secondly, when $u$ and $v$ are aligned, $\omega=1$ and $a_k=1$ implying $D_k(\omega)=1,$ which retrieves the bound from the Cauchy-Schwarz inequality. Finally, when $u$ and $v$ are orthogonal, as is usually approximately the case in high dimensions, $\omega=0$, $a_k=\sqrt{k}$ and
    \[
    D_k=\frac{1}{\sqrt{k}}\left(1-\frac{1}{k+1}\right)^{(k+1)/2}
    \uparrow \frac{1}{\sqrt{k}} \exp(-1/2),
    \]
    so we gain a factor of $e^{-1/2}/\sqrt{k}$ over the Cauchy-Schwarz inequality.
\end{remark}

Combining Lemma \ref{lemma.dotprod} with Theorem \ref{theorem.remainders} then gives the following, which we apply in practice for each of our regression models. 

\begin{corollary}\label{cor.DotBounds} Consider a single likelihood term $\ell_i(\theta)=\mathfrak{h}(\eta_i;y_i)$, where $\eta_i=x_i^\top \theta$, and define
  \[
  \omega:=\frac{(\theta-\thetahat)\cdot (\theta'-\theta)}{\|\theta-\thetahat\|\|\theta'-\theta\|}~~~\mbox{and}~~~
  \omega':=\frac{(\theta'-\thetahat)\cdot (\theta'-\theta)}{\|\theta'-\thetahat\|\|\theta'-\theta\|}.
  \]
  \CSadd{1. If $\|\mathfrak{h}''(\eta;y)\|\le K_1(y)$ for all $\eta$ and $y$ then
  $
\left|
\ell_i(\theta')-\ell_i(\theta)
-r_i^{(1)}(\theta,\theta';\thetahat)
\right|
\le c_i^{(1)} M^{(1)}(\theta,\theta')$, where $c_i^{(1)}=\|x_i\|^2 K_1(y_i)$ and
\[
M^{(1)}(\theta,\theta')=\|\theta'-\theta\|\max\left[\|\theta-\thetahat\|D_1(\omega),\|\theta'-\thetahat\|D_1(\omega')\right]~.
\]
2. If $\|\mathfrak{h}'''(\eta;y)\|\le L_1(y)$ for all $\eta$ and $y$ then 
  $
\left|
\ell_i(\theta')-\ell_i(\theta)
-r_i^{(2)}(\theta,\theta';\thetahat)
\right|
\le c_i^{(2)} M^{(2)}(\theta,\theta')$, where $c_i^{(2)}=
\frac{1}{2}\|x_i\|^3L_1(y_i)$ and
\begin{align*}
M^{(2)}(\theta,\theta')=
\|\theta'-\theta\|\left\{\frac{1}{6}\|\theta'-\theta\|^2
+
\|\theta-\thetahat\|^2D_2(\omega)
+
\|\theta'-\thetahat\|^2D_2(\omega')\right\}.
\end{align*}}
\end{corollary}
In moderate to high dimensions, most of the time, $\theta'-\theta$ is close to perpendicular to both $\theta-\thetahat$ and $\theta'-\thetahat$, so Corollary \ref{cor.DotBounds} gives a reduction of (typically) close to a factor of $2$ for the first-order bound and around $2.5$ for the second-order bound. In moderate to high dimensions, most $x$ are also close to perpendicular to the three vectors of interest, so the true remainder is typically much smaller than even this tighter bound. 

For the regression models in Sections \ref{section:simulation} and \ref{section:real_datasets}, we use the following values (justified in Appendix \ref{section:specifics_regression}):
\begin{itemize}
\item \underline{Logistic regression}: $\CSadd{K_1}(y)=1/4$ and $L\CSadd{_1}(y)=\sqrt{3}/18$; 
\item \underline{Probit regression}: $\CSadd{K_1}(y)=1$ and $L\CSadd{_1}(y)=0.3$;
\item\underline{Poisson regression} with an expectation of $\log[1+\lambda \exp(\eta)]$ for any fixed $\lambda$: $\CSadd{K_1}(y)=0.25+0.168y$ and $L\CSadd{_1}(y)=\sqrt{3}/18+0.061y$.
\item \CSadd{\underline{Robust  regression} using Student-t errors, so $Y-\eta\sim t_\nu$: $K_1(y)=(\nu+1)/\nu$ and $L_1(y)=(\nu+1)(3+2\sqrt{2})/(4\nu^{3/2})$.} 
\end{itemize}

For the Tuna algorithm with logistic regression, we may use the bound $|\ell_i(\theta')-\ell_i(\theta)|\le \|x_i\| \|\theta'-\theta\|$ \citep[e.g.,][]{cornish2019scalable,zhang2020asymptotically}. For Poisson regression with an expectation of $\log[1+\exp(\eta)]$, the bound is $|\ell_i(\theta')-\ell_i(\theta)|\le \|x_i\|\max(1,y_i)\|\theta'-\theta\|$\CSadd{, and for the robust regression, $|\ell_i(\theta')-\ell_i(\theta)|\le \|x_i\|\|\theta'-\theta\|/(2\sqrt{\nu})$. It is not possible to apply Tuna to probit regression since the first derivative of the log-likelihood is unbounded.}

\subsection{Computational cost and optimal tuning}
\label{sec.compCost}
In Appendix \ref{section:asymptotic_analysis_bounds}, we consider a sequence of posterior distributions in the multivariate-normal Bernstein-von Mises \CSadd{large-$n$} limit and examine the random walk Metropolis proposal \eqref{eqn.gen.RWM.prop} using a suitable scaling, $\lambda$, and variance matrix, $V$.  We show that \CSadd{as $d\to \infty$}, the computational cost per iteration of MH-SS is $\Theta(d^{3/2})$ when using $r_i^{(1)}$ and $\Theta(d^3/n^{1/2})$ when using $r_i^{(2)}$. We also show that whichever control variate is used, the corresponding costs when using the bounds in \cite{cornish2019scalable} are at least a factor of $d^{1/2}$ larger. 

Appendix \ref{section:tuning_guidelines} provides guidelines for selecting an optimal $\lambda$ in \eqref{eqn.gen.RWM.prop}. Taking the expected squared jumping distance (ESJD) as a measure of mixing efficiency for all algorithms, the overall efficiency of MH-SS-1 and MH-SS-2 is maximised at an acceptance rate of $\alpha \approx 0.452$, which corresponds to $\lambda_{opt} = 1.50$. The RWM's efficiency is maximised at $\alpha \approx 0.234$, yielding $\lambda_{opt} = 2.38$; we find that these values also hold for SMH-2. For SMH-1, the optimal scaling parameter is in the range between $0.5$ to $1.5$, and a scaling of $\lambda=1$ achieves at least $80\%$ of the optimal efficiency. However, the behaviour of the SMH-1 acceptance rate as $\lambda$ varies is different from that of the MH-SS and RWM algorithms, since the scaling must decrease more quickly with dimension than $\lambda/\sqrt{d}$ to maintain a reasonable acceptance rate.

\section{Simulation experiments}
\label{section:simulation}
In this section, we compare the MH-SS algorithm against the Tuna, SMH and MH algorithms, all using the same proposal form \eqref{eqn.gen.RWM.prop}. We apply them to a logistic regression model and a form of the Poisson regression model; see Appendix \ref{section:poisson_regression} for results on the latter model. For MH-SS and SMH, we consider first- and second-order control variates, which we denote by MH-SS-1 and MH-SS-2; with the same convention for SMH. We obtain an estimate of the posterior mode $\hat{\theta}$ through the stochastic gradient descent algorithm. This is needed for the control variate in MH-SS and SMH, but it is not necessary for Tuna and RWM.  
The Tuna algorithm increases $\Expect{B}$ by $\chi C^2 M(\theta,\theta')^2$, where $\chi$ is the additional tuning parameter. We follow \cite{zhang2020asymptotically}, choosing $\chi$  to increase the expected number of simulations by less than $1$ across most combinations of $(\theta,\theta')$. For reproducibility, we record $\lambda$ and the Tuna tuning parameter, $\chi$, in Table \ref{tab:values_chi_simulation} in Appendix \ref{appendix:experimental_details}. All  proposal distributions are of the form \eqref{eqn.gen.RWM.prop} with $V=-[H(\hat{\theta})]^{-1}$, as in Section \ref{subsec.NumCompbase}. For each method, $\lambda$ was chosen based on the maximisation of the mean squared jump distance (MSJD) divided by the average subsample size, $\mathbb{E}(B)$. Following Appendix \ref{section:tuning_guidelines}, we set $\lambda = 1.5$ for MH-SS-1 and MH-SS-2, $\lambda = 1$ for SMH-1, $\lambda = 2$ for SMH-2 and $\lambda = 2.38$ for the RWM. In all simulation experiments carried out in this section, all ESSs are at least $200$. To ensure the computational times are comparable, we implemented all methods in Python $3.9.6$.

\subsection{Logistic regression}
\label{sec.logisticRegressionSimExpts}

To empirically demonstrate the benefits of the proposed method, we initially test it on synthetic datasets generated from a succession of logistic regression models with varying data size, $n$, and parameter size, $d$. We sample each covariate (except the intercept) independently as $x_{i,j}\sim \Normal(0,1/d)$ and simulate independent $\beta_j\sim \Normal(0,1)$. These choices ensure the linear predictor, $x_i^\top \beta$, remains $\mathcal{O}(1)$ as $d$ increases and the probabilities obtained from the logistic function applied to the linear predictor are well distributed around $0.5$.

For each combination of $n$ and $d$, we run once each of the $6$ algorithms under consideration and calculate the effective sample size \cite[ESS;][]{liu1995blind} 
per second as a metric of efficiency. Additionally, we examine other metrics such as the average expected batch size, which we denote $\mathbb{E}(B)$; Appendix \ref{section:additional_simulation} presents the results in terms of the alternative efficiency metric, ESS$/\mathbb{E}(B)$.
To highlight any power-law relationships and enable all algorithms to be jointly displayed, results involving the number of observations, $n$, average expected batch size, $\mathbb{E}(B)$, and ESS per second are presented on a logarithmic scale.

\begin{figure}
\begin{center}
          \includegraphics[scale=0.5]{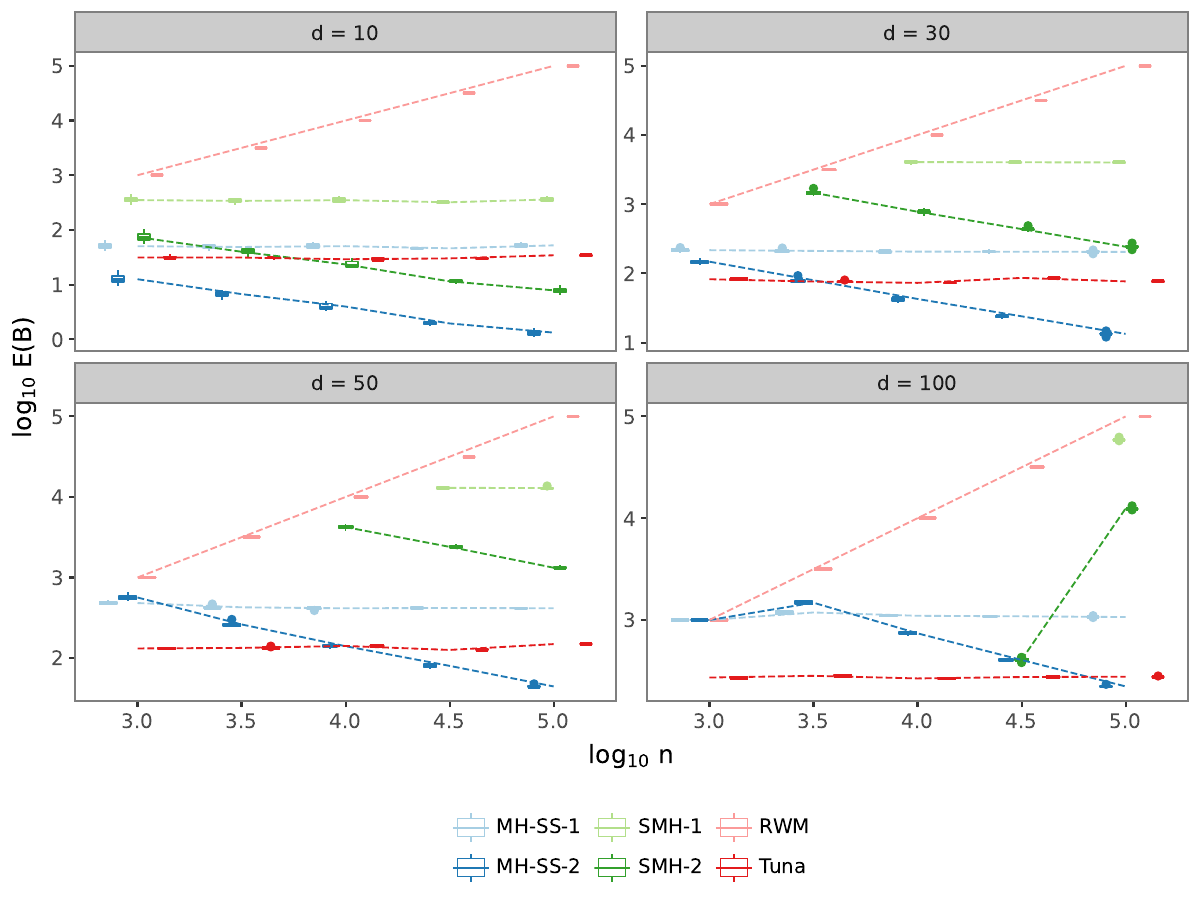}
     \captionsetup{justification=centering}
\end{center}
\vspace{-1cm}
\caption{Average batch size for MH-SS, SMH and RWM for the logistic regression model. For RWM, the average batch size is $n$. Both axes are presented in the logarithm base 10.}
\label{Expected_B_by_N}
\end{figure}

Figure \ref{Expected_B_by_N} shows how the average batch size varies by $n$ and $d$. As argued in Appendix \ref{section:asymptotic_analysis_bounds}, the algorithms based on the second-order control variates use fewer data values than their first-order counterparts and require even fewer as $n$ increases. Since the bounds we propose are tighter than those of SMH, the MH-SS algorithm uses less data. When $d=10$, SMH-2 sometimes uses fewer observations than MH-SS-1, but MH-SS-2 always uses the fewest; in all other dimensions considered, the MH-SS algorithms used fewer observations than the SMH algorithms. For example, when $d = 50$ and $n < 10^{4}$, SMH requires all $n$ observations, whereas MH-SS requires fewer than $n$ observations, even for $n = 10^3$. In all scenarios, Tuna and MH-SS-1 require a similar subsample size, the only exception being $d = 100$ where Tuna uses fewer observations. 

\begin{figure}[H]
\begin{center}
     \includegraphics[scale=0.5]{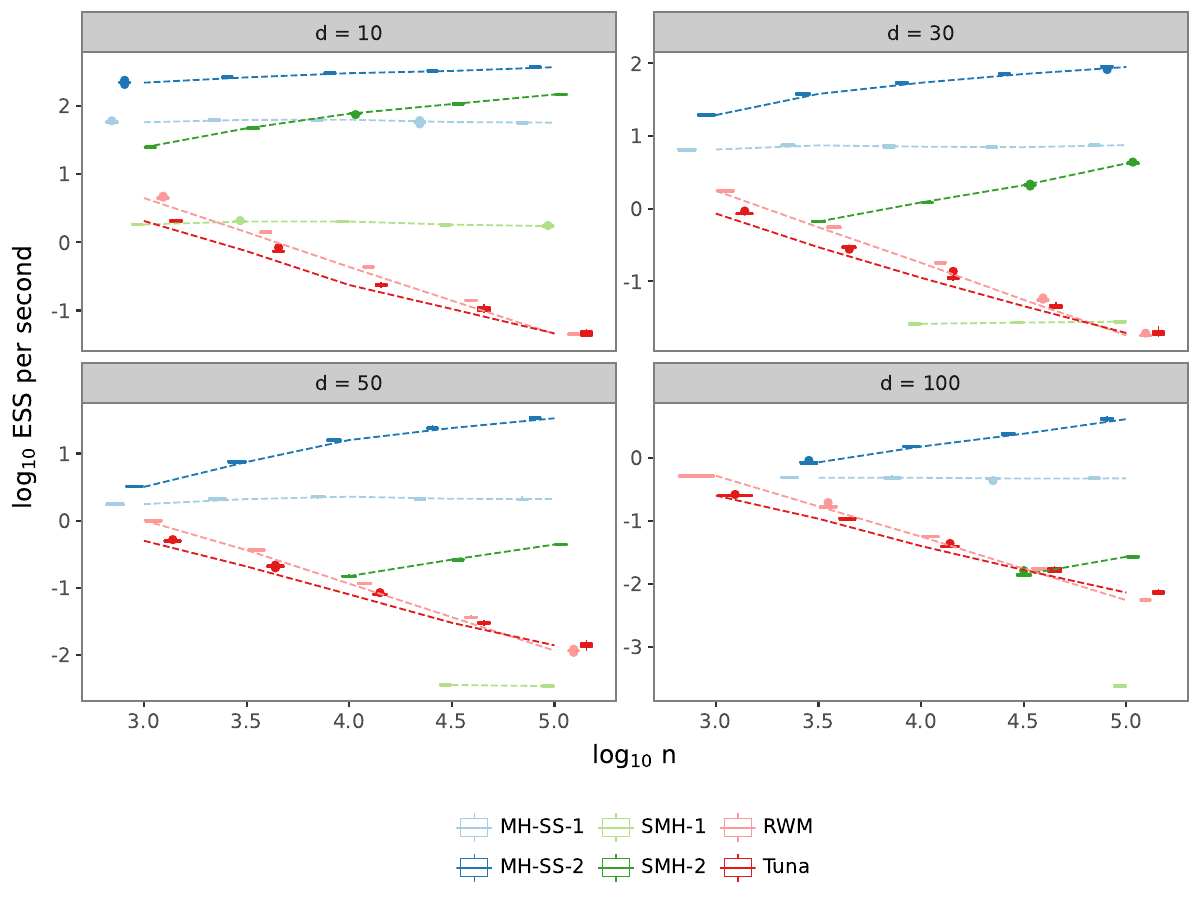}
     \captionsetup{justification=centering}
\end{center}
\vspace{-1cm}
\caption{ESS per second of MH-SS, SMH and RWM for the logistic regression model. Both axes are presented in the logarithm base 10. Some ESSs are omitted because $\mathbb{E}(B) \ge n$, which implies the use and efficiency of the RWM algorithm.}
\label{ESS_per_second}
\end{figure}

Figure \ref{ESS_per_second} compares the ESS per second across all methods. As expected, the ESS per second of MH-SS with first- and second-order control variates is greater than that of either SMH algorithm, except when $d=10$, where SMH-2 outperforms MH-SS-1. For $d = 10$, MH-SS-2 and SMH-2 exhibit similar efficiency for $n = 10^{5} = 100,000$, even though the former uses fewer observations. To understand this, we recall the results in the left-hand upper panel of Figure \ref{Expected_B_by_N}, where it is possible to see that $\mathbb{E}(B) \approx 10^0 = 1$ for MH-SS-2, while $\mathbb{E}(B) \approx 10^1 = 10$ for SMH-2. In practice, we observe that the run-time differences are tiny in cases where the average batch sizes are small, helping to explain the similarity in terms of efficiency. However, as $d$ increases, the efficiency gap becomes more pronounced because the SMH bounds are at least a factor of $d^{1/2}$ worse than the equivalent MH-SS bound. Even for $d = 30$, SMH-2 would be a competitive option against RWM only for $n > 10^{3.5}$, while SMH-1 demonstrates relatively poor performance for all considered dimensions.

Though Tuna and MH-SS-1 require a similar subsample size for all scenarios considered in Figure \ref{Expected_B_by_N}, except for $d = 100$, MH-SS with first- and second-order control variates outperforms Tuna in all dimensions in terms of efficiency. As stated in Section \ref{section:related_works}, Tuna tends to favour smaller subsample sizes in detriment of computational efficiency and, as argued in Section \ref{section:related_works}, its performance is comparable to the RWM algorithm. To recap, the heuristic proposed by \cite{zhang2020asymptotically} aims for an acceptance rate of $60\%$ and minimises the impact of the Tuna tuning parameter on the average batch size. As result, in practice Tuna takes tiny steps when exploring the posterior distribution. This means it uses small subsample sizes and thus has low per-iteration run-time;  however, such an approach is inefficient from the effective sample size viewpoint as small step sizes lead to high auto-correlation in the Markov chain.

\subsection{Decomposing the improvements over SMH}
\label{section:understanding_improvements}

Tuna uses no more information than the RWM, so it is unsurprising that their efficiencies are similar and that MH-SS's careful use of control variates leads to a substantial increase in efficiency. We now investigate the algorithms that use control variates, and separate the drivers of the superior performance of MH-SS over SMH.

Both SMH and our proposed method bound $|\ell_i(\theta^\prime)-\ell_i(\theta) - r_i^{(k)}(\theta, \theta^\prime; \widehat{\theta})| \leq c_i^{(k)} M^{(k)}(\theta, \theta^\prime)$, for $k = 1, 2$, though with different $c_i^{(k)}$ and $M^{(k)}(\theta, \theta^\prime)$, so we created a hybrid algorithm named SMH-NB. This hybrid approach combines the structure of SMH, which is detailed in Algorithm 1 of \cite{cornish2019scalable}, with our bounds, $c_i^{(k)}$ and $M^{(k)}(\theta,\theta')$, from Section \ref{sec:regModelBounds}. We compare these hybrid algorithms with the vanilla SMH and MH-SS to separate the benefit brought by these bounds from the other benefits of MH-SS. The scaling parameter in the proposal distribution of the SMH with new bounds is set to $\lambda = 0.5$ for SMH-1-NB and $\lambda = 1.5$ for SMH-2-NB. These values were chosen as they empirically maximise MSJD/$\mathbb{E}(B)$.

\begin{figure}[htbp]
\begin{center}
   \begin{subfigure}{0.5\textwidth}
        \caption{}
          \includegraphics[scale=0.25]{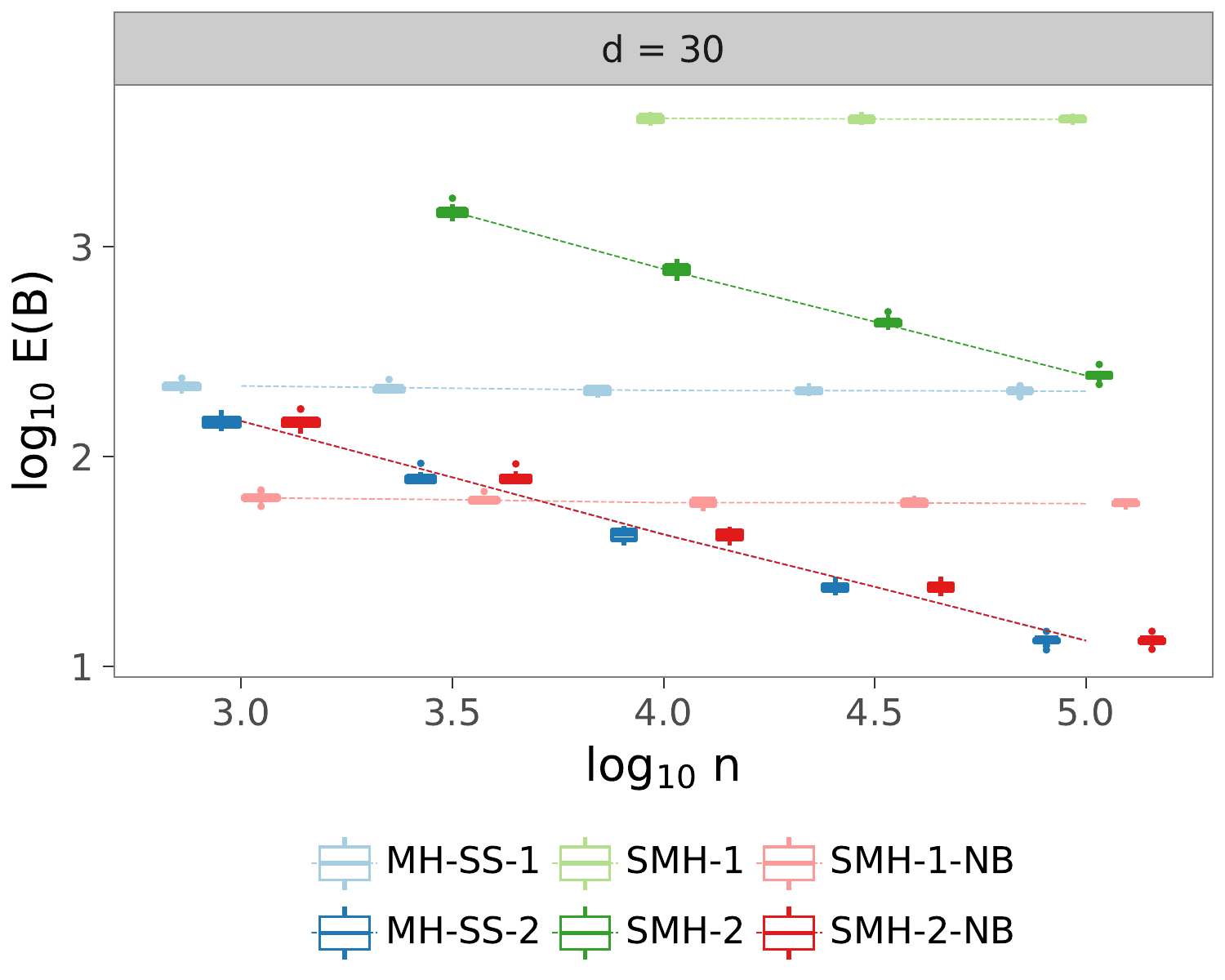}          
   \end{subfigure}
   \begin{subfigure}{0.4\textwidth} 
        \caption{}
          \includegraphics[scale=0.25]{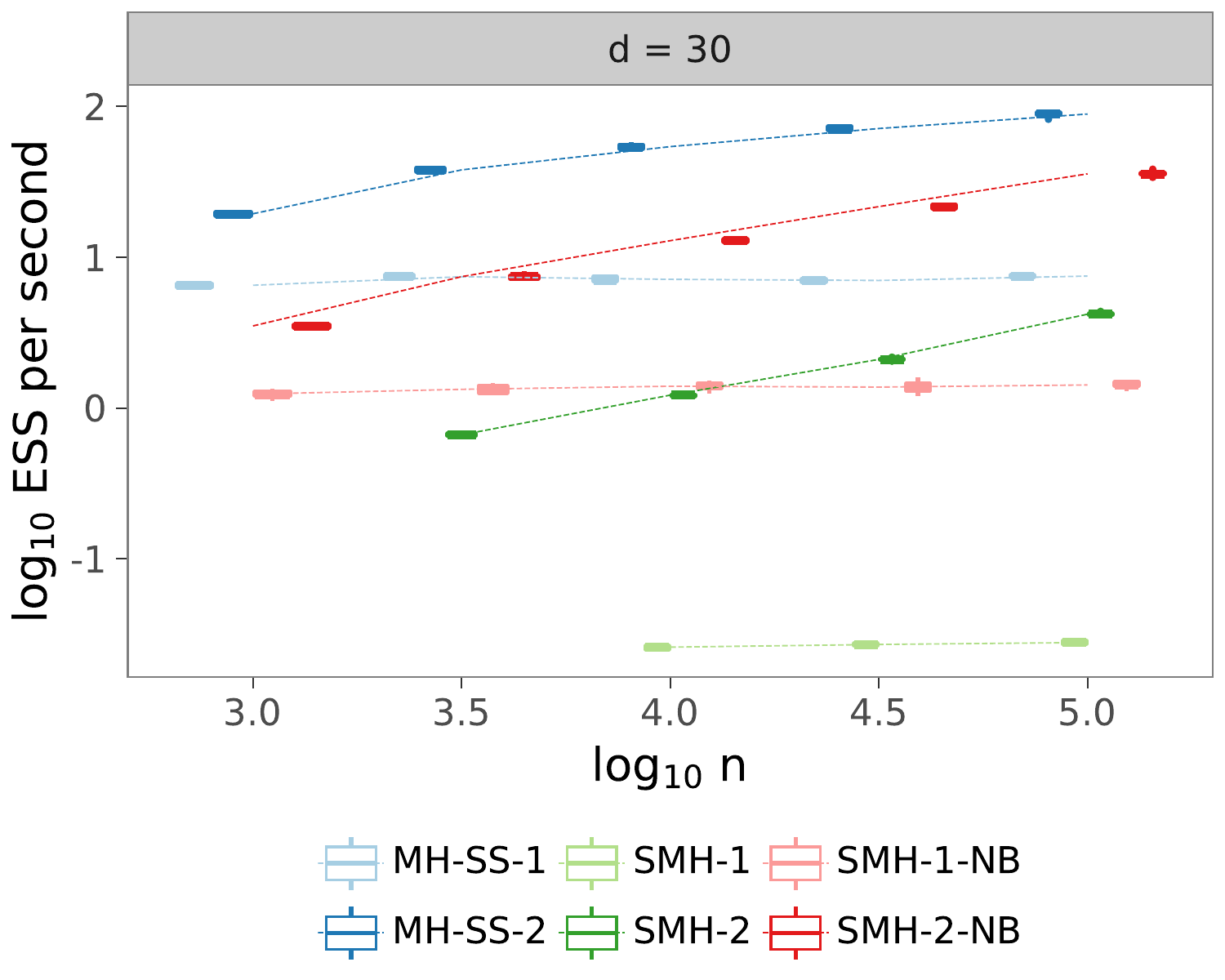}          
   \end{subfigure}
   \vspace{-0.3cm}
\caption{Average batch size and ESS per second for MH-SS, vanilla SMH and SMH with new bounds for the logistic regression model ($d = 30$).}
\label{fig:d_30_E_B_and_ESS_per_second}
\end{center}
\end{figure}

Panel (a) of Figure \ref{fig:d_30_E_B_and_ESS_per_second} demonstrates the substantial reduction in the subsample size that SMH can achieve when using our bounds. It also shows that, on average, SMH-NB with the first-order control variates requires smaller subsamples than MH-SS with the same approximation. This difference arises exclusively from the smaller scaling parameter in the proposal distribution of SMH-1-NB (i.e., $\lambda = 0.5$ for SMH-1-NB versus $\lambda = 1.5$ for MH-SS). MH-SS-2 and SMH-2-NB have approximately the same $\mathbb{E}(B)$ since they share the same $\lambda$. 

A smaller $\lambda$ implies smaller jumps from $\theta$ to $\theta^\prime$, negatively impacting efficiency metrics such as ESS since (when the acceptance rate is reasonable) a small $\lambda$ tends to increase the correlation between successive values in the chain, especially in high-dimensional settings. For example, in panel (b) of Figure \ref{fig:d_30_E_B_and_ESS_per_second}, both SMH-NB variants exhibit greater efficiency per second than their SMH counterparts. However, each is consistently outperformed by the corresponding MH-SS algorithm, even though it shares the same bound, demonstrating superior performance of the algorithm itself, as well as the bounds. 

In summary, the results presented in Figure \ref{fig:d_30_E_B_and_ESS_per_second} suggest that SMH requires less data per iteration and has improved efficiency if it adopts our tighter bounds. However, this is does  not make it as efficient as our proposed methods. Though SMH and MH-SS share similarities, such as the use of auxiliary variables and the assumption that the $\Delta_i$ are bounded, the final algorithms are different. SMH, and especially SMH-1, suffers from its acceptance rate being a product of individual acceptance probabilities. Moreover, SMH requires sampling up to $B - 1$ additional Bernoulli random variables per iteration.

\section{Results on real datasets}
\label{section:real_datasets}
Section \ref{section:simulation} showed that MH-SS outperforms the SMH, Tuna and RWM algorithms for different combinations of $n$ and $d$ using synthetic data. This section extends our comparison using real-world datasets and three regression models: logistic, probit and Poisson. 
It illustrate that MH-SS can be successfully applied to the probit model, as well as in more challenging settings with larger $n$ and correlated predictors which are not continuous and/or normally distributed. Tuna cannot be applied to the probit model, since it is not possible to bound the first derivative of the log-likelihood. We evaluated the algorithms in terms of computational efficiency, considering metrics such as ESS/$\mathbb{E}(B)$ and ESS per second; \CSadd{values reported are means and standard errors, estimated from $10$ replicate runs of each experiment.}
We also present the empirical acceptance rates and average batch sizes. In all results presented in this section, the ESSs are at least $150$ for all algorithms.

We first compare the methods on three binary datasets using the logistic and probit regression models only; see Appendices \ref{sec:usa_survey} and \ref{section:gas_detection} for results on two of the datasets. Our fourth and last comparison uses a Poisson regression model on a dataset with $n = 298, 290$ and $d = 28$. The first three datasets differ in many aspects, such as the numbers of observations and predictors as well the correlations between predictors. The later point is of particular interest as the simulation experiments only considered an intercept and continuous i.i.d. Gaussian predictors. Proposals are of the same form as in the Section \ref{section:simulation}.

We now point out aspects of the results common to all four analyses. Whether in terms of ESS per second or ESS/$\Expect{B}$, MH-SS-2 is always the most efficient algorithm. In terms of ESS per second, it is always at least an order of magnitude more efficient than any of the other algorithms. The second- and third-most efficient algorithms are always MH-SS-1 and SMH-2, though the order varies. MH-SS-2 always uses the fewest observations, but in all of the logistic regression models it is Tuna that uses the second fewest observations, and for the Poisson regression, it uses the third fewest observations. Despite the relatively low numbers of observations, Tuna always performs worse than either MH-SS-1 or SMH-2 (and often much worse than either) because the proposal scaling parameter is so small.

\subsection{High-energy particle physics}
\label{section:hepmass}

We apply the MCMC algorithms to the Hepmass dataset from the UCI machine learning repository. The dataset details the signatures of exotic particles obtained from a high-energy physics experiment. The binary response variable indicates whether a new particle of unknown mass is observed. To illustrate the applicability of the proposed methods, we compare the algorithms on a subsample of $ n = 1, 000, 000$ observations from the training set, using all $26$ continuous predictors available. For Tuna, we set $\chi=10^{-5}$ and $\lambda=0.013$.

\begin{table}[H]
\caption{Acceptance rate, average batch size, ESS per second and ESS/$\mathbb{E}(B)$ for the logistic and probit regression models applied to the Hepmass dataset.}
\label{tab_hepmass}
\centering
\setlength\extrarowheight{-7pt}
\begin{tabular}{cccccc}
\toprule
Model & Algorithm       & Acc. rate     & $\mathbb{E}(B)$   & ESS per second & ESS/$\mathbb{E}(B)$ \\ 
\hline
\multirow{6}{*}{Logistic}
& Tuna    & 0.591  & 295\,(2.06)       & 0.0023\,(0.0005)     & 0.0014\,(0.0002) \\
& MH-SS-1 & 0.447  & 766\,(4.46)       & 1.81\,(0.85)         & 1.25\,(0.03) \\
& MH-SS-2 & 0.456  & 31.2\,(0.27)      & \textbf{69.2}\,(22.4) & \textbf{31.1}\,(0.69) \\
& SMH-1   & 0.169  & 11,800\,(150)     & 0.016\,(0.003)       & 0.011\,(0.0002) \\
& SMH-2   & 0.322  & 493\,(6.46)       & \textbf{9.12}\,(4.84) & \textbf{2.43}\,(0.04) \\
& RWM     & 0.243  & 1,000,000         & 0.0033\,(0.0008)     & 0.0012\,(0.0006) \\
\hline
\multirow{5}{*}{Probit}
& MH-SS-1 & 0.413  & 1,031\,(3.82)     & 1.28\,(0.52)         & 1.01\,(0.04) \\
& MH-SS-2 & 0.422  & 18.4\,(0.21)      & \textbf{46.5}\,(9.03) & \textbf{55.9}\,(1.81) \\
& SMH-1   & 0.153  & 15,400\,(394)     & 0.021\,(0.003)       & 0.012\,(0.001) \\
& SMH-2   & 0.285  & 261\,(3.81)       & \textbf{21.5}\,(11.4) & \textbf{4.66}\,(0.18) \\
& RWM     & 0.199  & 1,000,000         & 0.034\,(0.003)       & 0.0015\,(0.00006) \\
\bottomrule
\end{tabular}
\end{table}

Table \ref{tab_hepmass} summarises the results of efficiency metrics for the logistic and probit regression models applied to the Hepmass dataset. The efficiencies of the RWM, Tuna and SMH are all around two (or more) orders of magnitude smaller than those of SMH-2 and MH-SS-1, which themselves are at least an order of magnitude smaller than that of MH-SS-2.

\subsection{Reported road casualties in the UK}
\label{section:road_casualties}
The UK Department for Transport publishes annual road safety statistics \citep{stats19}. The data include the accident's geographical coordinates, severity, speed limit of the road where the accident took place, details about the vehicles involved, weather conditions, road conditions, as well as time and date. We compare the algorithms on the consolidated UK road casualties data from $2020$ to $2022$. We  model at the individual level of $n = 298,290$ accidents the number of casualties based on 8 predictors, of which 2 are continuous. In total, the linear predictor in our Poisson model with mean $\log(1 + e^{\eta})$ has $d=28$ parameters (including the intercept). For the Tuna, we set $\chi=10^{-7}$ and $\lambda=0.19$.

As for the analyses of datasets with a binary response variable, we see (in Table \ref{tab_uk_road_accident}) that MH-SS-1 and MH-SS-2 on average require the smallest number of data points per MCMC iteration. Unsurprisingly, both versions of MH-SS present the best results in terms of computational efficiency. Though Tuna and MH-SS-1 have similar average batch sizes, again the latter is significantly more efficient as shown by its ESS per second and ESS/$\mathbb{E}(B)$.

\begin{table}[H]
\caption{Acceptance rate, average batch size, ESS per second and ESS/$\mathbb{E}(B)$ for a Poisson regression model applied to the UK road casualties dataset.}
\label{tab_uk_road_accident}
\centering
\setlength\extrarowheight{-7pt}
\begin{tabular}{cccccc}
\toprule
Model & Algorithm & Acc. rate & $\mathbb{E}(B)$ & ESS/sec & ESS/$\mathbb{E}(B)$ \\
\hline
\multirow{6}{*}{Poisson}
& Tuna    & 0.596  & 2,680\,(10.3)       & 0.0152\,(0.0048)      & 0.00227\,(0.00004) \\
& MH-SS-1 & 0.454  & 1,930\,(20.1)       & \textbf{1.43}\,(0.31) & \textbf{0.484}\,(0.015) \\
& MH-SS-2 & 0.457  & 391\,(8.93)         & \textbf{9.27}\,(2.41) & \textbf{2.33}\,(0.09) \\
& SMH-1   & 0.189  & 44,300\,(1,990)     & 0.0202\,(0.0047)      & 0.00501\,(0.00036) \\
& SMH-2   & 0.322  & 20,300\,(507)       & 1.01\,(0.15)          & 0.0548\,(0.0016) \\
& RWM     & 0.241  & 298,000             & 0.0118\,(0.0044)      & 0.00302\,(0.00011) \\
\bottomrule
\end{tabular}
\end{table}

\section{Discussion}
\label{section:discussion}
Posterior sampling via MCMC on datasets with a large number of observations poses a significant challenge, particularly to traditional methods like the Metropolis--Hastings (MH) algorithm, which requires the evaluation of all likelihood terms at each iteration. This has led Bayesian researchers to turn away from MCMC algorithms and to instead utilise approximate and biased inference schemes in order to approximate the posterior distribution. 
We have presented MH-SS, a novel exact Metropolis--Hastings-based algorithm designed to scale posterior sampling for large datasets through data subsampling. Our approach assumes that the log-likelihood differences can be tightly bounded and introduces an auxiliary Poisson variable which allows us to extend the Metropolis--Hastings ratio so that not all observations are needed, thereby reducing the computational effort per iteration.


Our approach shares features with the Tuna algorithm of \cite{zhang2020asymptotically}, using Poisson auxiliary random variables whose expectation depends on both the current and proposed value. However, as seen in the experiments and real-world applications of Sections \ref{section:simulation} and \ref{section:real_datasets}, Tuna typically performs relatively poorly because it requires sub-optimal scaling of the Metropolis--Hastings proposal in order to control the acceptance rate. Our more general formulation, combined with the use of control variates, leads to orders of magnitude greater efficiency. The scalable Metropolis--Hastings algorithm \citep{cornish2019scalable} uses the same generic form for the control variates as we do, but our bounds on the discrepancy between the control variate and the truth scale better with increasing dimension and this, together with the removal of the large product in the acceptance probability and another novel tightening, has a substantial impact on the overall efficiency. We have also provided guidelines on choosing the proposal scaling for MH-SS: choose $\lambda$ to target an empirical acceptance of around $0.452$. This guidance has been verified empirically and holds whether using first- or second-order control variates.


\CSadd{Although all of our worked examples are regression models with independent data, the methodology is not limited to such data. Consider regression models with data $y_{i,j}$ for the $i$th of $n>>1$ subjects, and a subject-specific random intercept (and, potentially, slope). MH-SS could subsample by subject and perform numerical integration over the random effects; since $n>>1$ including all random effects as latent variables could make the statespace too large for \emph{any} exact MCMC algorithm. Alternatively consider an AR($p$) process for data $y_{t}$, $t=p,\dots T$, which, given initial values of $y_{0},\dots, y_{p-1}$, are generated as:
$$
Y_{t}| (Y_{1:t-1}=y_{1:t-1})
\sim
\mathsf{Pois}(\log[1+\exp(\theta_0+\sum_{i=1}^{p}\theta_i y_{t-i})]),
~~~t=p,\dots,T.
$$
 This, and analogous binary- or robust- autoregressive models could be re-framed as standard regression model with a covariate vector of $x_t=(1,y_{t-p},\dots,y_{t-1})$.} 
  
\CSadd{We are grateful to an anonymous reviewer who pointed us to recent works \cite[]{salomone2020spectral,villani2024spectral} that transform a stationary time series to the frequency domain, creating the Whittle likelihood, a product of exponential densities, which would be amenable to MH-SS, provided suitable $c_i$ could be found. The accuracy of the Whittle likelihood depends on a central-limit-theorem argument, so MH-SS would be performing exact inference on an approximation to the true posterior, but this, nonetheless, is an interesting area for future research. In general, MH-SS can be applied to any model for which the bounds, $c_i$, exist and can be calculated in advance. The same applies to the SMH and TunaMH  methodologies in \cite{cornish2019scalable} and \cite{zhang2020asymptotically}, respectively.}

\CSadd{For some models, even with large amounts of data, difficulty in identifying one or more parameters might lead to a multi-modal posterior. In this case, before running the algorithm, one might gather information on each mode and, for each mode, create an appropriate control-variate function and bound. With some care, it is still possible to ensure detailed balance holds with respect to the true posterior. Appendix \ref{app.Multimodal} details this extension to MH-SS, and a simulation study that demonstrates the expected large increase in efficiency on a multimodal target.}

\bibliographystyle{apalike}
\bibliography{references}

\newpage
\begin{center}
{\large\bf SUPPLEMENTARY MATERIAL TO METROPOLIS--HASTINGS WITH SCALABLE SUBSAMPLING}
\end{center}

\appendix

\section{Proof of theoretical results for MH-SS}
\label{appendix_proofs}
\subsection{Proof that the acceptance ratio in \eqref{tuna_ratioNODA} is unbiased}

\label{sec.proofRatUnb}
Since $S_i\sim \mathsf{Pois}(\phi_i)$, directly from the probability generating function of a Poisson random variable, 
$$
\Expect{\left(\frac{\phi_i'}{\phi_i}\right)^{S_i}}
=
\exp\left\{\phi_i\left(\frac{\phi_i'}{\phi_i}-1\right)\right\}
=\exp\{\Delta_i\}.
$$
The result follows by the independence of the $S_i$ and substituting from \eqref{eqn.define.Delta}.

\subsection{Proof of Theorem \ref{thrm.optimal.gamma}}
\label{proof.thrm.optimal.gamma}

\emph{Proof:} Theorem \ref{thrm.optimal.gamma} relies on a preliminary result:

\begin{prop}\label{prop.monotone.E}
For any $c>0$, $\Delta \in \mathbb{R}$ and $\psi>0$, let
  $E(\psi):=\Expect{1\wedge c \left(\frac{\psi-\Delta}{\psi}\right)^S}$,
  where $S\sim \Poisson(\psi)$. Then $\frac{\md E}{\md \psi} \ge 0$.
\end{prop}

\emph{Proof of Proposition \ref{prop.monotone.E}:}
\[
E(\psi)
=
\exp(-\psi)\sum_{s=0}^\infty\frac{\psi^s}{s!}\left\{1\wedge c \left(\frac{\psi'}{\psi}\right)^s\right\}
=
\exp(-\psi)\sum_{s=0}^\infty\frac{1}{s!}\{\psi^s\wedge c(\psi-\Delta)^s\}.
\]
Now,
\[
a:=\frac{\md}{\md \psi}\left\{\psi^s\wedge c(\psi-\Delta)^s\right\}
=
s \psi^{s-1}1(\psi^s \le c(\psi-\Delta)^s)
+
sc (\psi-\Delta)^{s-1}1(\psi^s>  c(\psi-\Delta)^s).
\]
The quantity $a$ is either $s \psi^{s-1}$ or $sc (\psi-\Delta)^{s-1}$ so is at least as large as the minimum of the two. Hence, for $s>0$,
\[
\frac{\md}{\md \psi}\{\psi^s\wedge c(\psi-\Delta)^s\}
\ge s\{\psi^{s-1}\wedge c(\psi-\Delta)^{s-1}\}.
\]
Thus
\[
\frac{\md E}{\md \psi}
\ge
\exp(-\psi)\left[
  \sum_{s=1}^\infty\frac{s}{s!}\{\psi^{s-1}\wedge c(\psi-\Delta)^{s-1}\}
  -\sum_{s=0}^\infty\frac{1}{s!}\{\psi^{s}\wedge c(\psi-\Delta)^{s}\}
  \right]
=0.
\]
proving Proposition \ref{prop.monotone.E}. $\square$

The maximal value of $\psi$, therefore leads to the maximal value of $E(\psi)$.

Next, decompose the expectation over $S=(S_1,\dots,S_n)$ of the acceptance rate
\[
\Expect{\alpha_2(\theta,\theta';S_1,\dots,S_n)}
  =
  \Expect{\Expect{\Expect{\dots\Expect{\alpha_2(\theta,\theta';S_1,\dots,S_n)|S_2,\dots,S_n}\dots|S_{n-1},S_n}|S_n}}.
\]
Applying Proposition \ref{prop.monotone.E} recursively with $\psi=\phi_i$, $i=n,\dots,1$, shows that $\Expect{\alpha_2(\theta,\theta';S_1,\dots,S_n)}$ is maximised when $\psi$ is maximised since each of the intermediate expectations is maximised when $\psi$ is maximised.

Finally, defining $\Delta_i$ as in \eqref{eqn.define.Delta},
\[
\frac{\md }{\md \gamma}\phi_i
=
\max[0,\Delta_i]-\min[0,\Delta_i]-c_i M(\theta,\theta')
=|\Delta_i|-c_i M(\theta,\theta')\le 0,
\]
so $\Expect{\alpha_2(\theta,\theta',S)}$ is maximised when $\phi_i$ is maximised, which is at $\gamma=0$. $\square$

\subsection{Proof of Theorem \ref{theorem.remainders}}
\label{proof.theorem.remainders}

\subsubsection{Preliminaries}
\begin{lemma}[Mean-Value Theorem for $a:\mathbb{R}^d\to \mathbb{R}$] \label{lemma.MVMVT} Let $a: \mathbb{R}^d\to \mathbb{R}$ be once differentiable and let $x_A\in \mathbb{R}^d$ and $x_B\in \mathbb{R}^d$. Then $a(x_B)-a(x_A)=(x_B-x_A)^\top \nabla a(x')$ for some $x'$ on the line between $x_A$ and $x_B$.
\end{lemma}
\emph{Proof}: Let $x=(1-t)x_A+tx_B$ and set $b(t)=a(x(t))$. The Mean-Value Theorem for the univariate function $b$ gives $a(x_B)-a(x_A)=b(1)-b(0)=b'(t')$ for some $t'\in [0,1]$. But
\[
\frac{\md b}{\md t}
=
\frac{\md}{\md t}a\left((1-t)x_A+tx_B\right)
=
\sum_{i=1}^d \left.\frac{\md x_i}{\md t}\right|_{t=t'}~\left.\frac{\md a}{\md x_i}\right|_{t=t'} 
=
(x_B-x_A)^\top \nabla a(x'),
\]
where $x'=(1-t')x_A+t'x_B$. $\square$

Next, recall the \emph{trapezium rule}: $\int_a^b f(x) \md x= \frac{1}{2}\{f(a)+f(b)\}+\frac{1}{12}(b-a)^3f''(\xi)$, where $\xi\in [a,b]$. We will use a slightly different version.
\begin{prop}
  \label{prop.Trapezium}
If $f'(x)$ is Lipschitz with a Lipschitz constant of $L$, then
\[
\left|\int_a^b f(x) \md x - \frac{1}{2}(b-a)\{f(a)+f(b)\}\right|\le \frac{1}{12}(b-a)^3 L.
\]
\end{prop}

\emph{Proof}:
Let $I=\int_a^b(x-c)f'(x)\md x$, where $c=(a+b)/2$. Then, integrating by parts,
\[
I=[(x-c)f(x)]_a^b-\int_a^b f(x) \md x
=
\frac{1}{2}(b-a)\{f(a)+f(b)\}-\int_a^b f(x) \md x.
\]
However,
\[
\left| \int_a^b (x-c)f'(x) \md x\right|
=
\left| \int_a^b (x-c)\{f'(x)-f'(c)\} \md x\right|
\le
L\int_a^b (x-c)^2 \md x,
\]
which evaluates to the required expression. $\square$


\subsubsection{Proof of the theorem itself}
\emph{Proof}: We prove the implications \eqref{eqn.bounded.Hessian}$\Longrightarrow$\eqref{eqn.gen.rem.one} and \eqref{eqn.genLip}$\Longrightarrow$\eqref{eqn.gen.rem.two}; the other two results follow analogously.\\
\underline{Proof of \eqref{eqn.bounded.Hessian}$\Longrightarrow$\eqref{eqn.gen.rem.one}}: Lemma \ref{lemma.MVMVT} gives
$\ell(\theta')-\ell(\theta)=(\theta'-\theta)^\top\nabla \ell(\thetatil)$,
for some $\thetatil$ on the line between $\theta$ and $\theta'$. So
$\ell(\theta')-\ell(\theta)-r_1(\theta,\theta';\thetahat)=r'$, where 
$r':=(\theta'-\theta)^\top\left\{\nabla \ell(\thetatil)-\nabla \ell(\thetahat)\right\}$.

Let $a(x):=\{\nabla \ell(x)\}^\top(\theta'-\theta) $ and apply Lemma \ref{lemma.MVMVT} again, to see that
\[
r'=a(\thetatil)-a(\thetahat)=(\thetatil-\thetahat)^\top\nabla a(\theta^*)
=
(\thetatil-\thetahat)^\top H(\theta^*) (\theta'-\thetahat).
\]
for some $\theta^*$ on the line between $\thetahat$ and $\thetatil$. From \eqref{eqn.bounded.Hessian}, therefore,
$r' \le |x^\top (\theta'-\theta) |~|x^\top (\thetatil-\thetahat)|~K_1$.
Now, $\thetatil-\thetahat=\gamma (\theta-\thetahat) + (1-\gamma)(\theta'-\thetahat)$ for some $\gamma \in [0,1]$, so
\[
|x^\top (\thetatil-\thetahat)| =
\gamma |x ^\top (\theta-\thetahat)|
+(1-\gamma)|x^\top (\theta'-\thetahat)|
\le
\max\left(x ^\top (\theta-\thetahat),x ^\top (\theta'-\thetahat)\right),
\]
which leads to \eqref{eqn.gen.rem.one}.

\underline{Proof of \eqref{eqn.genLip}$\Longrightarrow$\eqref{eqn.gen.rem.two}}: Let $a=\theta$, $b=\theta'$ and $x\in [0,1]$. Then $f(z)=(\theta'-\theta)^\top g(\theta+z(\theta'-\theta))$ is the component in the $\theta'-\theta$ direction of the derivative of $\ell$ along the line joining $\theta$ to $\theta'$ and where $z$ gives the fraction of the total distance along the line. We also have
\[
f'(z) = (\theta'-\theta)^\top H\left(\theta + z\{\theta'-\theta\}\right)(\theta'-\theta),
\]
and from \eqref{eqn.genLip}
\[
|f'(z_2)-f'(z_1)|
\le
|x^\top (\theta'-\theta)|~|x^\top (\theta'-\theta)| ~|x^\top(\theta'-\theta)|~L_1~|z_2-z_1|.
\]
Thus $f'$ is Lipschitz. Proposition \ref{prop.Trapezium} applied with $a=0$ and $b=1$ then gives
\begin{equation}
  \label{eqn.applyTrapezium}
\left|\ell(\theta')-\ell(\theta)
-\frac{1}{2}(\theta'-\theta)
\left\{g(\theta)+g(\theta')
\right\}\right|
\le
\frac{L_1}{12} |x^\top (\theta'-\theta)|^3.
\end{equation}
Now, first order Taylor expansion of the scalar $(\theta'-\theta)^\top g(u)$ around $u=\thetahat$ gives
\begin{align*}
  (\theta'-\theta)^\top  g(\theta')
  &=
  (\theta'-\theta)^\top g(\thetahat)+
  (\theta'-\theta)^\top H(\thetahat)(\theta'-\thetahat)+(\theta'-\theta)^\top \{H(\eta')-H(\thetahat)\}(\theta'-\thetahat),\\
    (\theta'-\theta)^\top  g(\theta)
    &=
    (\theta'-\theta)^\top g(\thetahat)+(\theta'-\theta)^\top H(\thetahat)(\theta-\thetahat)+(\theta'-\theta)^\top \{H(\eta)-H(\thetahat)\}(\theta-\thetahat),
\end{align*}
where $\eta=t\theta+(1-t)\thetahat$ and $\eta'=t'\theta'+(1-t')\thetahat$ for $0\le t,t'\le 1$.
Thus
\[
\frac{1}{2}(\theta'-\theta)^\top\{g(\theta)+g(\theta')\}
-
r_2(\theta,\theta')=r'',
\]
where
\[
r''
:=
\frac{1}{2}(\theta'-\theta)\{H(\eta')-H(\thetahat)\}(\theta'-\thetahat)
+
\frac{1}{2}(\theta'-\theta)\{H(\eta)-H(\thetahat)\}(\theta-\thetahat).
\]
The triangle inequality combined with \eqref{eqn.genLip} then gives
\[
|r''|
\le
\frac{L_1}{2}|x^\top(\theta'-\theta)|+|x^\top(\theta-\thetahat)| ~|x^\top(\eta-\thetahat)| + \frac{L_1}{2}|x^\top(\theta'-\theta)|+|x^\top(\theta'-\thetahat)| ~|x^\top(\eta'-\thetahat)|.
\]
However, $\eta$ is on the line between $\theta$ and $\thetahat$, so $|x^\top(\eta-\thetahat)|\le |x^\top(\theta-\thetahat)|$, and similarly, $|x^\top(\eta'-\thetahat)|\le |x^\top(\theta'-\thetahat)|$. Combining with \eqref{eqn.applyTrapezium} via the triangle inequality gives \eqref{eqn.gen.rem.two}. $\square$

\subsection{Proof of Lemma \ref{lemma.dotprod}}
\label{proof.lemma.dotprod}
\emph{Proof}: Let $a>0$. If $u\cdot x<0$, set $u'\gets -u$ else $u'\gets u$. If $v\cdot x<0$ set $v'\gets -va||u||/||v||$ else $v'\gets av||u||/||v||$. Then, since the geometric mean is bounded by the arithmetic mean,
\[
\{(u'\cdot x)^k(v'\cdot x)\}^{1/(k+1)}
\le
\frac{1}{k+1}\{k u'\cdot x + v'\cdot x\}
=
\frac{1}{k+1}(ku'+v')\cdot x
\le
\frac{1}{k+1}||ku'+v'||~||x||,
\]
by the Cauchy-Schwarz inequality. For any vector $z$, define $\zhat:=z/\|z\|$. Thus,
\begin{align*}
\left\{(u'\cdot x)^k(v'\cdot x)\right\}^{2/(k+1)}
&\le
\frac{1}{(k+1)^{2}}\left\{k^2||u'||^2+||v'||^2+2k u'\cdot v'\right\}||x||^2\\
&=
\frac{\|u\|^2}{(k+1)^{2}}\left\{k^2+a^2+2ka \uhat' \cdot \vhat'\right\}\\
&\le
\frac{\|u\|^2}{(k+1)^{2}}\left\{k^2+a^2+2ka |\uhat' \cdot \vhat'|\right\},
\end{align*}
where the penultimate line follows as $||v'||=a||u'||$. However,
\[
(u'\cdot x)^k(v'\cdot x)
=
|u'\cdot x|^k|v'\cdot x|
=
a\frac{\|u\|}{\|v\|}|u\cdot x|^k|v\cdot x|
\]
and $|\uhat'\cdot \vhat'|=|\uhat\cdot \vhat|=|\omega|$, so
\[
\left\{|u\cdot x|^k|v\cdot x|\right\}^{2/(k+1)}
\le
\|u\|^{2k/(k+1)}\|v\|^{2/(k+1)}\|x\|^2 \frac{h_k(a)}{(k+1)^2},
\]
where
\[
h_k(a):=\frac{k^2+a^2+2ka|\omega|}{a^{2/(k+1)}}.
\]
This is true for all $a$, and so is true at the minimiser of $h_k(a)$, which solves $a^2+(k-1)a|\omega|=k$ and is $a=a_k(\omega)$ as defined in the statement of the lemma. At $a=a_k(\omega)$,
\[
k^2+a^2+2ka|\omega|
=
k^2+(k+1) a_k(\omega)|\omega| +k
=
(k+1)\left\{k+a_k(\omega)|\omega|\right\}.
\]
Hence, the right hand side of the inequality is 
\[
\|u\|^{2k/(k+1)}\|v\|^{2/(k+1)}\|x\|^2 \frac{h_k(a_k(\omega))}{(k+1)^2}
=
\|u\|^{2k/(k+1)}\|v\|^{2/(k+1)}\|x\|^2 \frac{k+a_k(\omega)|\omega|}{(k+1)a_k(\omega)^{2/(k+1)}}.
\]
Raising to the power of $(k+1)/2$ gives the required result. $\square$

\subsection{Specifics for regression models}
\label{section:specifics_regression}

\begin{theorem} \label{thrm.regn.model.bounds}With a single observation $y$, for logistic regression $|\hell''(\eta)|\le K_{logit}(y)=1/4$ and $|\hell'''(\eta)|\le L_{logit}(y)=\sqrt{3}/18$, for probit regression $|\hell''(\eta)|\le K_{probit}(y)=1$, and for Poisson regression with an expectation of $\log\left(1+\lambda \exp(\eta)\right)$, $|\hell''(\eta)|\le K_{Pois}(y)= 0.25+0.168 y$.
\end{theorem}

\begin{remark}
Plotting $\hell'''(\eta)$ for probit regression and the relevant portion of $\hell'''(\eta)$ for Poisson regression over a fine grid shows that for probit regression, $|\hell'''(\eta)|< L_{probit}(y)=0.30$ and for Poisson regression, $|\hell'''(\eta)|<L_{Pois}(y)={\sqrt{3}}/{18}+0.061y$; we do not formally prove these bounds.
\end{remark}

\begin{remark}
    \CSadd{The bounds for robust regression quoted in the main text arise from derivative bounds derived in \cite[][Appendix G.2]{cornish2019scalable}.}
\end{remark}

\emph{Proof}: \underline{Logistic regression},
$\hell(\eta)=y\log[p/(1-p)]+\log(1-p)=y\eta-\log(1+\exp(\eta))$, so $\hell'(\eta)=y-\frac{\exp(\eta)}{1+\exp(\eta)}$ and
\[
-\hell''(\eta)=\frac{\exp(\eta)}{\{1+\exp(\eta)\}^2}=p(1-p).
\]
Hence, $0\le -\hell''(\eta)\le 1/4$. Furthermore
\[
\hell'''(\eta) = \frac{\exp(\eta)-\exp(2\eta)}{\{1+\exp(\eta)\}^3}.
\]
This is maximised at $\exp(\eta)=2-\sqrt{3}$ and minimised at $\exp(\eta)=2+\sqrt{3}$. Substitution gives $-\sqrt{3}/18\le \hell'''(\eta)\le \sqrt{3}/18$.
$\square$

\emph{Aside}: Notice that $|\hell'(\eta)|=|y-p|\le 1$. This leads to the Tuna bound (no control variates).

\underline{Probit regression}, $\hell(\eta)=y \log \Phi(\eta) + (1-y)\log \Phi(-\eta)$,
\begin{align*}
\hell'(\eta)&= y \frac{\phi(\eta)}{\Phi(\eta)}-(1-y) \frac{\phi(\eta)}{\Phi(-\eta)},\\
-\hell''(\eta)&= y\phi(\eta)\frac{\phi(\eta)+\eta\Phi(\eta)}{\Phi(\eta)^2}
+(1-y)\phi(\eta)\frac{\phi(\eta)-\eta\Phi(-\eta)}{\Phi(-\eta)^2}.
\end{align*}
For $\hell''(\eta)$, $(y,\eta)\leftrightarrow (1-y,-\eta)$ switches the first and second terms, and only one of these terms is ever non-zero, so it is sufficient to bound the first term.
$$
T(\eta)=\phi(\eta)\frac{\phi(\eta)+\eta\Phi(\eta)}{\Phi(\eta)^2}.
$$

We now show that $0\le T(\eta)\le 1$ by showing that it is decreasing and finding the limits at $-\infty$ and $\infty$; from this, $0\le -\hell''(\eta)\le 1$.

Firstly, set $g(\eta)=\phi(\eta)/\Phi(\eta)\ge 0$. Calculus gives:
\[
T'=\frac{\phi(\eta)}{\Phi(\eta)^3}\left[1-(\phi(\eta) +\eta \Phi(\eta))(2\phi(\eta)+\eta \Phi(\eta))\right]
=
g(\eta)\left[1-(g(\eta)+\eta)(2g(\eta)+\eta)\right].
\]
Equation (4) and Section 3 of \cite{sampford1953some} shows that $g(-\eta)[(g(-\eta)-\eta)(2g(-\eta)-\eta)-1]>0$ for all $\eta$; hence $T'(\eta)<0$ for all $\eta$.

Next, we write $T(\eta)=T_1(\eta)T_2(\eta)$, where $T_1(\eta)=\frac{\phi(\eta)/\eta}{\Phi(\eta)}$ and $T_2(\eta)=\frac{\eta\phi(\eta)+\eta^2\Phi(\eta)}{\Phi(\eta)}$. Applying l'Hopital's rule to $T_1$ gives
\[
\lim_{\eta \to -\infty}T_1(\eta)
=
\lim_{\eta \to -\infty} \frac{-\phi(\eta)-\phi(\eta)/\eta^2}{\phi(\eta)}=-1.
\]
Applying l'Hopital to $T_2$ gives
\[
\lim_{\eta \to -\infty}T_2(\eta)
=
\lim_{\eta \to -\infty} \frac{-\eta^2\phi(\eta)+\phi(\eta)+\eta^2\Phi(\eta)+2\eta\Phi(\eta)}{\phi(\eta)}
=
1+2\lim_{\eta \to -\infty}\frac{1}{T_1(\eta)}=-1.
\]
So $\lim_{\eta\to -\infty}T(\eta)=1$.

Finally, as $\eta\to \infty$, $\Phi(\eta)\to 1$ and both $\phi(\eta)$ and $\eta \phi(\eta)\to 0$, so $T(\eta)\to 0$. $\square$

\emph{Note}: \cite{bohning1999lower} gives the same bound, but does not prove the monotonicity between the limits as $\eta \to \pm \infty$.

\underline{Poisson regression}: Setting $\eta'=\eta+\log \lambda$, $\md \hell/\md \eta=\md \hell /\md \eta' \times \md \eta'/\md \eta= \md \hell/ \md \eta'$, so without loss of generality we may set $\lambda=1$. Now, $\hell=y\log \log [1+\exp(\eta)]-\log(1+\exp(\eta)]-\log(y!)$, 
\begin{align}
\label{eqn.Pois.helldash}
\hell'(\eta)
&=
y \frac{1}{\log[1+\exp(\eta)]}\frac{\exp(\eta)}{1+\exp(\eta)}-\frac{\exp(\eta)}{1+\exp(\eta)},\\
\label{eqn.Pois.helldashdash}
-\hell''(\eta)
&=
\frac{\exp(\eta)}{\{1+\exp(\eta)\}^2}
\left[y\left\{\frac{\exp(\eta)}{\{\log(1+\exp(\eta))\}^2}-\frac{1}{\log(1+\exp(\eta))}\right\}+1\right].
\end{align}

Set
\[
0\le S(\eta):=\frac{\exp(\eta)}{\{1+\exp(\eta)\}\log(1+\exp(\eta))}\le 1,
\]
as $\log(1+x)\ge x/(x+1)$ for all $x>-1$. The coefficient of $y$ in $-\hell''(\eta)$ can be written as $S(\eta)R(\eta)$, where
\[R(\eta)=S(\eta)-1/(1+\exp(\eta))\ge 0,\]
since $\log(1+x)\le x$.

Firstly, by Taylor expansion, $\log(1+\exp(x))\ge \exp(x)-\frac{1}{2}\exp(2x)$, so, for $\eta<0$,
\[
R(\eta)\le \frac{1}{\{1+\exp(\eta)\}\{1-\frac{1}{2}\exp(\eta)\}}-\frac{1}{1+\exp(\eta)}
=
\frac{\exp(\eta)}{2+\exp(\eta)-\exp(2\eta)}\le \frac{1}{2}\exp(\eta).
\]
Thus, for $\eta\le -\log 3$, $S(\eta)R(\eta)\le R(\eta)\le R(-\log 3)\le 1/6$.

Secondly, $\log(1+x)>\log(x)$, so
\[
R(\eta)\le \frac{\exp(\eta)}{\eta\{1+\exp(\eta)\}}-\frac{1}{1+\exp(\eta)}
=
\frac{1}{\eta\{1+\exp(-\eta)\}}-\frac{1}{1+\exp(\eta)}\\
\le \frac{1}{\eta}.
\]
Hence, for $\eta\ge 6$, $S(\eta)R(\eta)\le R(\eta)\le R(6)\le 1/6$.

Next, we numerically maximise $S(\eta)R(\eta)$ on the interval $[-\log 3,6]$ to obtain $SR<0.168$.

Finally, the term that does not involve $y$ is positive, with a magnitude of at most $1/4$, which leads to the bound.

\emph{Aside}: To obtain the bounds for the Tuna algorithm \citep{zhang2020asymptotically}, it is needed to bound $|\hell'(\eta)|$. Applying l'Hopital's rule to the coefficient of $y_i$ in $\hell'(\eta)$ \eqref{eqn.Pois.helldash}, $S(\eta)$, gives
\begin{equation}
\lim_{\eta \to \infty}S(\eta) = 0
~~~\mbox{and}~~~
\lim_{\eta \to -\infty}S(\eta) = 1, \nonumber
\end{equation}
and the limits of the second term in $\hell'(\eta)$ are $0$ and $-1$ as $\eta$ goes to $-\infty$ and $\infty$, respectively. So, 
\begin{equation}
\lim_{\eta \to \infty}\hell'(\eta) = -1
~~~\mbox{and}~~~
\lim_{\eta \to -\infty}\hell'(\eta) = y. \nonumber
\end{equation}
Using $\log(1+\exp(\eta))<\exp(\eta)$ for all $\eta$, the coefficient of $y_i$ in \eqref{eqn.Pois.helldashdash} is always positive. Thus, $\hell''(\eta)<0$ for all $\eta$ and $\hell'(\eta)$ is decreasing and hence lies between its limits, $-1$ and $y$. So, $|\hell'(\eta)| \leq \mbox{max}(1,y)$. $\square$

\section{Tuning the MH-SS proposal, and alternatives}
\label{section:tuning_guidelines}

In this section, we provide guidelines for the choice of the scaling parameter $\lambda$ in the special case of a random-walk Metropolis proposal \eqref{eqn.gen.RWM.prop}.
In our analysis, we set $V\equiv V_d = -[H(\hat{\theta})]^{-1}$, where $[H(\hat{\theta})]^{-1}$ represents the inverse of the Hessian at an estimate $\widehat{\theta}$ near the mode. In the large-$n$ Bernstein-von Mises limit, this is equivalent to exploring a $d$-dimensional standard Gaussian posterior for $\psi= \sqrt{V_d}^{-1}\theta$, with a proposal of $\psi'\sim \Normal\left(\psi,(\lambda^2/d) I_d\right)$ and, as such, fits squarely within the framework of previous analyses of the high-dimensional limit \cite[]{gelman1997weak, roberts2001optimal, sherlock2009optimal, SheThiGol2021}. 

\cite{SheThiGol2021} investigate the tuning of delayed-acceptance random walk Metropolis algorithms. However, while the MH-SS algorithms do have a delayed-acceptance component, they have two specific properties that, respectively aid and necessitate a new treatment: the extreme accuracy/inaccuracy of the approximate target and an approximately linear relationship between the proposal scaling and the computational cost. 

The computational cost per iteration is approximately proportional to the number of data points used and the expectation of this is proportional to $M(\theta,\theta')$. With the proposal \eqref{eqn.gen.RWM.prop}, $\|\psi'-\psi\|_2=\Theta(1)$, whereas both $\|\psi-\widehat{\psi}\|_2$ and $\|\psi'-\widehat{\psi}\|_2$ are $\Theta(d^{1/2})$, where $\widehat{\psi}=\sqrt{V_d}^{-1}\widehat{\theta}$. Provided $V_d$ is well-conditioned, $\|\theta'-\theta\|_2$ is, therefore, also a factor of $d^{1/2}$ smaller than either $\|\theta-\thetahat\|_2$ or $\|\theta'-\thetahat\|_2$; see also the analysis in Appendix \ref{section:asymptotic_analysis_bounds}. Theorem \ref{theorem.remainders} then shows that for both first- and second-order control variates, $M(\theta,\theta')\propto \lambda$. This justifies the more formal:

\begin{assumption}
\label{ass.scaling.A}
The computational cost per iteration of both MH-SS-1 and MH-SS-2 is proportional to $\lambda$.
\end{assumption}

\begin{figure}[htbp]
\begin{center}
   \begin{subfigure}{0.52\textwidth}
        \caption{}
          \includegraphics[scale=0.37]{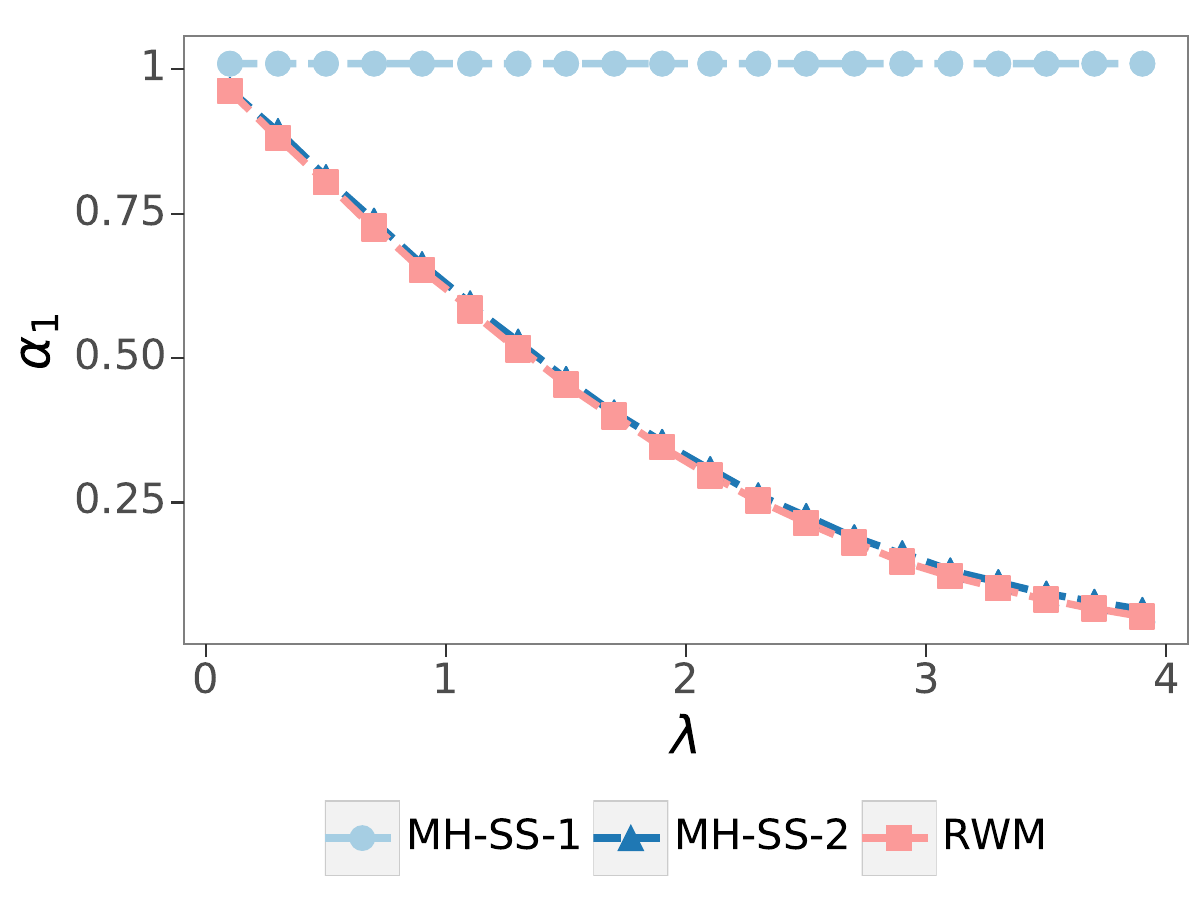}
   \end{subfigure}
   \begin{subfigure}{0.47\textwidth} 
        \caption{}
          \includegraphics[scale=0.37]{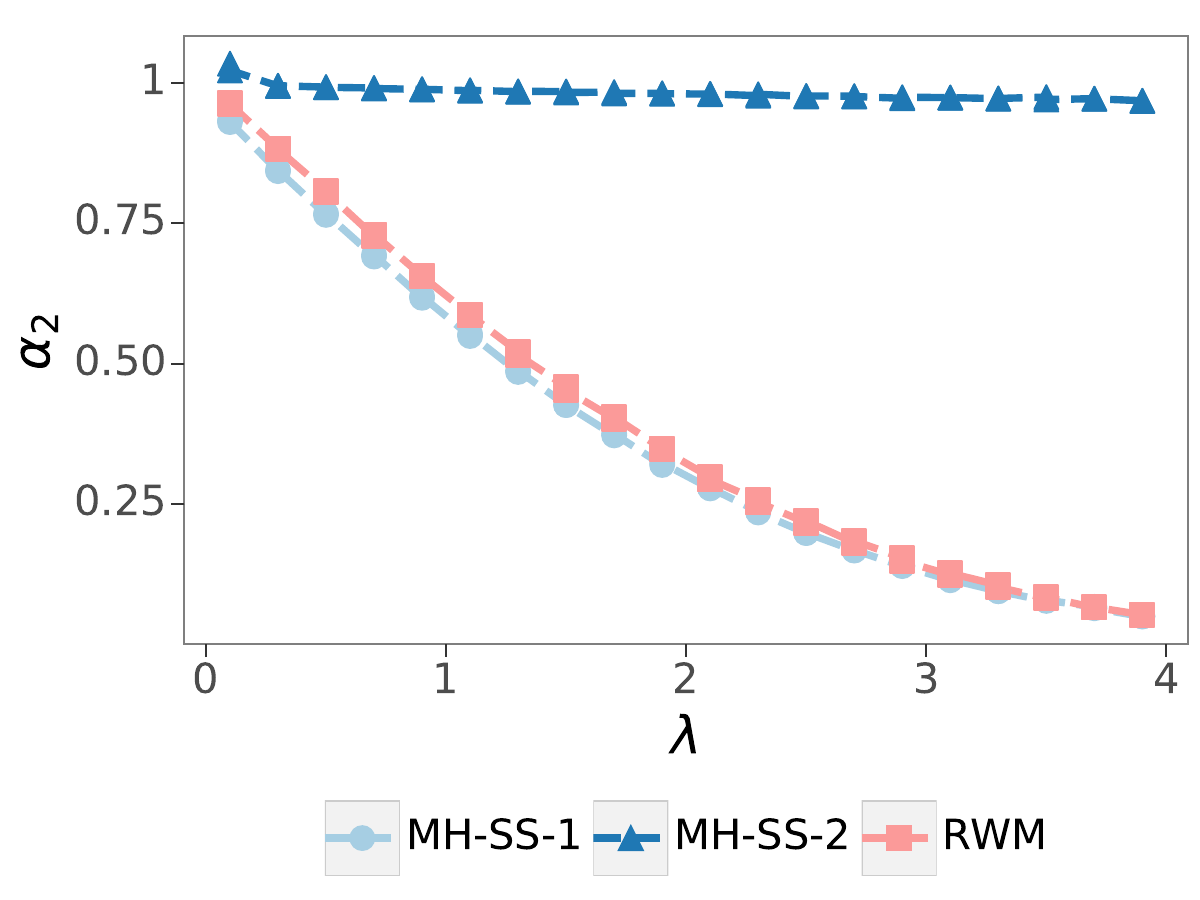}
   \end{subfigure}
   \vspace{-1cm}
\caption{Acceptance rates: $\alpha_1$ in \eqref{eqn.StageOneAcc} on the left and $\alpha_2$ in \eqref{eqn.StageTwoAcc} on the right for a logistic regression target in dimension $d=100$ with $n=30,000$ observations, and with covariates and true coefficients simulated as described in Section \ref{section:simulation}.}
\label{fig:optimal_scaling_acc_rates_alphas_1_and_2}
\end{center}
\end{figure}

In Appendix \ref{section:optimal_scaling}, we argue for the following assumption, which we have observed in all of our empirical studies (see, for example, Figure \ref{fig:optimal_scaling_acc_rates_alphas_1_and_2}):

\begin{assumption}
\label{ass.scaling.B}
For first-order control variates, $\Expect{\alpha_1(\theta,\theta')}\approx 1$ and for values of $\theta'$ that pass the screening stage, $\Expect{\alpha_{2, MHSS}(\theta,\theta')}\approx \Expect{\alpha_{RWM}(\theta,\theta')}$.
For second-order control variates, $\Expect{\alpha_1(\theta,\theta')}\approx \Expect{\alpha_{RWM}(\theta,\theta')}$ and for values of $\theta'$ that pass the screening stage, $\Expect{\alpha_{2,MHSS}(\theta,\theta')}\approx 1$.
\end{assumption}

Thus, whichever form of control variate is used, we anticipate that the expected acceptance rate for a given scaling will be approximately the same as the expected acceptance rate for the RWM algorithm with that scaling. High-dimensional asymptotics for the Metropolis--Hastings algorithm with a RWM proposal \eqref{eqn.gen.RWM.prop} \cite[e.g.,][]{gelman1997weak, roberts2001optimal, sherlock2009optimal} provide a limiting expected acceptance rate  of  $2\Phi(-J\lambda/2)$ for some constant, $J$, that depends on the roughness of the posterior; with our choice of $V_d$, $J=1$. The expected squared jumping distance of the $i$th component of $\theta$ is
\[
\ESJD(\lambda)=\Expect{\left(\theta^{(j)}_i-\theta^{(j+1)}_i\right)^2},
\]
where $\theta^{(j)}\sim \pi$ is the $j$th value in the MCMC chain and $\theta^{(j+1)}$ is the $j+1$th. This is a natural measure of the efficiency of the algorithm and, in the case of a limiting diffusion, is exactly the right measure of the efficiency in the high-dimensional limit as it is proportional to the speed of the limiting diffusion. In this limit, the ESJD is proportional to the product of the expected acceptance probability and the square of the scaling parameter, $\ESJD\propto 2\lambda^2 \Phi(-\lambda/2)$. Maximising this quantity leads to the well-known optimal acceptance rate of $0.234$ of the RWM algorithm.

Given Assumption \ref{ass.scaling.A}, however, the overall efficiency for MH-SS algorithms is proportional to
\[
\frac{\ESJD(\lambda)}{\lambda}
\propto 2\lambda \Phi(-\lambda/2).
\]
This is maximised at $\lambda_{opt}\approx 1.50$ which gives an acceptance rate of $2\Phi(-\lambda_{opt}/2)\approx 0.452$. 

\begin{figure}[htbp]
\begin{center}
   \begin{subfigure}{0.52\textwidth}
        \caption{}
          \includegraphics[scale=0.37]{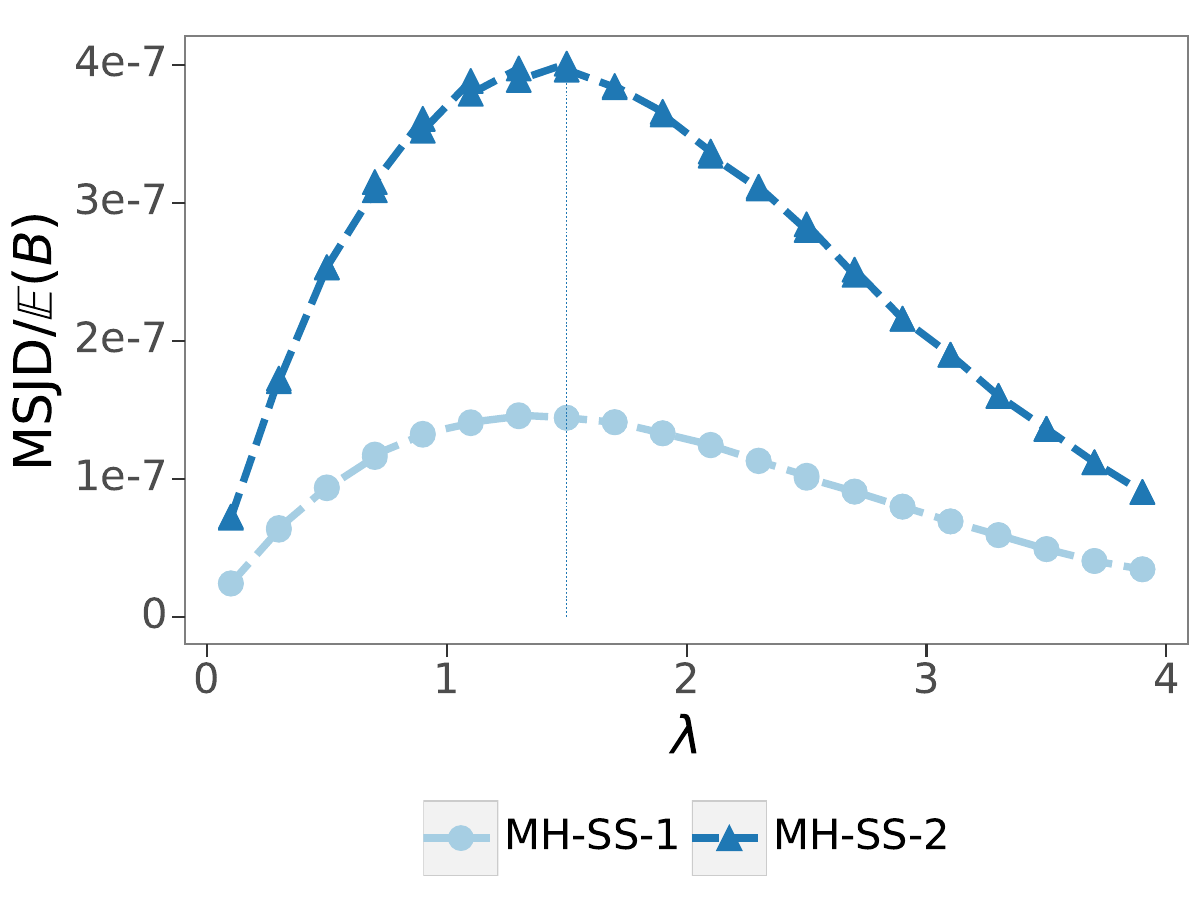}
   \end{subfigure}
   \begin{subfigure}{0.47\textwidth} 
        \caption{}
          \includegraphics[scale=0.37]{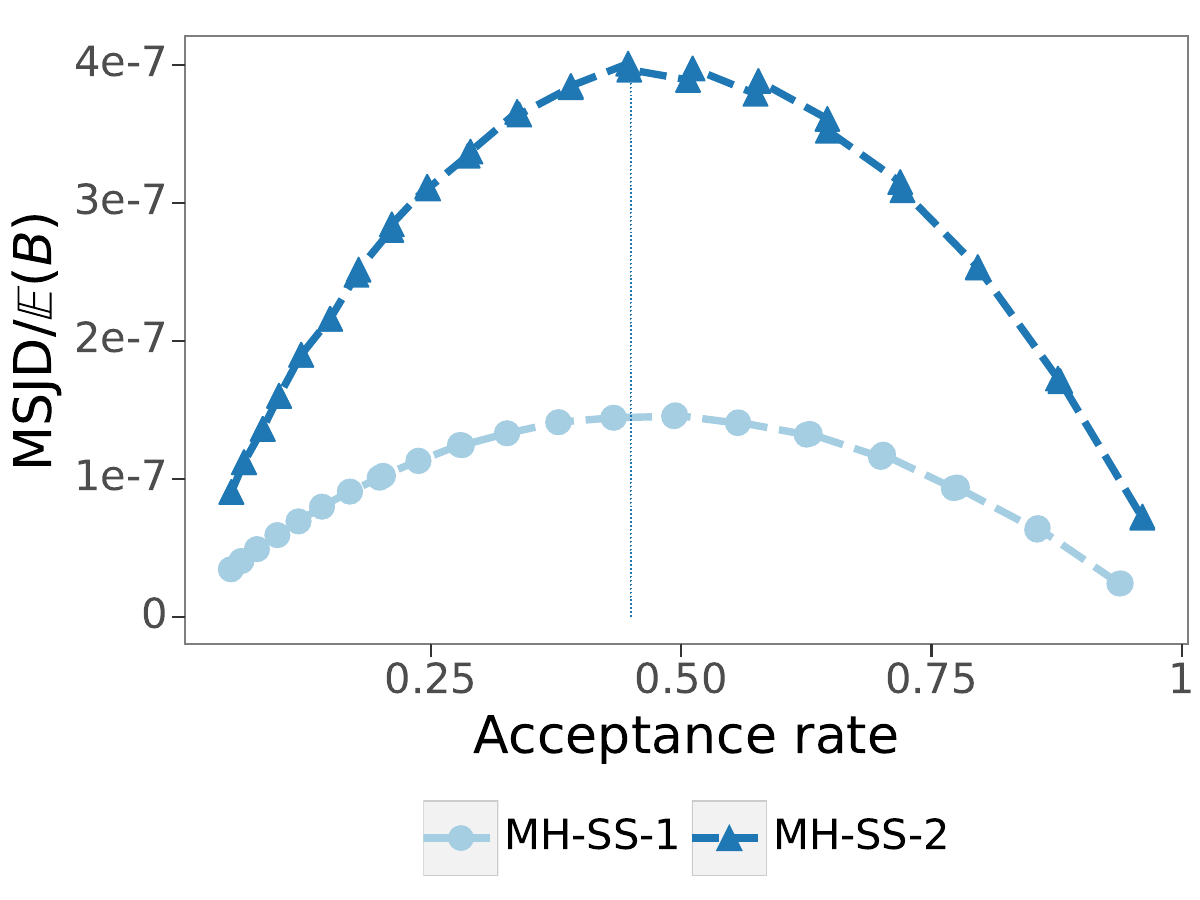}
   \end{subfigure}
\vspace{-1cm}
\caption{Optimal scaling for the MH-SS algorithms based on synthetic data from a logistic regression model with $n=30,000$ observations and $d=100$ covariate parameters, including an intercept. The efficiency metric (MSJD/$\mathbb{E}(B)$) is plotted as function of (a) the scaling parameter ($\lambda$) and (b) the empirical acceptance rate. 
}
\label{fig:optimal_scaling_acc_rate}
\end{center}
\end{figure}

We found this advice to hold true in all of our empirical studies, an example of which is given in Figure \ref{fig:optimal_scaling_acc_rate}.

The same argument applies to any algorithm that uses our control variates, provided that for any reasonable scaling, $\lambda$, the overall acceptance rate is close to that for the RWM. The SMH  bounds used in \cite{cornish2019scalable} are reproduced in Appendix \ref{sec.SMHbounds} and have only a weak relative dependence on $\lambda$. Consequently, we might expect SMH-2 to be optimally scaled when the acceptance rate is around $0.234$. Figure \ref{fig:additiona_results} in Appendix \ref{section:additional_simulation} verifies both of these points for a particular logistic regression scenario. It also shows that the behaviour of the SMH-1 acceptance rate on $\lambda$ is very different to that of the RWM. We find that the scaling must decrease more quickly with dimension than $\lambda/\sqrt{d}$ to maintain a reasonable acceptance rate as dimension increases. Although there is no obviously optimal acceptance rate that we might expect to hold across different scenarios, we observed empirically through simulation experiments with various numbers of observations $n$ (from $30,000$ to $100,000$) and dimensions $d$ (from $10$ to $100$) that the optimal scaling parameter for SMH-1 is in the range between $0.5$ to $1.5$. Furthermore, for each combination of $n$ and $d$, a scaling of $\lambda=1$ achieved at least $80\%$ of the optimal efficiency for that combination. Consequently, in Sections \ref{section:simulation} (simulation experiments) and \ref{section:real_datasets} (real-world applications), we set $\lambda = 1$ for SMH-1. 

\subsection{Alternative MH Proposals}
\label{app.general.proposals}
For random walk-based proposals, $q$, such as $\theta'\sim \mathsf{N}(\theta,V)$ for some variance matrix $V$, or for the Crank-Nicholson proposal with pre-conditioning based on $\widehat{\pi}$ that is suggested in \cite{cornish2019scalable},  Algorithm \ref{mh_ss_algorithm} can be applied as written. For proposals, such as that of the Metropolis-adjusted Langevin algorithm (MALA), that use $\nabla \log \pi(\theta)$, naively, the costs of proposing $\theta'$ and of evaluating both $q(\theta'|\theta)$ and $q(\theta|\theta')$ are all $\mathcal{O}(n)$. In this case a cheap, unbiased low-variance estimator of $\nabla \log \pi$ at both the proposed and current parameter values can be obtained by subsampling and using control variates; for example: 
\[
\widehat{\nabla}\log \pi(\theta)
=
\sum_{i=1}^n g_i(\thetahat) + \frac{n}{m}\sum_{j=1}^m \{g_{i_j}(\theta)-g_{i_j}(\thetahat)\}, 
\]
where $i_1,\dots,i_m$ are the indices of the random subsample chosen at this iteration for estimating the gradients and $\sum_{i=1}^n g_i(\thetahat)$ is pre-computed.

\section{Asymptotic analysis of bounds}
\label{section:asymptotic_analysis_bounds}

We first analyse the tightness of the bounds from Theorem \ref{theorem.remainders} and then incorporate the computational cost of evaluating the bounds under the assumption that the cost of evaluating the likelihood $\ell_i(\theta)$ is $\mathcal{O}(d)$. We compare the costs to those of the SMH algorithm \citep{cornish2019scalable}.

In regression examples with each $\beta_{j}\sim \mathsf{N}(0,1)$, $j=1,\dots,d$, independently, we sampled covariates
\begin{equation}
  \label{eqn.xDist}
  x_{i,j}\sim\mathsf{N}(0,1/d),~i=1,\dots,n,~j=2,\dots,d,
\end{equation}
independently, with $x_{i,1}=1$. The variance was chosen so that the linear predictor, $x_i^\top\beta$ remains $\mathcal{O}(1)$ as $d$ increases.

We find that the overall costs of evaluating our algorithm's accept-reject probability using $r^{(1)}$ is $\mathcal{O}(d^{3/2})$, which is independent of $n$, whereas for $r^{(2)}$ the cost is $\mathcal{O}(d^3 n^{-1/2})$. The cost of the SMH algorithm depends on the distribution of the covariates used. If (as in our simulations) these are Gaussian, then each iteration of the SMH-1 algorithm costs $\mathcal{O}\left(d^2\log d\right)$, and for SMH-2 the cost is $\mathcal{O}\left(d^{7/2}\{\log d\}^{3/2}n^{-1/2}\right)$; the terms in $\log d$ vanish if the covariates are bounded and the terms become larger if the covariates have heavier-than-Gaussian tails. For reference, the RWM costs $\mathcal{O}(dn)$: reductions in the cost per-data point come at the expense of increasing cost with dimension.

\textbf{Assumptions}: We assume the following:
\begin{itemize}
    \item 
The Bernstein-von Mises limit applies and so
\begin{equation}
  \label{eqn.BVM}
  \theta \sim \mathsf{N}\left(\thetatil,\frac{1}{n} \IE^d(\theta_*)^{-1}\right),
\end{equation}
where $\theta_*$ is the true parameter value and $\thetatil$ is the posterior mode. In regression models, for example, if the covariate vector $X_i\sim \nu$, for some distribution, $\nu$, then $\IE^d(\theta_*)=\Expects{X\sim \nu}{\IE^d(\theta_*;X)}$, the expectation over covariate vectors of the expected Fisher information conditional on the covariates. 
\item We consider the RWM, with a proposal of $\theta'\sim \mathsf{N}(\theta,\lambda^2_{n,d}V^d)$ for some fixed (as $n$ increases) positive-definite matrix, $V^d$.
\item We assume that $\Sigma^d:=\IE^d(\theta_*)^{-1}$ is well behaved as $d\to \infty$, satisfying the conditions in \eqref{eqn.traceIE} and that the user ensures that $V^d$ is similarly well-behaved, satisfying \eqref{eqn.traceVd}:
\begin{align}
  \frac{1}{d}\mathsf{trace}(\Sigma^d)\to t_\pi<\infty,
  &
  ~\frac{1}{d}\sum_{i=1}^d \left(\Sigma^d_{i,i}\right)^{1/2}\to t^*_{\pi}>0
   \label{eqn.traceIE}
~~~\mbox{and}~~~
  \frac{1}{d^2}\sum_{i=1,j=1}^{d,d}\left(\Sigma^d_{i,j}\right)^2 \to 0,\\
  \label{eqn.traceVd}
  \frac{1}{d}\mathsf{trace}(V^d)\to t_v<\infty,
  &
  ~\frac{1}{d}\sum_{i=1}^d \left(V^d_{i,i}\right)^{1/2}\to t^*_v>0
  ~~~\mbox{and}~~~
  \frac{1}{d^2}\sum_{i=1,j=1}^{d,d}(V^d_{i,j})^2 \to 0.
\end{align}
\item To keep the acceptance rate bounded as $n$ and $d$ become large, we choose
\begin{equation}\label{eq.RWMscaling}
\lambda_{n,d}=\frac{\ell}{n^{1/2}d^{1/2}}.
  \end{equation}
The function of $n$ follows directly from \eqref{eqn.BVM}, whereas the function of $d$ follows from \cite{gelman1997weak} and the many subsequent articles on optimal scaling of the random walk Metropolis algorithm.
\item We assume that $\thetahat$, the approximation to the mode, found via a stochastic-gradient algorithm, satisfies:
\begin{equation}
\label{eqn.thetatilthetahat}
  ||\thetahat-\thetatil||_2\le c~\frac{d^{1/2}}{n^{1/2}},
\end{equation}
for some $c<t_{\pi}$ and not depending on $n$ or $d$. This assumption is weak: in practice, for a reasonable approximation to the mode, we would expect $c<<t_{\pi}$.
\end{itemize}

\begin{prop}
  \label{prop.Sizes}
  Subject to assumptions \eqref{eqn.BVM}, \eqref{eqn.traceIE}, \eqref{eqn.traceVd}, \eqref{eq.RWMscaling} and \eqref{eqn.thetatilthetahat}:
  \begin{align*}
  \|\theta'-\theta\|_2=\Theta(n^{-1/2}),
  ~~~
  \|\theta-\thetahat\|_2=\Theta(d^{1/2}n^{-1/2})&
  ~~~\mbox{and}~~~
    \|\theta'-\thetahat\|_2=\Theta(d^{1/2}n^{-1/2}),\\
  \Expect{\|\theta'-\theta\|_1}=\Theta(d^{1/2}n^{-1/2}),
  ~~~
  \Expect{\|\theta-\thetahat\|_1}=\Theta(dn^{-1/2})&
  ~~~\mbox{and}~~~
    \Expect{\|\theta'-\thetahat\|_1}=\Theta(dn^{-1/2}),\\
     \|\theta'-\theta\|_1=\mathcal{O}(d^{1/2}n^{-1/2}),
  ~~~
  \|\theta-\thetahat\|_1=\mathcal{O}(dn^{-1/2})&
  ~~~\mbox{and}~~~
    \|\theta'-\thetahat\|_1=\mathcal{O}(dn^{-1/2}), 
  \end{align*}
  \CSadd{where results in the first and last line hold in probability.}
\end{prop}
\emph{Proof}:
We will use the following repeatedly: for any vector $A\sim \mathsf{N}(0,W)$ for some positive definite matrix $W$: 
\[
\Expect{\sum_{i=1}^dA_i^2}=\mathsf{trace}(W),
~
\Var{\sum_{i=1}^dA_i^2}=2\sum_{i,j=1}^{d,d} W_{i,j}^2,
~
\Expect{\sum_{i=1}^d|A_i|}=\sqrt{\frac{2}{\pi}}\sum_{i=1}^d \sqrt{W_{i,i}}.
\]
From \eqref{eqn.traceVd} and \eqref{eq.RWMscaling}, as $d\to \infty$, for any fixed $n$,
\begin{align*}
\Expect{n||\theta'-\theta||_2^2}&= n\lambda^2_{n,d} \mathsf{trace}(V_d)\to \ell^2 t_v,\\
\Var{n||\theta'-\theta||_2^2}&
=2n^2 \lambda_{n,d}^4\sum_{i,j=1}^n V_{d,i,j}^2 \to 0,
\end{align*}
so $n ||\theta'-\theta||_2^2\to \ell^2 t_v$ \CSadd{in probability}, giving the bound for $\|\theta'-\theta\|_2$. Analogously, \CSadd{from \eqref{eqn.traceIE}, $\Expect{n\|\theta-\thetatil\|^2/d}=t_{\pi}$ and}  $n\|\theta-\thetatil\|_2^2/d\to t_\pi$ \CSadd{in probability}. By the triangle inequality,
\[
\frac{n^{1/2}}{d^{1/2}}\|\theta-\thetatil\|_2
-
\frac{n^{1/2}}{d^{1/2}}\|\thetatil-\thetahat\|_2
\le
\frac{n^{1/2}}{d^{1/2}}\|\theta-\thetahat\|_2
\le  
\frac{n^{1/2}}{d^{1/2}}\|\theta-\thetatil\|_2
+
\frac{n^{1/2}}{d^{1/2}}\|\thetatil-\thetahat\|_2.
\]
In the limit as $d\to \infty$, and subject to \eqref{eqn.traceIE} and \eqref{eqn.thetatilthetahat}, we obtain
\[
t_{\pi}-c \le
\frac{n^{1/2}}{d^{1/2}}\|\theta-\thetahat\|_2
\le t_{\pi}+c,
\]
giving the bounds for $\|\theta-\thetahat\|_2$.
The triangle inequality then gives 
\[
\|\theta-\thetahat\|_2
-
\|\theta'-\theta\|_2
\le
\|\theta'-\thetahat\|_2
\le
\|\theta-\thetahat\|_2
+
\|\theta'-\theta\|_2,\]
giving the bounds for $\|\theta'-\thetahat\|_2$. \CSadd{Thus, we have proved the sizes stated in the first line of the proposition.}

For the $L_1$ norms, by Jensen's inequality, 
\[\|A\|_1=\sum_{i=1}^d |A_i|=d\frac{1}{d}\sum_{i=1}^d|A_i|\le d\sqrt{\frac{1}{d}\sum_{i=1}^d A_i^2}=d^{1/2}\|A\|_2.
\]
We already know the limiting $L_2$ norms of $\theta'-\theta$, $\theta-\thetahat$ and $\theta'-\thetahat$ so we know that the limiting $L_1$ norms can be no more than a factor $d^{1/2}$ larger, \CSadd{giving the sizes in the final line of the proposition statement}. However,
\[
\frac{n^{1/2}}{d^{1/2}}
\Expect{\|\theta'-\theta\|_1}
=
\frac{n^{1/2}}{d^{1/2}}
\sqrt{\frac{2}{\pi}}
\frac{\ell}{n^{1/2}d^{1/2}}
\sum_{i=1}^d
\sqrt{V_{ii}}
=
\sqrt{\frac{2}{\pi}}
\frac{\ell}{d}
\sum_{i=1}^d
\sqrt{V_{ii}}
\to
\ell\sqrt{\frac{2}{\pi}}t^*_v.
\]
Similarly,  
\[
\frac{n^{1/2}}{d}
\Expect{\|\theta-\thetahat\|_1}
=
\frac{n^{1/2}}{d}
\sqrt{\frac{2}{\pi}}\frac{1}{\sqrt{n}}
\sum_{i=1}^d
\sqrt{\Sigma_{ii}}
=
\sqrt{\frac{2}{\pi}}
\frac{1}{d}
\sum_{i=1}^d
\sqrt{V_{ii}}
\to 
\sqrt{\frac{2}{\pi}} t^*_\pi.
\]
The triangle inequality shows that $\Expect{\|\theta'-\thetahat\|_1}$ behaves similarly to $\Expect{\|\theta-\thetahat\|_1}$. 
$\square$

\subsection{Our bounds}
\label{app.ourBounds}

\CSadd{We now consider the generic bounds of Corollary \ref{cor.DotBounds}.
Directly from \eqref{eqn.xDist},  $\|x_i\|_2=\Theta(1)$, from which both $c_i^{(1)}$ and $c_i^{(2)}$ are $\Theta(1)$. Further,  Proposition \ref{prop.Sizes} gives $M^{(1)}(\theta,\theta')=   \Theta(\|\theta'-\theta\|_2)\Theta(\|\theta-\thetahat\|_2)=\Theta(d^{1/2}/n)$ and $M^{(2)}(\theta,\theta')=   \Theta(\|\theta'-\theta\|_2)\Theta(\|\theta-\thetahat\|_2^2)=\Theta(d/n^{3/2})$.
In particular, then, for a single observation,
\begin{align*}
    \left|
    \ell_i(\theta')-\ell_i(\theta)
    -r_i^{(1)}(\theta,\theta';\thetahat)
    \right|
    &\le 
    c_i^{(1)} M^{(1)}(\theta,\theta')=\Theta\left(d^{1/2}n^{-1}\right),\\
    \left|
    \ell_i(\theta')-\ell_i(\theta)
    -r_i^{(2)}(\theta,\theta';\thetahat)
    \right|
    &\le   
    c_i^{(2)}M^{(2)}(\theta,\theta')=\Theta\left(dn^{-3/2}\right),
\end{align*}
\CSadd{where the size ($\Theta(d^{1/2}n^{-1})$ and $\Theta(dn^{-3/2})$) apply in probability in the limit of large $n$ and $d$.} 
The above bounds are proportional to the probability that a given observation will be considered for use in the accept-reject stage.} There are $n$ observations, and evaluation of each $r_i^{(1)}$ is $\Theta(d)$, so the overall costs of $r^{(1)}$ is $n\times \Theta(d^{1/2}n^{-1})\times \Theta(d) = \mathcal{O}(d^{3/2})$. In general, evaluation of each $r_i^{(2)}$ has a cost of $\Theta(d^2)$. Hence the cost for $r^{(2)}$ $n\times \Theta(d n^{-3/2})\times \Theta(d^2)= \Theta(d^3/n^{1/2})$.

\subsection{SMH bounds}
\label{sec.SMHbounds}

For the SMH algorithm, Equation (13) of \cite{cornish2019scalable} has
\begin{align*}
    \left|
    \ell_i(\theta')-\ell_i(\theta)
    -r_i^{(1)}(\theta,\theta';\thetahat)
    \right|
    &\le
    \frac{1}{2}\left(||\theta-\thetahat||_1^2+||\theta'-\thetahat||_1^2\right) \Ubar_{2,i},\\
    \left|
    \ell_i(\theta')-\ell_i(\theta)
    -r_i^{(2)}(\theta,\theta';\thetahat)
    \right|
    &\le
    \frac{1}{6}\left(||\theta-\thetahat||_1^3+||\theta'-\thetahat||_1^3\right) \Ubar_{3,i},    
    \end{align*}
    where
    \[
    \Ubar_{2,i}=\max_{j,k=1,\dots,d} \sup_{\theta \in \Theta}\left|\frac{\partial^2 U_i}{\partial \theta_j \partial \theta_k}\right|
    ~~~\mbox{and}~~~
    \Ubar_{3,i}=\max_{j,k,l=1,\dots,d} \sup_{\theta \in \Theta}\left|\frac{\partial^3 U_i}{\partial \theta_j \partial \theta_k\theta_l}\right|
    \]

Repeating the total cost analysis, including the $\mathcal{O}(d)$ cost of evaluating $r^{(1)}$ and the $\mathcal{O}(d^2)$ cost for $r^{(2)}$, for the scalable MH bounds, we obtain costs of
$\mathcal{O}(d^3n^{-1})\sum_{i=1}^n\Ubar_{2,i}$ and $\mathcal{O}(d^5 n^{-3/2})\sum_{i=1}^n\Ubar_{3,i}$ for first and second-order control variates, respectively.

For the logistic regression model, Appendix G.1 of the SMH paper \citep{cornish2019scalable} derives $\Ubar_{2,i}=\frac{1}{4}\max_{1\le j\le d}|x_{i,j}|^2$ and $\Ubar_{3,i}=\frac{1}{6\sqrt{3}}\max_{1\le j\le d}|x_{i,j}|^3$. Thus, if $d^{1/2}x_{i,j}$ is bounded, so are $d~\Ubar_{2,i}$ and $d^{3/2}\Ubar_{3,i}$. If, on the other hand, each $x_{i,j}$ is Gaussian then $\Ubar_{2,i}=\mathcal{O}(d^{-1}\log d)$ and $\Ubar_{3,i}=\mathcal{O}\left(d^{-3/2}\{\log d\}^{3/2}\right)$. In this case, we have total computational costs of $\mathcal{O}\left(d^2\log d\right)$ for SMH-1 and $\mathcal{O}\left(d^{7/2}\{\log d\}^{3/2}n^{-1/2}\right)$ for SMH-2.

\section{Additional results: simulation experiments}
\label{section:additional_simulation}

\subsection{Poisson regression}
\label{section:poisson_regression}
We now compare the MH-SS, Tuna, SMH and RWM algorithms using a set of synthetic data generated from a Poisson regression model with expectation $\log(1+e^{x^\top \beta})$. We simulate synthetic datasets with $d = 30$, $n = (10^{4.5}, 10^{5}) = (31,622, 100,000)$ and sample the predictors and coefficients as for the logistic regression scenarios. For Tuna, we set (in the notation of \cite{zhang2020asymptotically}), $\chi=2\times 10^{-5}$ in both cases and $\lambda=0.09$ for $n = 31,622$ and $\lambda= 0.05$ for $n = 100,000$. Following the suggestions of \cite{zhang2020asymptotically}, these combinations lead to acceptance rates of approximately $60\%$ and minimise the impact of $\chi$ on the average batch size.

Table \ref{tab_poisson} shows that MH-SS-1 and MH-SS-2 have the largest ESS per second. The values in bold denote the two largest efficiency metrics. Once again, the empirical results for $\mathbb{E}(B)$ show that the SMH bounds scale poorly in moderate to high dimensions and that our proposed bounds are tighter. For instance, MH-SS-1 requires $14$ times fewer observations than SMH-1, for $n = 31,622$, and it is slightly more efficient than SMH-2 in both scenarios. We note that the computational efficiency of SMH-1 is one of the lowest among the compared algorithms because, as pointed out in Section \ref{section:related_works}, it degrades quickly as $d$ increases. 
Though Tuna requires on average the smallest subsample sizes per MCMC iteration, its efficiency is similar to that of the RWM. As already discussed, this low efficiency is due to the scaling parameter, $\lambda$, of the proposal distribution for Tuna being smaller than for any other algorithm, generating highly correlated posterior samples.

\begin{table}[H]
\caption{Acceptance rate, average batch size, ESS per second and ESS/$\mathbb{E}(B)$ for the Poisson regression model applied to synthetic data in $d=30$.}
\label{tab_poisson}
\centering
\setlength\extrarowheight{-5pt}
\begin{tabular}{ccccc}
\hline
Model & Acc.\ rate & $\mathbb{E}(B)$ & ESS/sec & ESS/$\mathbb{E}(B)$ \\
\hline
\multicolumn{5}{c}{$n = 31{,}622$} \\ \hline
Tuna    & 0.605 & 100\,(3.46)        & 0.0418\,(0.0052)   & 0.342\,(0.038) \\
MH-SS-1 & 0.425 & 203\,(9.63)        & \textbf{4.31}\,(0.45)  & 37.9\,(1.32) \\
MH-SS-2 & 0.451 & 19.2\,(1.39)       & \textbf{60.9}\,(3.21)  & \textbf{447}\,(26.2) \\
SMH-1   & 0.167 & 2{,}890\,(126)     & 0.0592\,(0.0036)   & 0.558\,(0.021) \\
SMH-2   & 0.318 & 256\,(21.2)        & 4.29\,(0.43)        & \textbf{38.3}\,(2.57) \\
RWM     & 0.237 & 31{,}600           & 0.0498\,(0.0021)   & 0.313\,(0.0035) \\
\hline
\multicolumn{5}{c}{$n = 100{,}000$} \\ \hline
Tuna    & 0.604 & 111\,(2.15)        & 0.0257\,(0.0068)   & 0.241\,(0.058) \\
MH-SS-1 & 0.423 & 195\,(3.85)        & \textbf{6.57}\,(0.53)  & 38.4\,(1.46) \\
MH-SS-2 & 0.457 & 10.5\,(0.33)       & \textbf{92.4}\,(10.8)  & \textbf{803}\,(48.3) \\
SMH-1   & 0.163 & 2{,}770\,(55.4)    & 0.0588\,(0.0085)   & 0.577\,(0.024) \\
SMH-2   & 0.320 & 128\,(6.88)        & 6.52\,(1.19)        & \textbf{69.8}\,(4.81) \\
RWM     & 0.238 & 100{,}000          & 0.0109\,(0.0022)   & 0.105\,(0.0007) \\
\hline
\end{tabular}
\end{table}

\subsection{Additional results for logistic regression}
\begin{figure}[htbp]
\begin{center}
     \includegraphics[scale=0.72]{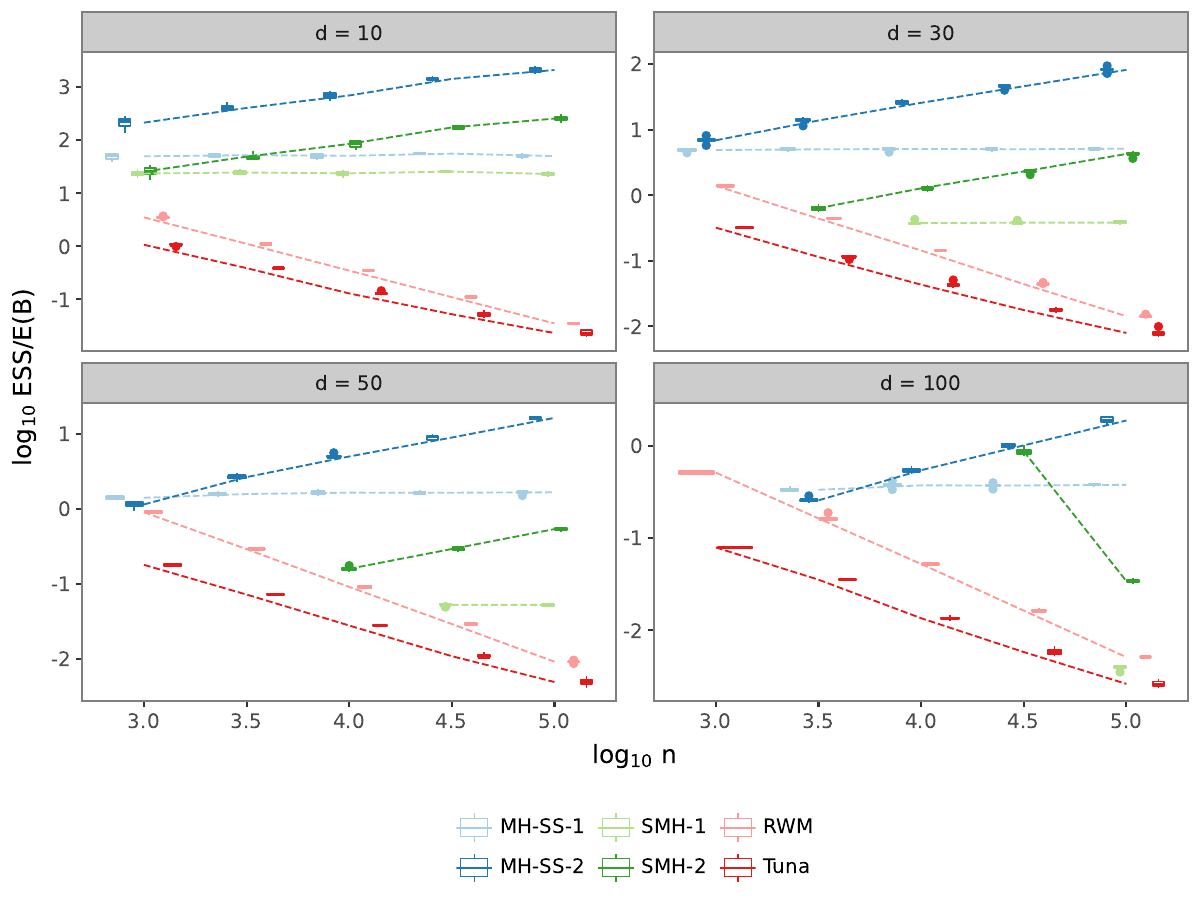}     
     \captionsetup{justification=centering}
\end{center}
\caption{ESS divided by $\mathbb{E}(B)$ for MH-SS, SMH and RWM for the logistic regression model. Both axes are presented in the logarithm base 10. Some ESSs are omitted because $\mathbb{E}(B) \ge n$, which implies the use of the RWM algorithm.}
\label{ESS_over_B}
\end{figure}

In this section, we present additional comparisons using ESS/$\mathbb{E}(B)$ as the efficiency metric. Figure \ref{ESS_over_B} shows that the MH-SS algorithms consistently outperform SMH, Tuna and the RWM in terms of ESS/$\mathbb{E}(B)$. Notably, the efficiency gap between MH-SS-1 and SMH-1 is more pronounced compared to MH-SS-2 and SMH-2, especially as $d$ increases. This discrepancy can be attributed to two main factors. Firstly, the bounds on the log-likelihood differences in MH-SS-1 are significantly tighter than those in SMH-1, as demonstrated in Section \ref{section:theoretical_results} and Appendix \ref{section:asymptotic_analysis_bounds}. Tighter bounds result in a smaller $\mathbb{E}(B)$. Secondly, the acceptance rates of SMH-1 deteriorate drastically as $d$ increases, as discussed in Section \ref{section:related_works}, whereas they tend to remain stable for SMH-2. As efficiency metrics like ESS take into account the autocorrelation of the posterior samples, lower acceptance rates lead to smaller ESS. In summary, the results for SMH-1 are influenced by a combination of loose bounds and poor acceptance rates, especially in high-dimensional settings.

\subsection{Comparison of implementations}
The results presented throughout the paper consider an implementation of SMH that evaluates the likelihood at most $B \leq n$ times in each MCMC iteration. If the proposal $\theta^\prime$ is accepted, it indicates that the likelihood was exactly evaluated $B$ times. Conversely, if $\theta^\prime$ is rejected, it suggests that the likelihood was most likely evaluated fewer than $B$ times because SMH is designed to reject a proposal based on sequential point-wise likelihood evaluations; for more details, see Algorithm 1 in \cite{cornish2019scalable}. In contrast, both MH-SS and Tuna are originally designed to evaluate the likelihood exactly $B$ times before deciding whether to accept or reject the proposal. 

The sequential point-wise rejection process employed by SMH implies that likelihood evaluations cannot be computed in a vectorised fashion in programming languages like Python and R. Consequently, while loops are required to handle these evaluations. We point out that, for the sake of fair run-time comparisons, the results presented in the paper for RWM, SMH, MH-SS and Tuna were all obtained from implementations based on for-loops. In practice, however, we find that vectorised operations in Python are often faster compared to those relying on for-loops. 

In an effort to enhance the run-time efficiency of SMH, we implemented a vectorised approach for evaluating the likelihood. For fairness in the comparisons, when vectorising the loops in SMH-1 and SMH-2, we also vectorised the computations in Tuna, MH-SS-1, MH-SS-2 and in the RWM. In the `vectorised SMH' algorithm, the likelihood is evaluated exactly $B$ times regardless of whether $\theta^\prime$ is accepted or rejected. We note that the impact of SMH vectorization is exclusively observed in metrics related to run-times, such as ESS per second. Specifically, only the denominator of the ESS per second changes, as the ESS remains the same. Metrics like $\mathbb{E}(B)$ and ESS/$\mathbb{E}(B)$ also remain the same. Though this modification resulted in a significant computational speed-up, particularly for SMH-1 as it tends to evaluate more likelihood terms, it was not enough to make SMH more computationally efficient than MH-SS. 

\begin{figure}[htbp]
\begin{center}
   \begin{subfigure}{0.47\textwidth}
        \caption{For-loop implementations}
          \includegraphics[scale=0.53]{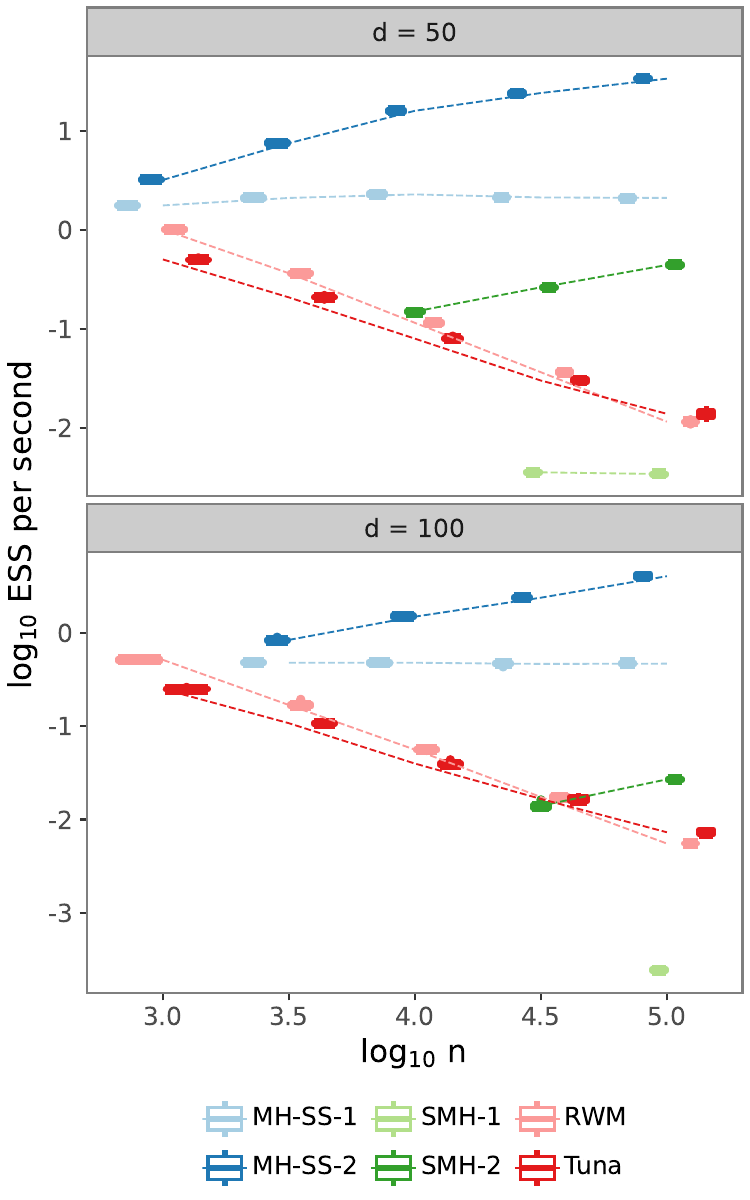}
   \end{subfigure}
   \begin{subfigure}{0.5\textwidth} 
        \caption{Vectorised implementations}
          \includegraphics[scale=0.53]{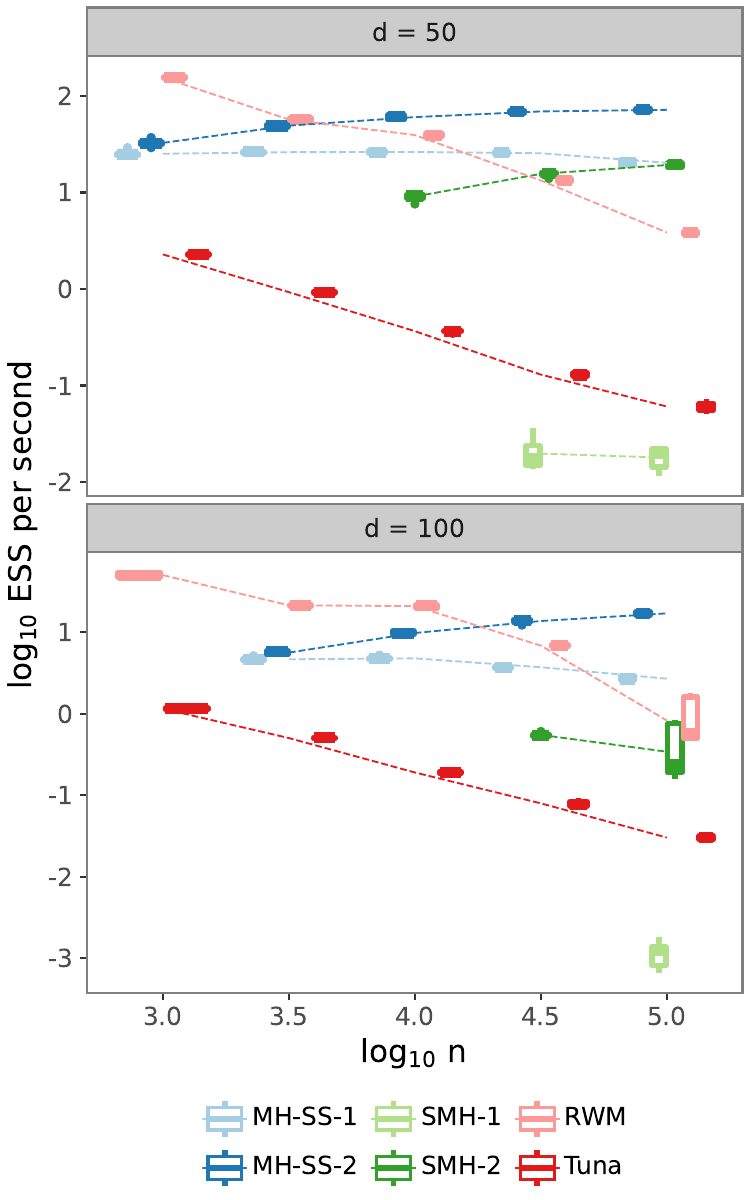}
   \end{subfigure}
\caption{ESS per second for SMH and MH-SS on a logistic regression model. Except for the RWM, all algorithms have either for-loop-based implementations (panel(a)) or are all vectorised (panel (b)). Both axes are presented in the logarithm base 10. Some ESSs are omitted because $\mathbb{E}(B) \ge n$, which implies the use of the RWM algorithm.}
\label{ESS_per_second_d_50_100}
\end{center}
\end{figure}

Figure \ref{ESS_per_second_d_50_100} shows the results of ESS per second obtained from simulation experiments with a logistic regression model. Panel (a) exhibits results based on foor-loop implementations, while panel (b) shows the vectorised versions. We stress that within each panel, the RWM, SMH, MH-SS and Tuna implementations are either all based on for-loops or all vectorised. Overall, comparing panels (a) and (b), we can see that the vectorised versions of SMH-1 and SMH-2 are significantly faster than their for-loop counterparts. For instance, for $d = 50$ and $n = 10^5$, the efficiency of SMH-1 and SMH-2 shows a 15- and 39-fold improvement, respectively, as their ESS per second jumped from $10^{-2}$ and $10^{-0.3}$ to $10^{-0.8}$ and $10^{1.3}$, respectively. Unsurprisingly, the gains in terms of run-times for the vectorised MH-SS are smaller, especially for MH-SS-2 with large $n$, as it requires smaller subsamples.

\subsection{Empirical results of the optimal scaling for SMH}
In addition to the optimal scaling results for MH-SS presented in Appendix \ref{section:tuning_guidelines}, we also performed simulation experiments to find the optimal acceptance rate for the scalable Metropolis--Hastings algorithm \citep{cornish2019scalable} in order to guarantee a fair comparison against MH-SS, Tuna and RWM in the Sections \ref{section:simulation} and \ref{section:real_datasets}. This section presents the efficiency results for a set of values for the scaling parameter $\lambda$ ranging from $0.1$ to $4$ with increments of $0.2$. Our empirical analysis focuses primarily on the results of the vanilla SMH-1 and SMH-2, though we also show SMH's efficiency when equipped with the bounds introduced in Section \ref{sec.bounds} (i.e., SMH-NB).

\begin{figure}[H]
\begin{center}
   \begin{subfigure}{0.52\textwidth}
        \caption{}
          \includegraphics[scale=0.37]{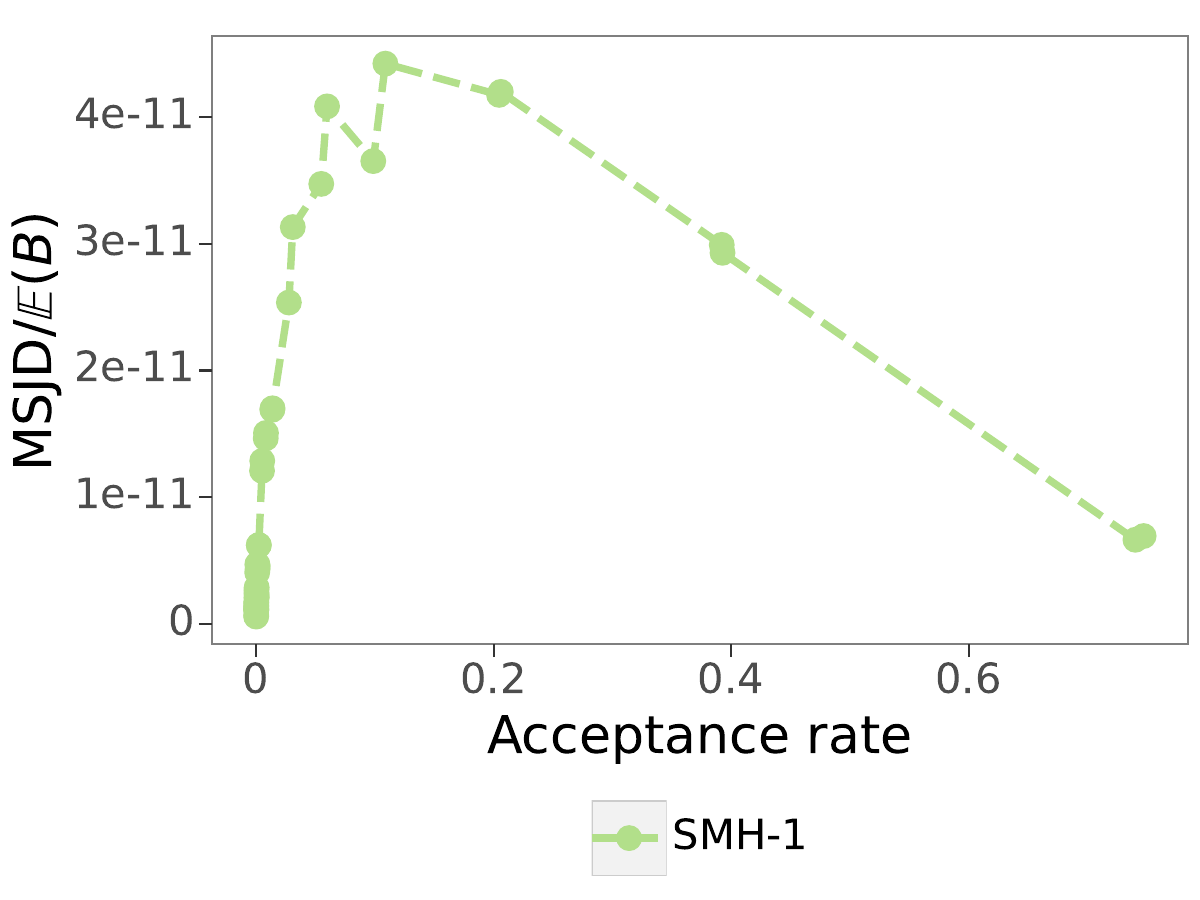}
   \end{subfigure}
   \begin{subfigure}{0.47\textwidth} 
        \caption{}
          
          \includegraphics[scale=0.37]{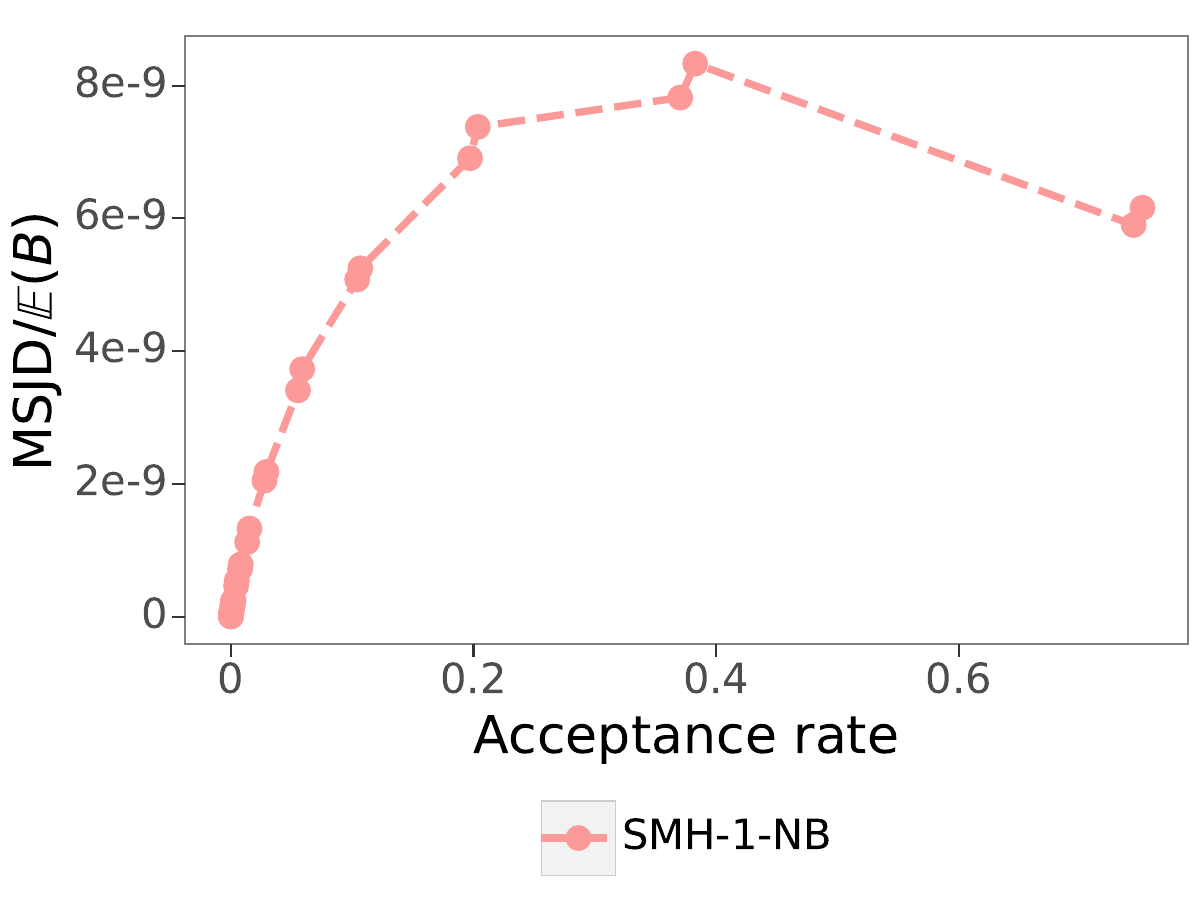}
   \end{subfigure}
   \begin{subfigure}{0.52\textwidth}
        \caption{}
          \includegraphics[scale=0.37]{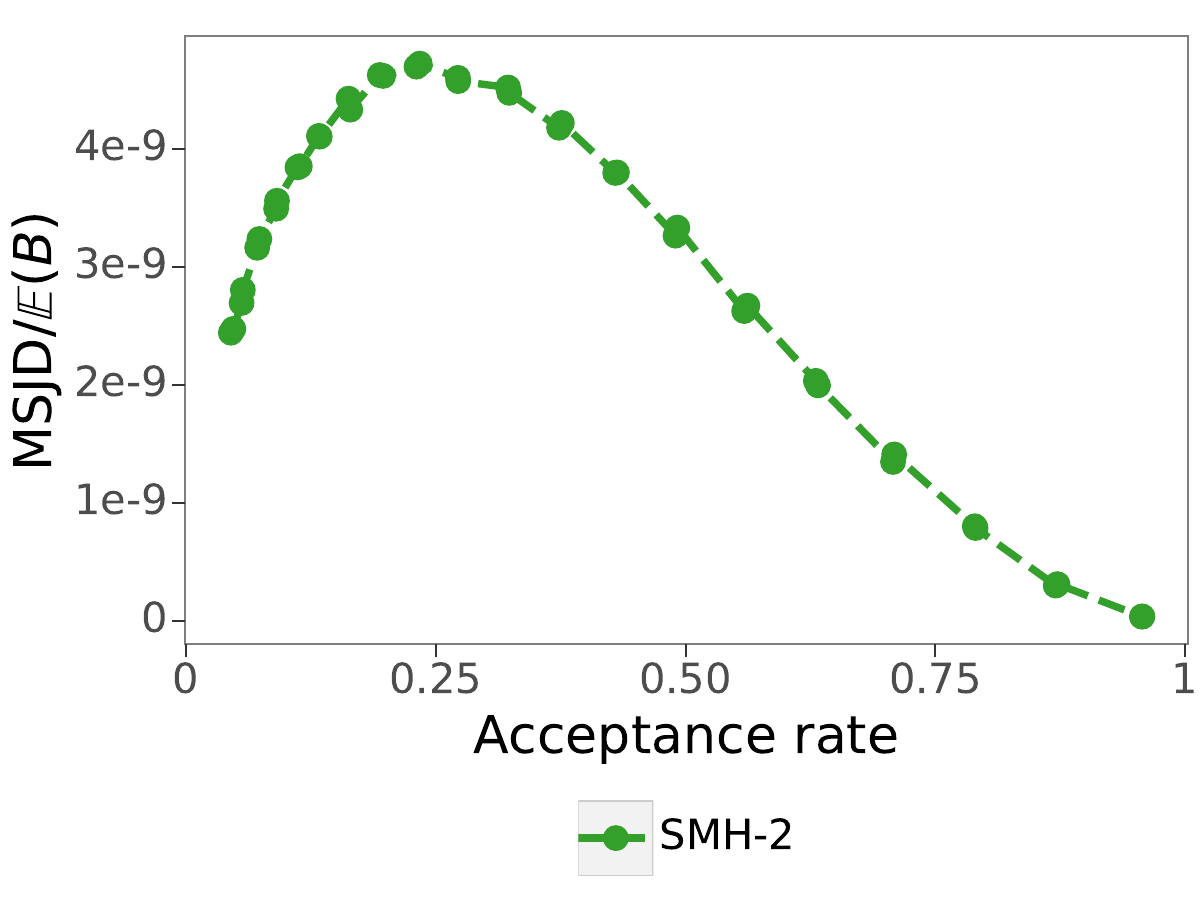}
   \end{subfigure}
   \begin{subfigure}{0.47\textwidth} 
        \caption{}
          
          \includegraphics[scale=0.37]{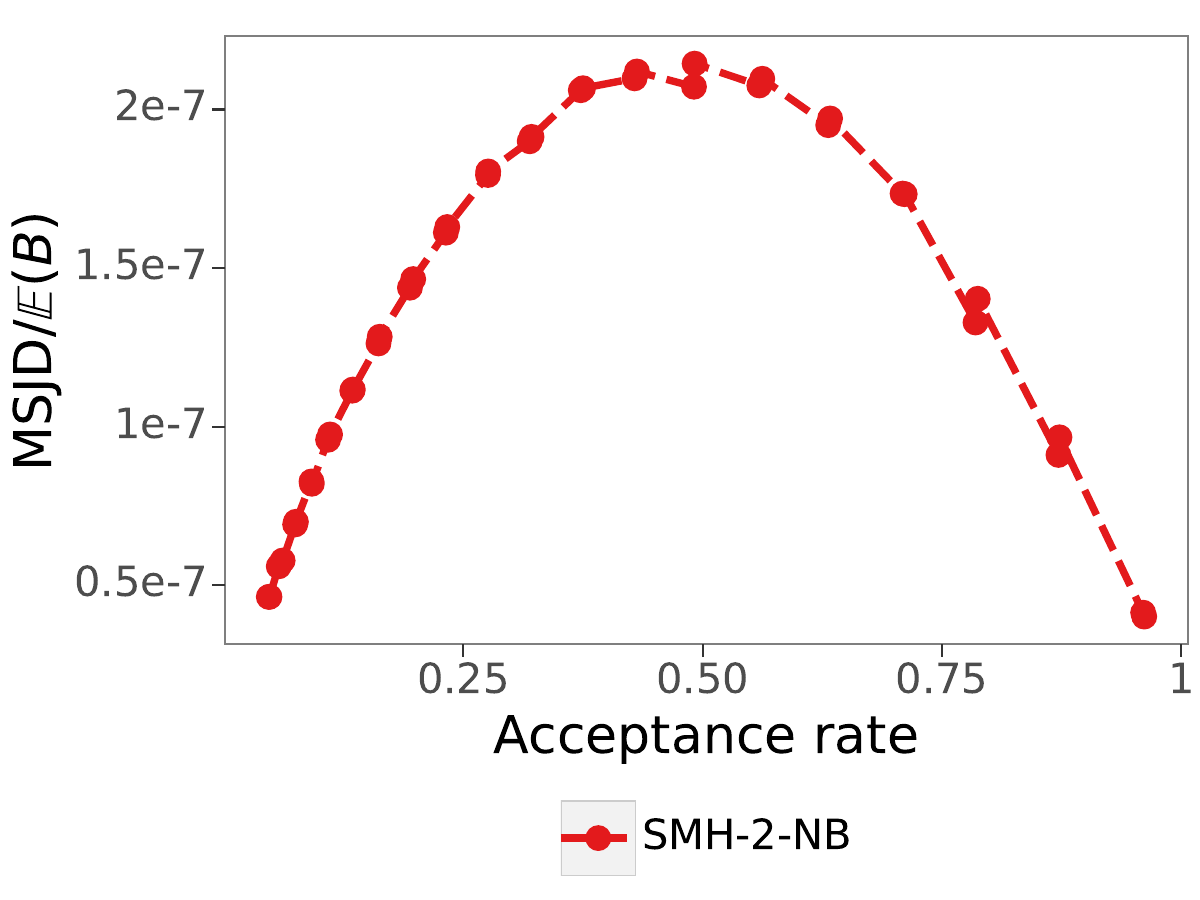}
   \end{subfigure}
   \begin{subfigure}{0.47\textwidth} 
        \caption{}
        \includegraphics[scale=0.37]{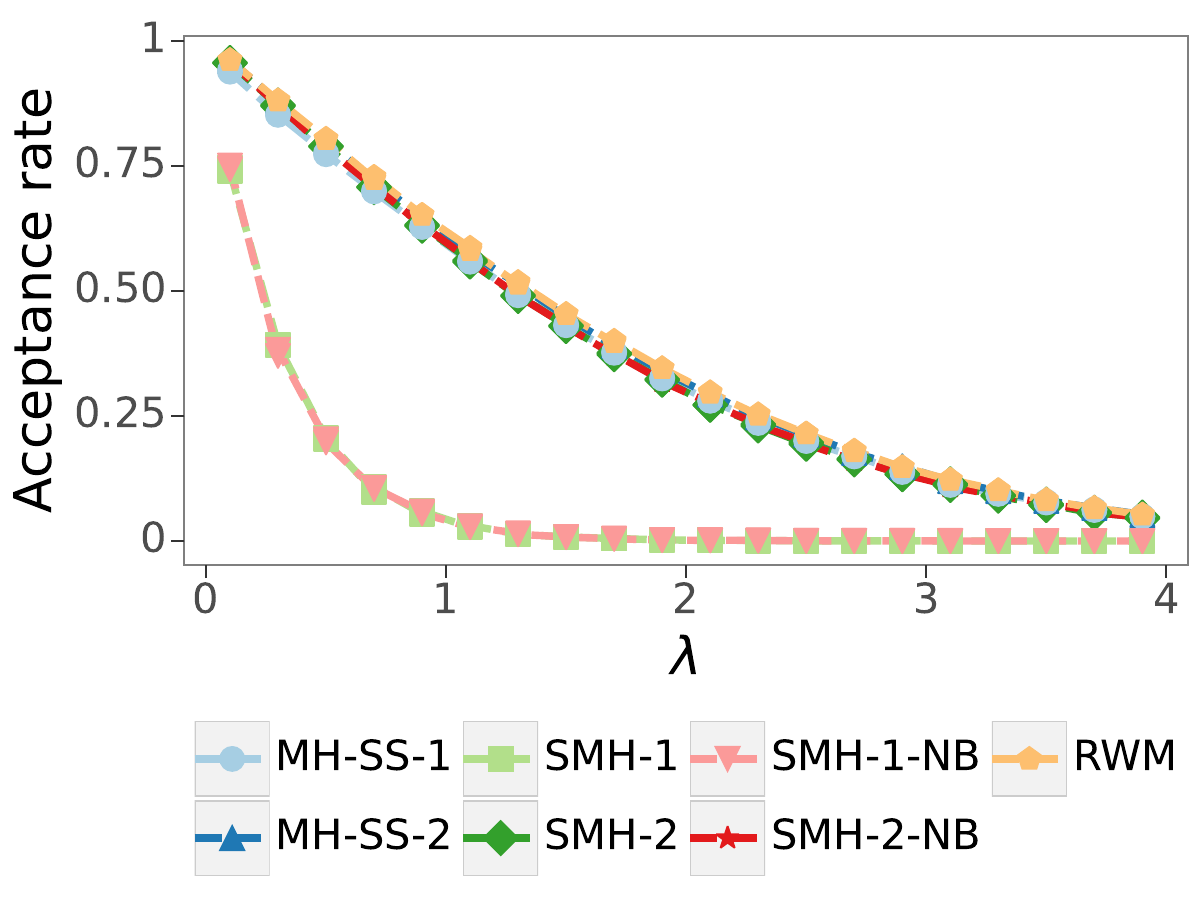}
          
   \end{subfigure}

\caption{Acceptance rates and mean squared jumping distance (MSJD) over average batch size ($\mathbb{E}(B)$) based on a simulation experiment with a logistic regression model with dimension $d = 100$ and $n=100,000$.}
\label{fig:additiona_results}
\end{center}
\end{figure}

As stated in Appendix \ref{section:tuning_guidelines} and now displayed in panel (e) of Figure \ref{fig:additiona_results}, SMH-2 has an optimal scaling around that of the RWM, which in turn leads to an acceptance rate close to $0.234$ (panel (c)). Panel (e) also displayes that all algorithms, except SMH-1 and SMH-1-NB, have similar acceptance rates as a function of the scaling parameter, $\lambda$. For these algorithms, we find that the scaling must decrease more quickly with dimension than $\ell/\sqrt{d}$ to maintain a reasonable acceptance rate as dimension increases. For SMH-1, however, we found that the optimal acceptance rate is less obvious. For example, for synthetic data generated from a logistic regression with $n = 100,000$ and $d = 100$, panel (a) of Figure \ref{fig:additiona_results} suggests that acceptance rates from $8\%$ to $20\%$ maximise SMH-1's efficiency. However, we observed in other experiments, not shown here for brevity, that the relation between efficiency and optimal acceptance rate might change as $n$ and $d$ vary.

\subsection{Experimental details of the Tuna tuning hyperparameter}
\label{appendix:experimental_details}

To aid reproducibility of the results in the paper, Tables \ref{tab:values_chi_simulation_related_works} and \ref{tab:values_chi_simulation} in this section present the values of the Tuna tuning parameter as well as scaling parameter $\lambda$ following the heuristic proposed by \cite{zhang2020asymptotically}.

\begin{table}
\centering
\caption{Values of the hyperparameter $\chi$ in the Tuna algorithm used in the experiments of Section \ref{section:related_works} considering $n = 31,622$.}
\label{tab:values_chi_simulation_related_works}
\begin{tabular}{ccc}
\toprule
$d$ & $\chi$ & $\lambda$\\ \hline
$5$  &  $1e^{-3}$ & $0.03$      \\ 
$10$  &  $5e^{-4}$ & $0.04$      \\ 
$20$  &  $1e^{-4}$ & $0.06$      \\ 
$30$  &  $5e^{-5}$ & $0.065$     \\ 
$40$  &  $3e^{-5}$ & $0.08$     \\ 
$50$  &  $2e^{-5}$ & $0.09$     \\ 
$60$ &  $1e^{-5}$ & $0.1$     \\
\bottomrule
\end{tabular}
\end{table}

\begin{table}
\centering
\caption{Values of the hyperparameter $\chi$ in the Tuna algorithm used in the simulation experiments of Section \ref{section:simulation}.}
\label{tab:values_chi_simulation}
\begin{tabular}{cccc}
\toprule
$n$    & $d$ & $\chi$ & $\lambda$\\ \hline
\multirow{4}{*}{$1,000$}
 & $10$  &  $1e^{-3}$ & $0.14$      \\ 
 & $30$  &  $2e^{-4}$ & $0.21$     \\ 
 & $50$  &  $6e^{-5}$ & $0.25$     \\ 
 & $100$ &  $1e^{-5}$ & $0.35$     \\
\hline

\multirow{4}{*}{$3,162$} 
 & $10$  & $9e^{-4}$ & $0.08$       \\
 & $30$  & $2e^{-4}$ & $0.11$        \\ 
 & $50$  & $6e^{-5}$ & $0.15$     \\ 
 & $100$ & $1.5e^{-5}$ & $0.22$      \\
\hline

 \multirow{4}{*}{$10,000$} 
 & $10$  & $2e^{-3}$ & $0.04$       \\
 & $30$  & $2e^{-4}$ & $0.06$      \\ 
 & $50$  & $6e^{-5}$ & $0.09$     \\ 
 & $100$ & $1.5e^{-5}$ & $0.12$       \\ 
\hline
\multirow{4}{*}{$31,622$} 
 & $10$ & $1e^{-3}$ & $0.025$       \\ 
 & $30$ & $1e^{-4}$ & $0.04$      \\ 
 & $50$ & $6e^{-5}$ & $0.045$      \\ 
 & $100$& $1.5e^{-5}$ & $0.07$       \\
\hline
\multirow{4}{*}{$100,000$} 
 & $10$  & $1e^{-3}$ & $0.015$       \\ 
 & $30$  & $2e^{-4}$ & $0.02$       \\ 
 & $50$  & $5e^{-5}$ & $0.03$       \\ 
 & $100$ & $1.5e^{-5}$ & $0.04$       \\
\bottomrule
\end{tabular}
\end{table}

\section{Additional results: real-world applications}
\label{sec.additionalRealWorld}

\subsection{U.S. Current Population Survey}
\label{sec:usa_survey}

We illustrate the applicability of our method on the 2018 United States Current Population Survey. 
The CPS is a monthly household survey carried out by the U.S. Census Bureau and the U.S. Bureau of Labor Statistics that gathers information on the labour force for the population of the U.S. The data contain variables such as income, education, occupation, participation in welfare programs and health insurance. Using a sample of $n = 500,000$ survey participants, we wish to model whether or not an individual has a total pre-tax personal income above $\$25,000$ for the previous year, based on $10$ predictors. We fit the logistic and probit models with an intercept term and normalise the covariates to improve numerical stability, ensuring each has a mean of zero and variance of one. As some of the variables are categorical, the linear predictor contains $d=31$ parameters (including the intercept). For the Tuna algorithm, we set $\chi = 10^{-5}$ and $\lambda = 0.019$.

Table \ref{tab_CPS} presents the results for acceptance rate, average batch size, ESS per second and ESS divided by $\mathbb{E}(B)$ for each of the algorithms. Whether for the logistic or probit models, SMH-1 is not a viable option compared to the RWM. Although SMH-1 uses approximately $10\%$ of the data, its efficiency per second is less than that of the RWM. This is due to its relatively low scaling parameter, $\lambda$, and the additional computational overheads compared with the vanilla RWM, due to it being a more complex algorithm, including sampling extra random variables and performing vector/matrix multiplications. We stress that in all of our implementations, operations that can be pre-computed to speed up both SMH and MH-SS are performed accordingly. 

\begin{table}[H]
\caption{Acceptance rate, average batch size, ESS per second and ESS/$\mathbb{E}(B)$ for the logistic and probit regression models applied to a sample of $n = 500{,}000$ from the U.S. CPS.}
\label{tab_CPS}
\centering
\setlength\extrarowheight{-5pt}
\begin{tabular}{cccccc}
\toprule
Model & Algorithm & Acc.\ rate & $\mathbb{E}(B)$ & ESS/sec & ESS/$\mathbb{E}(B)$ \\
\hline
\multirow{6}{*}{Logistic}
& Tuna    & 0.594 & 284\,(3.82)      & 0.00321\,(0.00045) & 0.00183\,(0.00021) \\
& MH-SS-1 & 0.441 & 989\,(12.8)      & \textbf{2.38}\,(0.85) & \textbf{0.920}\,(0.020) \\
& MH-SS-2 & 0.456 & 88.4\,(2.22)     & \textbf{28.1}\,(7.87) & \textbf{10.6}\,(0.39) \\
& SMH-1   & 0.158 & 49{,}100\,(1{,}930) & 0.0177\,(0.0059) & 0.0192\,(0.0067) \\
& SMH-2   & 0.319 & 8{,}290\,(208)   & 1.61\,(0.43) & 0.138\,(0.0037) \\
& RWM     & 0.239 & 500{,}000        & 0.00418\,(0.00062) & 0.00227\,(0.00048) \\
\hline
\multirow{5}{*}{Probit}
& MH-SS-1 & 0.404 & 1{,}540\,(12.4)  & 1.62\,(0.51) & \textbf{0.911}\,(0.024) \\
& MH-SS-2 & 0.411 & 71.1\,(1.76)     & \textbf{31.4}\,(8.99) & \textbf{9.29}\,(0.39) \\
& SMH-1   & 0.140 & 57{,}800\,(2{,}500) & 0.0276\,(0.021) & 0.0034\,(0.0004) \\
& SMH-2   & 0.276 & 5{,}080\,(77.1)  & \textbf{3.05}\,(2.01) & 0.128\,(0.0031) \\
& RWM     & 0.193 & 500{,}000        & 0.189\,(0.117) & 0.0021\,(0.00004) \\
\bottomrule
\end{tabular}
\end{table}

\subsection{Detection of gas mixtures}
\label{section:gas_detection}

We now apply the various MCMC sampling algorithms to a dataset available in the UCI machine learning repository 
which gathers recordings of multiple chemical sensors used to detect gas mixtures at varying concentrations. Our goal is to predict for a sample of size $n = 250,000$ whether the concentration of Ethylene in the air, measured in parts per million, is above zero. Though we consider $d = 7$ continuous predictors only (including an intercept), each of which corresponds to a different model sensor, some of them are highly correlated. In practice, this correlation is expected to increase $\mathbb{E}(B)$ for SMH and MH-SS, thereby potentially reducing their computational efficiency relative to RWM. For the Tuna algorithm, we set $\chi = 7 \times 10^{-6}$ and $\lambda = 0.025$.

Table \ref{tab_gamma_telescope} presents the efficiency metrics for all algorithms applied to the logistic and probit models. Our analysis in Appendix \ref{section:asymptotic_analysis_bounds} shows that the MH-SS bounds scale better with dimension than those of SMH. It is no coincidence, therefore, that it is in this real example, which has, by some margin,  the lowest dimension of any of our real-data examples, that the performance of SMH-2 is closest to that of MH-SS-2. We also note that, Tuna (which can only be applied for the logistic regression) is not competitive with any of the algorithms, except for the RWM.

\begin{table}[H]
\caption{Acceptance rate, average batch size, ESS per second and ESS/$\mathbb{E}(B)$ for the logistic and probit regression models applied to the gas sensor dataset.}
\label{tab_gamma_telescope}
\centering
\setlength\extrarowheight{-5pt}
\begin{tabular}{cccccc} 
\toprule
Model & Algorithm & Acc.\ rate & $\mathbb{E}(B)$ & ESS/sec & ESS/$\mathbb{E}(B)$ \\
\hline
\multirow{6}{*}{Logistic} 
& Tuna    & 0.609 & 321\,(3.47)       & 0.0282\,(0.0051)     & 0.0061\,(0.0005) \\
& MH-SS-1 & 0.478 & 945\,(10.8)       & 6.88\,(0.82)         & 2.33\,(0.04) \\
& MH-SS-2 & 0.477 & 165\,(2.31)       & \textbf{83.6}\,(14.9) & \textbf{13.8}\,(0.34) \\
& SMH-1   & 0.437 & 2{,}260\,(42.0)   & 15.9\,(0.95)         & 0.655\,(0.017) \\
& SMH-2   & 0.349 & 680\,(13.6)       & \textbf{46.3}\,(5.02) & \textbf{3.63}\,(0.10) \\
& RWM     & 0.274 & 250{,}000         & 0.0359\,(0.0029)     & 0.0127\,(0.0013) \\
\hline
\multirow{6}{*}{Probit} 
& MH-SS-1 & 0.415 & 1{,}340\,(17.9)   & 4.46\,(0.73)         & 1.74\,(0.03) \\
& MH-SS-2 & 0.414 & 103\,(1.92)       & \textbf{74.1}\,(20.5) & \textbf{23.1}\,(0.69) \\
& SMH-1   & 0.379 & 2{,}720\,(36.1)   & 10.2\,(0.49)         & 0.569\,(0.116) \\
& SMH-2   & 0.287 & 382\,(7.96)       & \textbf{74.5}\,(15.4) & \textbf{6.33}\,(0.16) \\
& RWM     & 0.231 & 250{,}000         & 0.321\,(0.044)       & 0.0289\,(0.0053) \\
\bottomrule
\end{tabular}
\end{table}

\subsection{\CSadd{A note on the Poisson regression of Section} \ref{section:road_casualties}}

The casualty counts in the Poisson regression example of Section \ref{section:road_casualties},  might be spatially correlated along a given route and across intersections. Given GIS information, it might be possible to obtain "along-route" distances between observations; however exact inference for a Gaussian-process-based Poisson model with $n\approx  300k$ observations is infeasible for any MCMC algorithm. Instead, a conditional-autoregressive model could be fitted, based on the $p$ nearest neighbors along a route. MH-SS could be implemented in a similar manner to the autoregressive model in the discussion.
\section{Justification of optimal-scaling Assumption \ref{ass.scaling.B}}
\label{section:optimal_scaling}
The use of first-order control variates, $r_i^{(1)}$, is equivalent to using the approximation 
 \[
 \log \phat^{(1)}(y_{1:n}|\theta):=(\theta-\thetahat)^\top \sum_{i=1}^n g_i(\thetahat)+c=(\theta-\thetahat)^\top\nabla \log p(y_{1:n}|\thetahat)+c,
 \]
 whereas the use of second-order control variates, $r_i^{(2)}$, is equivalent to 
 \[
 \log \phat^{(2)}(y_{1:n}|\theta):=-\frac{1}{2}(\theta-\thetahat)^\top \left\{\sum_{i=1}^n H_i(\thetahat) \right\} (\theta-\thetahat)+c\approx p(y_{1:n}|\theta),
 \]
 for large $n$, by the Bernstein-von Mises Theorem.

If $\thetahat$ is a reasonable approximation to the mode, then $\nabla \log \pi(\thetahat)\approx 0$ (indeed, this would be one definition of a reasonable approximation).  In the Bernstein-von Mises limit, $\|\theta-\thetahat\|_2=\mathcal{O}(n^{-1/2}d^{1/2})$, and it is straightforward to show that provided $\left.\nabla \log p(y_{1:n}|\theta)\right|_{\thetahat}=\mathcal{O}(n^{1/2})$, $\log \phat^{(1)}(y_{1:n}|\theta')-\log \phat^{(1)}(y_{1:n}|\theta)=\mathcal{O}(1)$, typically. 

Implicit in the justification of \eqref{tuna_ratioNODA} is that, from the moment generating function of $S_i\sim\Poisson(\phi_i)$, 
\[
\Expect{\left(\phi_i'/\phi_i\right)^{S_i}}
=
\exp\left[\phi_i\left(\frac{\phi'_i}{\phi_i}-1\right)\right]
=
\exp[\phi_i'-\phi_i]
=
\exp[\ell_i(\theta')-\ell_i(\theta)-r_i(\theta,\theta';\thetahat)].
\]
Thus, the acceptance ratio for $\alpha_{2,MHSS}(\theta,\theta')$ in \eqref{eqn.StageTwoAcc} has an expectation of 
\begin{align*}
\Expect{
\exp\left\{\sum_{i\in \mathcal{I}}S_i\{\log \phi_i' - \log \phi_i\}\right\}
}
&=
\Expect{
\prod_{i=1}^n \left\{\frac{\phi_i'}{\phi_i}\right\}^{S_i}
}
=\frac{p(y_{1:n}|\theta')}{p(y_{1:n}|\theta)}\exp\left[-\sum_{i=1}^n r_i(\theta,\theta';\thetahat)\right]
.
\end{align*}
Despite our best efforts, whether first- or second-order control variates are used, the upper bounds from Corollary \ref{cor.DotBounds} are typically very loose. As argued directly after the statement of the corollary, this is because in moderate to high dimensions, $\theta'-\theta$, $\theta-\thetahat$ and $\theta'-\thetahat$ are close to perpendicular to most covariate vectors $x_i$. However, we cannot use this fact to obtain better bounds without $\mathcal{O}(n)$ effort at each iteration. Typically, therefore, $c_iM(\theta,\theta')>> |\ell_i(\theta')-\ell_i(\theta)-r_i(\theta,\theta';\thetahat)|$. With $\gamma=0$, \eqref{eqn.define.phi} and \eqref{eqn.define.phip} tell us that 
$\phi_i=c_i M(\theta,\theta')(1-\delta_i)$ and $\phi_i'=c_i M(\theta,\theta')(1-\delta_i')$ with $\delta_i=-\min[0,\Delta_i]/\{c_iM(\theta,\theta')\}<<1$ and $\delta'_i=-\min[0,\Delta'_i]/\{c_iM(\theta,\theta')\}<<1$. Again from the moment generating function of a Poisson $\phi_i$ random variable:
\[
\Expect{\left(\phi_i'/\phi_i\right)^{2S_i}}
=
\exp\left[
\frac{{\phi'_i}^2}{\phi_i}-\phi_i
\right]
\approx 
\exp\left[2\phi_i'-2\phi_i\right]
=
\Expect{\left(\phi_i'/\phi_i\right)^{S_i}}^2,
\]
where the approximation arises from a first-order Taylor expansion. Thus, to first order,
\[
\Expect{
\left(\prod_{i=1}^n \left\{\frac{\phi_i'}{\phi_i}\right\}^{S_i}\right)^2
}
=
\Expect{
\prod_{i=1}^n \left\{\frac{\phi_i'}{\phi_i}\right\}^{S_i}
}^2
\]
and so $\Var{\exp\left\{\sum_{i\in \mathcal{I}}S_i\{\log \phi_i' - \log \phi_i\}\right\}}=0$ to first order. Heuristically, this justifies the assumption that for large $d$ and typical $\theta\sim \pi$, 
\[
\exp\left[\sum_{i\in \mathcal{I}}S_i\{\log \phi_i' - \log \phi_i\}\right]
\approx
\frac{p(y_{1:n}|\theta')}{p(y_{1:n}|\theta)}
\exp\left[-\sum_{i=1}^n r_i(\theta,\theta';\thetahat)\right].
\]
The above and the arguments about $\phat^{(1)}$ and $\phat^{(2)}$ that precede it then imply Assumption \ref{ass.scaling.B}.

\section{\CSadd{Multimodal targets}}
\label{app.Multimodal}
\subsection{MH-SS extension}
For certain models, some structural identifiability issue might render a posterior multimodal even when large amounts of data are available. Canonical examples include mixture models and hidden-Markov models. In such cases, the control variates created from mode $\thetahat^{(1)}$ would be inappropriate while the chain is close to a different mode, $\thetahat^{(2)}$. The obvious solution is for MH-SS to choose the control variate associated with the nearest mode to the current value, $\theta$; however, such an algorithm would not satisfy detailed balance with respect to the true posterior distribution and so would lead to biased inferences. Notice that the entire proof of Proposition \ref{propDB} is conditional on a given (arbitrary) $\theta$ and $\theta'$. Thus, provided the same control variates and bounds are used in the expressions $\alpha(\theta,\theta';s,\thetahat)$ and $\alpha(\theta',\theta;s,\thetahat)$ in \eqref{eqn.generalDB}, the proof follows exactly as it stands. Thus, for the detailed-balance condition to hold, we need \emph{the same} control variates and bounds to be chosen whether $\theta$ is the current value and $\theta'$ the proposal or $\theta'$ is the current value and $\theta$ is the proposal.

Our extension of MH-SS to the multimodal scenario chooses the control variates according to the value of  $\overline{\theta}:=(\theta+\theta')/2$. When $\overline{\theta}$ is between the modes, we will use the average of the control variates
\[
\overline{r}_i(\theta,\theta';\thetahat_1,\thetahat_2):=\frac{1}{2}\left\{r_i(\theta,\theta';\thetahat_1)+r_i(\theta,\theta';\thetahat_2)
\right\},
\]
and the natural bound derived as follows. The triangle inequality and \eqref{eqn.basicBound} give
\begin{align*}
|\ell_i(\theta')-\ell_i(\theta)-\overline{r}_i(\theta,\theta';\thetahat_1,\thetahat_2)|
&\le
\frac{1}{2}\left|\ell_i(\theta')-\ell_i(\theta)-r_i(\theta,\theta';\thetahat_1)\right|\\
&~~~+
\frac{1}{2}\left|\ell_i(\theta')-\ell_i(\theta)-r_i(\theta,\theta';\thetahat_2)\right|\\
&\le
c_i\overline{M}(\theta,\theta';\thetahat_1,\thetahat_2),
\end{align*}
where $\overline{M}(\theta,\theta';\thetahat_1,\thetahat_2):=\left\{M(\theta,\theta';\thetahat_1)+M(\theta,\theta';\thetahat_2)\right\}/2$.
For simplicity, we decide on which control variate to use according to the projection of $\overline{\theta}$ onto the line between $\thetahat_1$ and $\thetahat_2$:
\[
D:=(\overline{\theta}-\thetahat_1)
\times \frac{\thetahat_2-\thetahat_1}{\|\thetahat_2-\thetahat_1\|}.
\]
Specifically, for $i=1,\dots,n$,
\begin{equation}
r_i(\theta,\theta';\thetahat_1,\thetahat_2)
=
\left\{
\begin{array}{ll}
r_i(\theta,\theta';\thetahat_1)&\mbox{if }D<\|\thetahat_2-\thetahat_1\|/3,\\
r_i(\theta,\theta';\thetahat_2)&\mbox{if }D>2\|\thetahat_2-\thetahat_1\|/3,\\
\overline{r}_i(\theta,\theta';\thetahat_1,\thetahat_2)&\mbox{otherwise}.\\
\end{array}
\right.
\end{equation}
Similarly, $M(\theta,\theta';\thetahat_1,\thetahat_2)$ is $M(\theta,\theta';\thetahat_1)$, $M(\theta,\theta';\thetahat_2)$ or $\overline{M}(\theta,\theta';\thetahat_1,\thetahat_2)$ for the same three ranges of $D$.

We first describe the new MH-SS algorithm. We then describe a simple mixture model that leads to a bimodal posterior even with large amounts of data, and we construct a short simulation study on the posterior distribution for this model.

The set up for the multimodal MH-SS is the same as Algorithm \ref{mh_ss_algorithm_set_up} except that Steps 5, 6 and 7 find approximations to both modes, individuals gradients (and Hessians if CV2) and sums of these, respectively.  

Compared with Algorithm \ref{mh_ss_algorithm}, the multimodal Algorithm has an additional step between Steps 3 and 4, to decide on $M(\theta,\theta';\thetahat_1,\thetahat_2)$ and the set $r_i(\theta,\theta';\thetahat_1,\thetahat_2)$ ($i=1,\dots,n)$; Steps 4, 7 and 16 then use these. With first-order covariates and a flat prior, the delayed acceptance step adds complication for no benefit, since if $\thetahat_1$ and $\thetahat_2$ were exact modes (rather than approximate) the Stage 1 acceptance probability would be $1$. Thus, in the experiments described below we stripped out the DA step, both for MHSS and MH.

\subsection{Model and bound}
\label{sec.bimodal.model}
We consider data arising from a true robust-regression model of
$$
Y_i\stackrel{iid}{\sim} \mathsf{t}_{\nu}(x_i^\top \beta,1);
$$
see \cite{cornish2019scalable}.
However, we only observe covariates $x_{i,2},\dots,x_{i,d}$; we do not observe the first covariate and only know that for $i=1,\dots,n$, $x_{i,1}=2W_i-1$, where $W_i\stackrel{iid}{\sim} \mathsf{Bernoulli}(0.5+\delta)$ for some known $\delta$. This creates a mixture model with a log likelihood of
$$
\sum_{i=1}^n \log\left\{(0.5+\delta) \mathsf{t}_\nu(y_i;\beta_1+x_{i,2:d}^\top \beta_{2:d},1)+(0.5-\delta)\mathsf{t}_{\nu}(y_i;-\beta_1+x_{i,2:d}^\top \beta_{2:d},1)\right\}.
$$

Define $x_-$ to be the vector of covariates except the $\pm 1$ associated with $\beta_1$, and let $\beta_-:=(\beta_{2},\dots,\beta_d)$. From this, we can define $\eta_-:=x_-^\top \beta_-$. 

The log likelihood is, therefore,
\begin{align}
\label{eqn.Mixture}
\ell(\beta)&=
\log \{(0.5+\delta)f(y-(\beta_1+\eta_-))+ (0.5-\delta)f(y-(-\beta_1+\eta_-))\}\\
&=
\log\{c_1f(\beta_1+z)+c_2 f(\beta_1-z)\}.\nonumber
\end{align}
where $c_1:=0.5+\delta$, $c_2=0.5-\delta$ and $z=y-\eta_-$.

\begin{theorem}
\label{thrm.t.mixture}
   Let $\ell$ be as defined in \eqref{eqn.Mixture}, with $f(x)\propto(1+x^2/\nu)^{-(\nu+1)/2}$. Let $H(\beta)$ be the Hessian matrix, with $H_{j,k}(\beta)=\partial^2\ell/\partial \beta_j \partial \beta_k$. Set $a=(a_1,a_-)^\top$, $b=(b_1, b_-)^\top$ and $x=(x_1,x_-)^\top$. Then, for $\nu\ge \sqrt{8}$,
   \begin{equation}
       \label{MixtureHessBound}
       |a^\top H b| \le (|a_1| + |x_-^\top a_-|)(|b_1|+|x_-^\top b_-|) B_\nu,
       ~~~\mbox{where}
       ~~~B_\nu:=\frac{(\nu+1)^2}{2\nu} + \frac{\nu+1}{4\nu(\nu+3)};
   \end{equation}
   for $\nu<\sqrt{8}$, the same result holds but with $B_{\nu}:=(\nu+1)(\nu+5)/(4\nu)$.
\end{theorem}
This is not the required format for $K_1$ in \eqref{eqn.bounded.Hessian} of Theorem \ref{theorem.remainders}, but it does allow us to find a $K_2$ from Theorem \ref{theorem.remainders}. Specifically, let the vector $|a|:=(|a_1|,|a_2|,\dots,|a_d|)^\top$. Then,
\[
|a_1|+|x^\top a_-|
\le
|a|^\top |x|
\le 
\|a\| \|x\|.
\]
So, Theorem \ref{thrm.t.mixture} tells us that  $|a^\top H b|\le \|a\| \|x\|^2 K_\nu\|b\|$. Thus $K_2=\|x\|^2 B_\nu$.

\subsection{Simulations and Results}

We simulated $n=10^6$ points from the model in Section \ref{sec.bimodal.model}, with $d=3$, $\beta=(0.06,0,0)$, $\delta=0.005$, $\nu=10$ and $x_{i,j}\stackrel{iid}{\sim} \mathsf{N}(0,0.06^2)$, $i=1,\dots,n$, $j=2,3$. The choice of $\delta$ ensures both modes are still important but one has a little more information than the other. The value for $\beta_1$ gives modes that are separated but for which the MCMC is able to travel between them, thus creating a feasible experiment. We place uniform, improper priors on all of the $\beta_j$, $j=1,\dots,d$.

We used a proposal variance of $\lambda^2 I_3$. We chose $\lambda$ to target an acceptance rate of 30\% for the MH algorithm, since it is known \cite[e.g.][]{sherlock2009optimal} that this is approximately optimal in $d=3$, and we aimed for $45\%$ for MHSS. We ran MH for $3\times 10^5$ iterations and MHSS for $3\times 10^6$ iterations. 

Table \ref{table.bimodal.settings} shows the scaling parameter, acceptance rate, number of iterations, CPU time (including one-off set-up costs) and average value of $B$ for each algorithm. On average, each iteration of MHSS uses less than 0.3\% of the data and takes less than a 200th of the CPU time of the MH. 

\begin{table}
\begin{center}
\caption{Settings and summaries for RWM and MHSS: proposal scaling, number of iterations, CPU time (in hours), empirical acceptance rate and average number of data points used per iteration.\label{table.bimodal.settings}}
\begin{tabular}{l|ll|lll}
Algorithm& $\lambda$&$n_{its}$& CPU hrs&$\alpha$&$\overline{B}$\\
\hline
MH&0.028&$3\times 10^5$&35.639&0.300&$10^6$\\
MHSS&0.008&$3\times 10^6$&1.685&0.438&2876\\
\end{tabular}
\end{center}
\end{table}

Table \ref{table.bimodal.ESS} provides the effective sample sizes (ESSs) and ESSs per CPU hour for each algorithm. For $\beta_1$ the ESS reflects the number of times the algorithm switches between the two modes. Even though the RWM has more mode switches per 1000 iterations (mostly reflecting the much larger scaling parameter), the MHSS has more switches per CPU hour.

\begin{table}
\begin{center}
\caption{Effective sample size (ESS) and ESS/CPU hour for each parameter in each run.\label{table.bimodal.ESS}}
\begin{tabular}{l|lll|lll}
Algorithm&$\beta_1$& $\beta_2$& $\beta_3$& $\beta_1$/hr& $\beta_2$/hr& $\beta_3$/hr\\
\hline
MH&4573&30790&30632&125&840&836\\
MHSS&1211&61320&62939&719&36392&37353
\end{tabular}
\end{center}
\end{table}

\begin{figure}[H]
\begin{center}
       \includegraphics[scale=0.30]{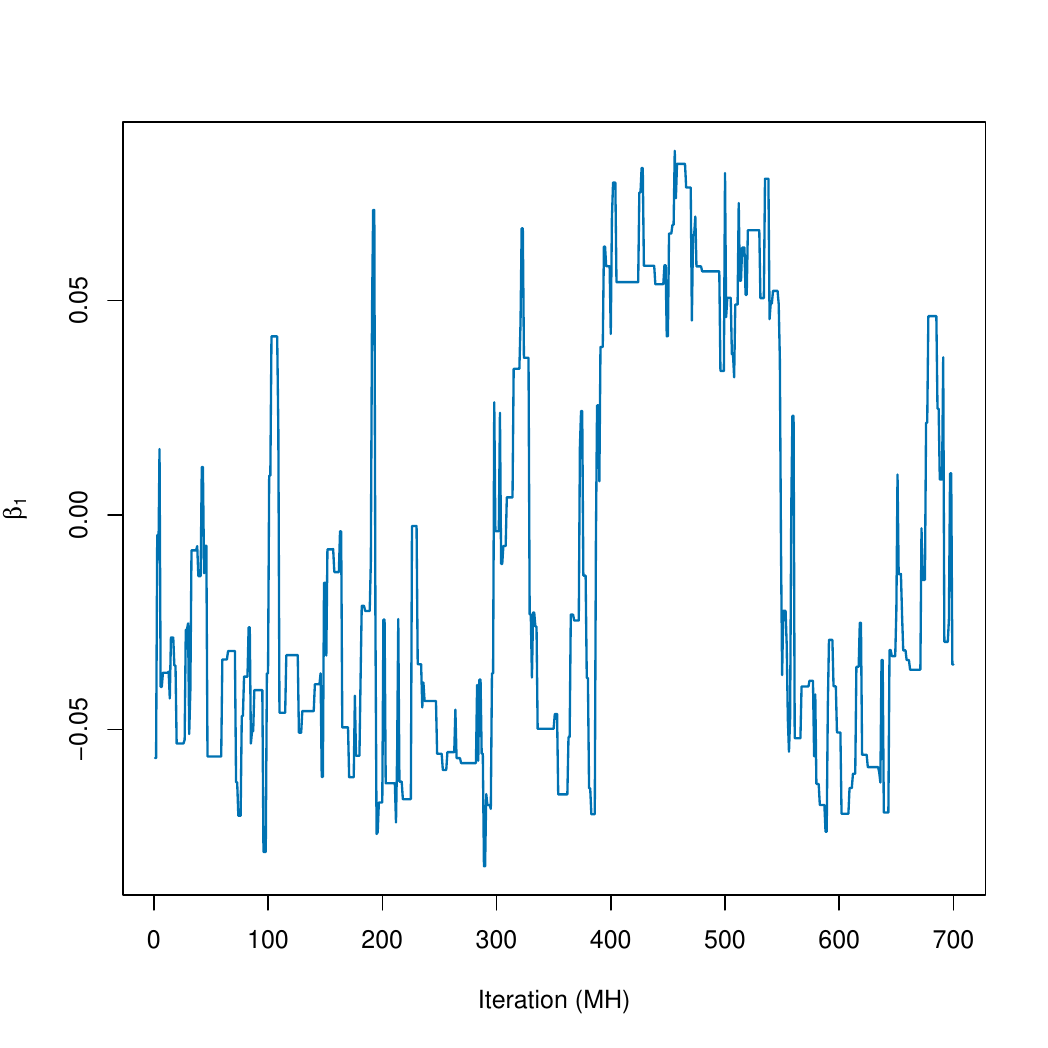}
       \includegraphics[scale=0.30]{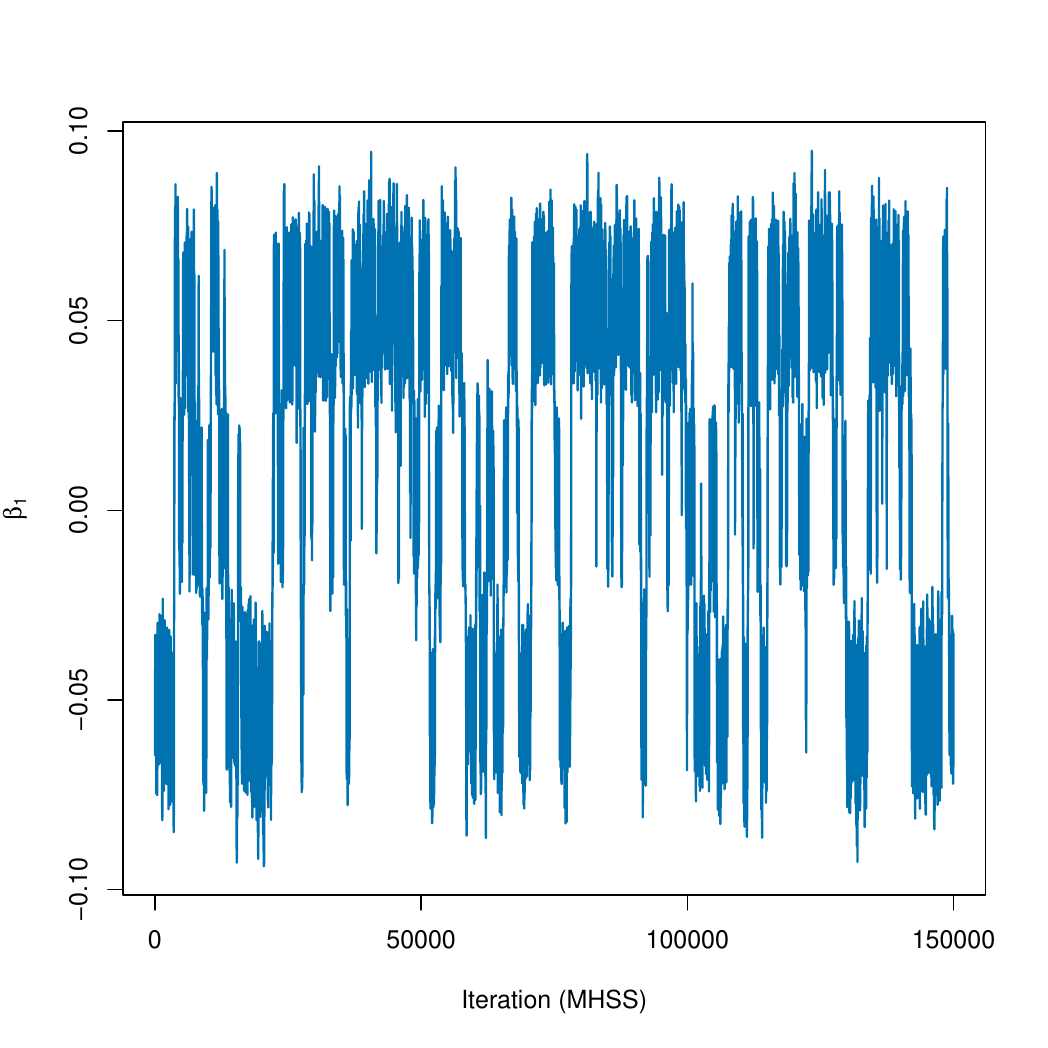}
       \includegraphics[scale=0.30]{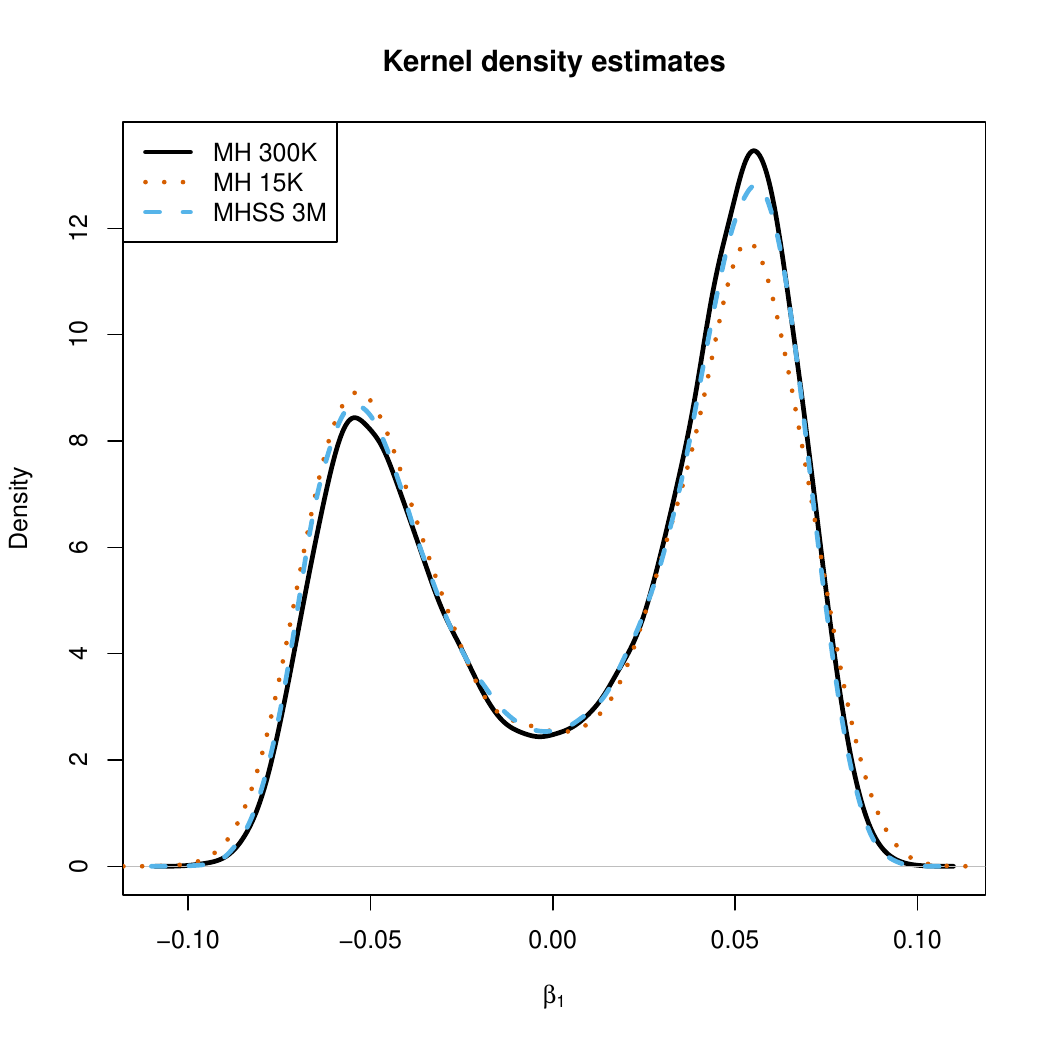}
     \captionsetup{justification=centering}
\end{center}
\vspace{-0.5cm}
\caption{Traceplots for $\beta_1$ over the first 5 CPU minutes of iterations for MH and MHSS, and kernel density estimates of the posterior for $\beta_1$ from the $3\times 10^5$ iterations of MH (black solid), the $3\times 10^6$ iterations of MHSS (blue dashed) and a spearate run of $14,184$ iterations of MH (red dotted), which took the same CPU times as the MHSS run.}
\label{fig:multimodal_results}
\end{figure}

The top left panel of Figure \ref{fig:multimodal_results} provides the trace plot for the first 700 iterations ($\approx 5$ minutes) of the MH algorithm; the next panel shows the first 150K iterations ($\approx 5$ CPU minutes) of MHSS. Not only does MHSS switch modes more often but it explores the individual modes more thoroughly. The right-hand panel plots three kernel density estimates of the marginal posterior density for $\beta_1$. The curve from $3\times 10^5$ iterations of MH may be used as a baseline against which to compare the estimates from $3\times 10^6$ iterations from MHSS and a separate run of  $14,184$ iterations from MH, which took the same CPU time as the MHSS run. Unsurprisingly, the MHSS run provides a better approximation to the marginal density than the MH run with the same CPU time.

\subsection{Proof of Theorem \ref{thrm.t.mixture}}
We may re-write \eqref{eqn.Mixture} as
\[
\ell(\beta)=g(\beta_1,z)
:=
\log\{c_1f(\beta_1+z)+c_2 f(\beta_1-z).\}
\]
We will bound the three derivatives $g_{\beta_1,\beta_1}$, $g_{\beta_1,z}$ and $g_{z,z}$.

First, notice that for any function $u(x)$,
\[
\frac{\md^2 \log u}{\md x^2}
=\frac{1}{u}\frac{\md^2 u}{\md x^2}-\left\{\frac{1}{u}\frac{\md u}{\md x}\right\}^2.
\]
Next, suppose $u(x)=u_1(x)+u_2(x)$, where for $j\in \{1,2\}$, $u_j(x)\ge 0$, $|u'_j(x)|\le K_A u_j(x)$ and $|u_j''(x)|\le K_B u_j(x)$. Then
\begin{align*}
    |u_1'(x)+u_2'(x)|&\le |u_1'(x)|+|u_2'(x)|
\le K_A \{u_1(x)+u_2(x)\},\\
|u_1''(x)+u_2''(x)|&\le |u_1''(x)|+|u_2''(x)|\le K_B\{u_1(x)+u_2(x)\}.
\end{align*}
Thus,
\[
\left|\frac{1}{u}\frac{\md^2 u}{\md x^2}\right|\le K_A
~~~\mbox{and}~~~
\left|\frac{1}{u}\frac{\md u}{\md x}\right|\le K_B,
\]
and, hence
\[
\left|\frac{\md^2 \log u}{\md x^2}\right|
\le
K_B+K_A^2.
\]
Similarly, for a function $u(x,y)$,
\[
\frac{\partial^2 \log u}{\partial x\partial y}
=\frac{1}{u}\frac{\partial^2 u}{\partial x \partial y}-\left\{\frac{1}{u}\frac{\partial u}{\partial y}\right\}.
\]
Supposing that $u(x,y)=u_1(x,y)+u_2(x,y)$, where for $j\in 1,2$, , $|u_{j,y}(x,y)|\le K_A u_j(x,y)$ and $|u_{j,xy}(x,y)|\le K_B u_j(x,y)$ (with subscripts of $x$ and $y$ denoting partial derivatives), then similar calculations show that
\[
\left|\frac{\partial^2 \log u}{\partial x\partial y}\right|
\le
K_B+K_A^2.
\]

\begin{lemma}
\label{lemma.tderivs}
    Let $f(x)$ be as defined in Theorem \ref{thrm.t.mixture}. Then
    \[
\left|\frac{f'(x)}{f(x)}\right|
\le
\frac{\nu+1}{2\sqrt{\nu}}
~~~\mbox{and, for}~\nu\ge \sqrt{8},~~~
\left|\frac{f''(x)}{f(x)}\right|
\le\frac{(\nu+1)(\nu+2)^2}{4\nu(\nu+3)};
    \]
    otherwise $\left|f''(x)/f(x)\right|\le (\nu+1)/\nu$.
\end{lemma}

Lemma \ref{lemma.tderivs} is proved in Appendix \ref{sec.prove.lemma.tderivs}.

Writing $u_1=c_1f(\beta_1+z)$ and $u_2=c_2 f(\beta_1-z))$ we see that for either function, whether we differentiate with respect to $\beta_1$ or $z$ the magnitude of the derivative does not change; only the sign may change. Thus
\[
\max\left\{|g_{\beta_1,\beta_1}|,|g_{z,z}|,|g_{\beta_1,z}|\right\}
\le
\frac{(\nu+1)(\nu+2)^2}{4\nu(\nu+3)}
+\left\{\frac{\nu+1}{2\sqrt{\nu}}\right\}^2
=B_\nu,
\]
after some simplification.

Finally, we notice that for $j\ge2$ and $k\ge 2$
\[
\frac{\partial^2 \ell}{\partial \beta_1^2}=g_{\beta_1,\beta_1},
~~~
\frac{\partial^2 \ell}{\partial \beta_1 \partial \beta_k}
=
g_{\beta_1,z}x_k
~~~\mbox{and}~~~
\frac{\partial^2 \ell}{\partial \beta_j \partial \beta_k}
=x_j x_k g_{z,z}
\]
Thus
\begin{align*}
|a^\top H b| &=|a_1 b_1 g_{\beta_1,\beta_1}
+
a_1 g_{\beta_1,z} x_-^\top b_-
+
a_-^\top x_- g_{z,\beta_1} b_1 + a_-^\top x_- g_{z,z} x_-^\top b_- |\\
&\le
|a_1b_1| |g_{\beta_1,\beta_1}|
+ 
|a_1 x_-^\top b_-| ~|g_{\beta_1,z}|
+
|a_-^\top x_- b_1|~|g_{z,\beta_1}|+ |a_-^\top x_-|~|b_-^\top x_-|~ |g_{z,z}|\\
&\le
\{|a_1|~|b_1|+ 
|a_1| |b_-^\top x_-|
+
|a_-^\top x_-||b_1|
+ |a_-^\top x_-|~|b_-^\top x_-|\}B_\nu,
\end{align*}
as required.

\subsubsection{Proof of Lemma \ref{lemma.tderivs}}
\label{sec.prove.lemma.tderivs}

Multiplicative constants will cancel, so it is sufficient to consider $f(x)=\left(1+x^2/\nu\right)^{-(\nu+1)/2}$. After some calculus, we have
\begin{align*}
    f'(x)&=-\frac{\nu+1}{\nu}x\left(1+\frac{x^2}{\nu}\right)^{-(\nu+3)/2},\\
    f''(x) &=
    \frac{\nu+1}{\nu}
    \left\{
    \frac{\nu+2}{\nu}x^2-1
    \right\}\left(1+\frac{x^2}{\nu}\right)^{-(\nu+5)/2}.
\end{align*}
Thus
$$
\frac{f'(x)}{f(x)}
=
-(\nu+1)\frac{x}{\nu+x^2}.
$$
This reaches its global maximum and minimum values at $x=\pm \sqrt{\nu}$, giving the first part of Lemma \ref{lemma.tderivs}.

Further,
$$
\frac{f''(x)}{f(x)}
=
\frac{\nu+1}{\nu}
\left\{
\frac{\nu+2}{\nu}x^2-1
\right\}
\left(1+\frac{x^2}{\nu}\right)^{-2}.
$$
Setting $u:=x^2/\nu$, we need to find the maximum absolute value of
$$
d(u):=\frac{\nu+1}{\nu}\frac{(\nu+2)u-1}{(1+u)^{2}},
$$
for non-negative $u$. This function is negative and increasing at $u=0$ and positive and decreasing from its maximum, so its absolute value is either $|d(0)|=(\nu+1)/\nu$ or the function's maximum. For $d'(u_*)=0$ we require 
$$
(1+u_*)^2(\nu+2)=2\{(\nu+2)u-1\}(1+u_*)
\implies
u_*=\frac{\nu+4}{\nu+2}.
$$
Finally, 
$$
d(u_*)=\frac{(\nu+1)(\nu+2)^2}{4\nu(\nu+3)}
=
\frac{(\nu+2)^2}{4(\nu+3)}\times \frac{\nu+1}{\nu},
$$
which shows that the maximum absolute value is $d(u_*)$ for $\nu\ge \sqrt{8}$ and $(\nu+1)/\nu$ otherwise.

\section{\CSadd{Alternative tuning for Tuna}}

\label{sec.alternative.acc.tuna}
\begin{figure}[H]
\begin{center}
     \includegraphics[scale=0.35]{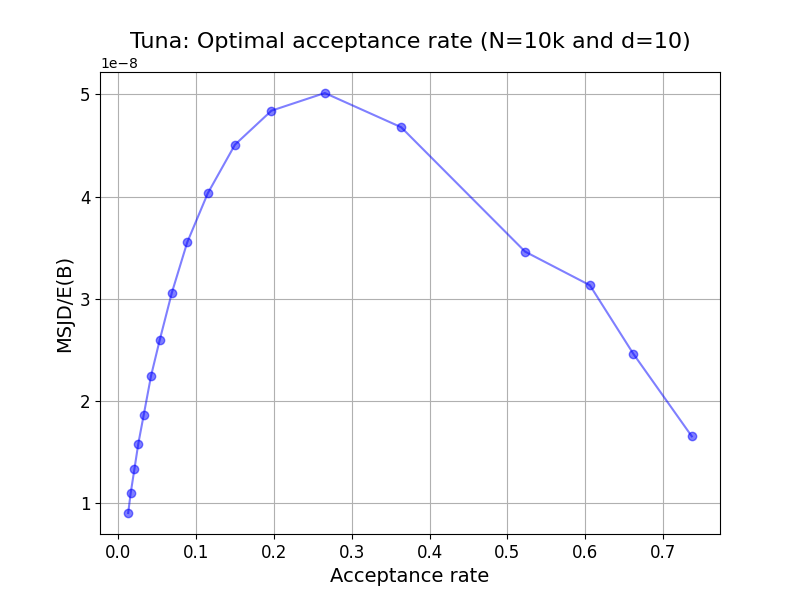} 
     \includegraphics[scale=0.35]{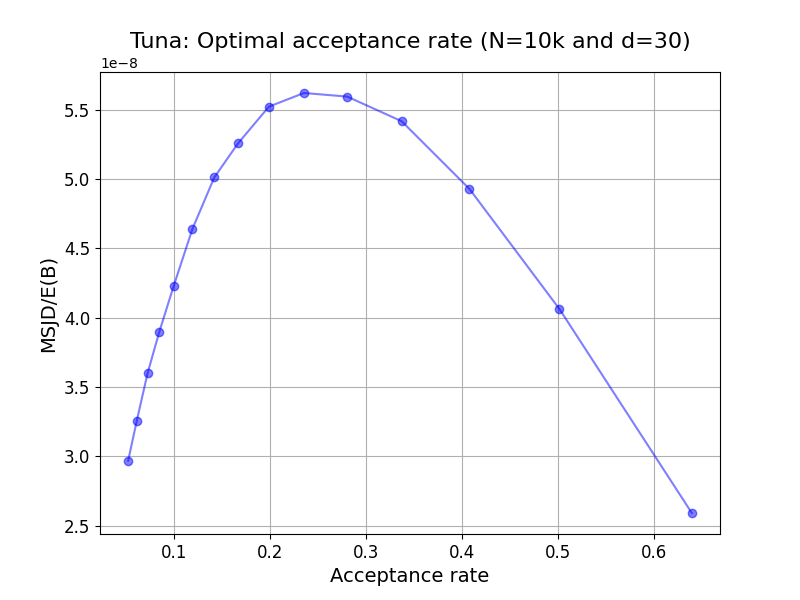}
     \includegraphics[scale=0.35]{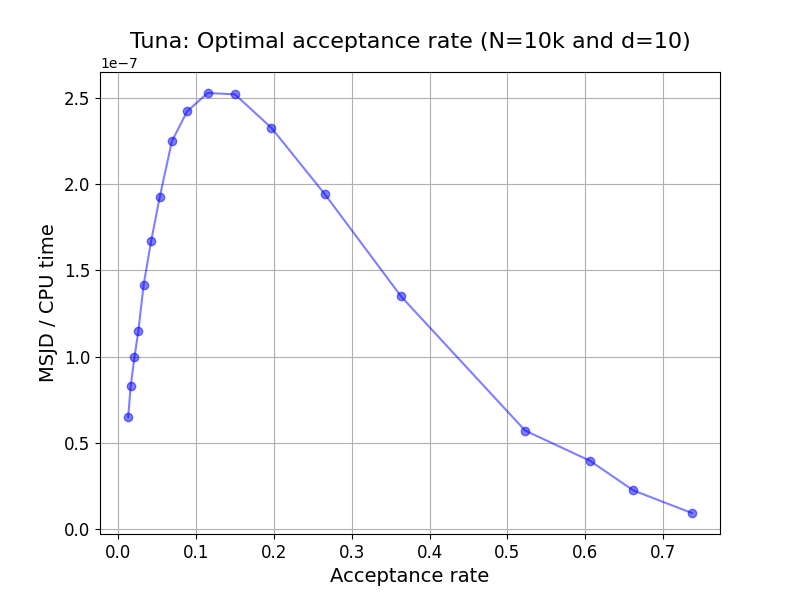} 
     \includegraphics[scale=0.35]{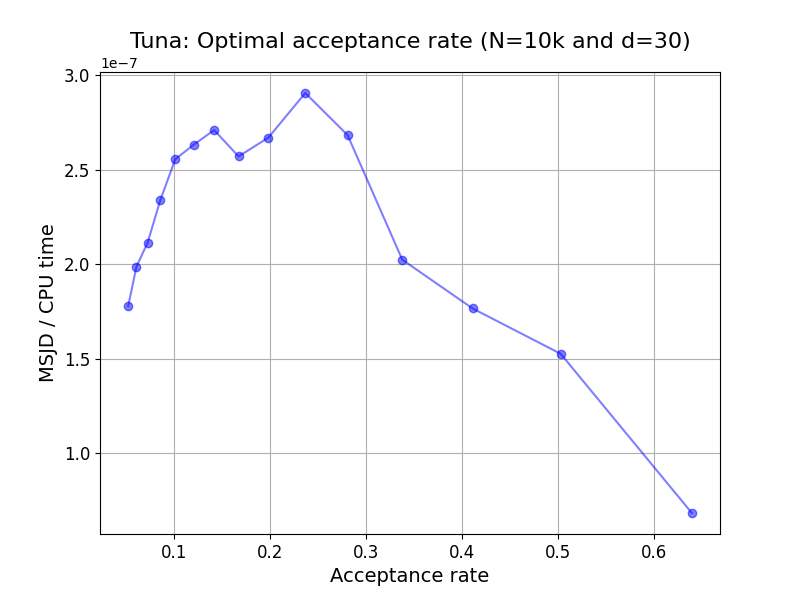}     
     \captionsetup{justification=centering}
\end{center}
\caption{Simulation experiment with a logistic regression model with dimension $d = 10, 30$ and $n = 10, 000$. Mean squared jumping distance (MSJD) over average batch size ($\mathbb{E}(B)$) and MSJD over CPU time are both plotted against empirical acceptance rate. The proposal distribution is $q(\theta^\prime | \theta) = \mbox{N}(\theta, \frac{\lambda^2}{d} V_d)$.}
\label{fig:rev1com3_optimal_acc_rate}
\end{figure}

For each of two simulated logistic-regression data sets, we repeatedly re-ran Tuna, varying the scaling of the proposal distribution from run to run. For each data set, Figure \ref{fig:rev1com3_optimal_acc_rate} plots efficiency, in terms of MSJD/E[B] and in terms of MSJD/CPU time, against the empirical acceptance rate, $\widehat{\alpha}$. In all cases efficiency is maximised at a lower acceptance rate than $0.60$. In three of the four plots, the optimal acceptance rate is around $0.24$ and in one it is a little lower than $0.24$, but the efficiency at $0.24$ is around $80\%$ of the maximum efficiency; this suggests a target of $\widehat{\alpha}\approx 0.24$ might be prefereable. In terms of MSJD/$\Expect{B}$, the improvement in efficiency when targeting $\widehat{\alpha}=0.24$ rather than $\widehat{\alpha}=0.60$ is less then a factor of two; in terms of MSJD/$\Expect{B}$, the ratio is closer to a factor of four.

\begin{figure}[H]
\begin{center}
     \includegraphics[scale=0.6]{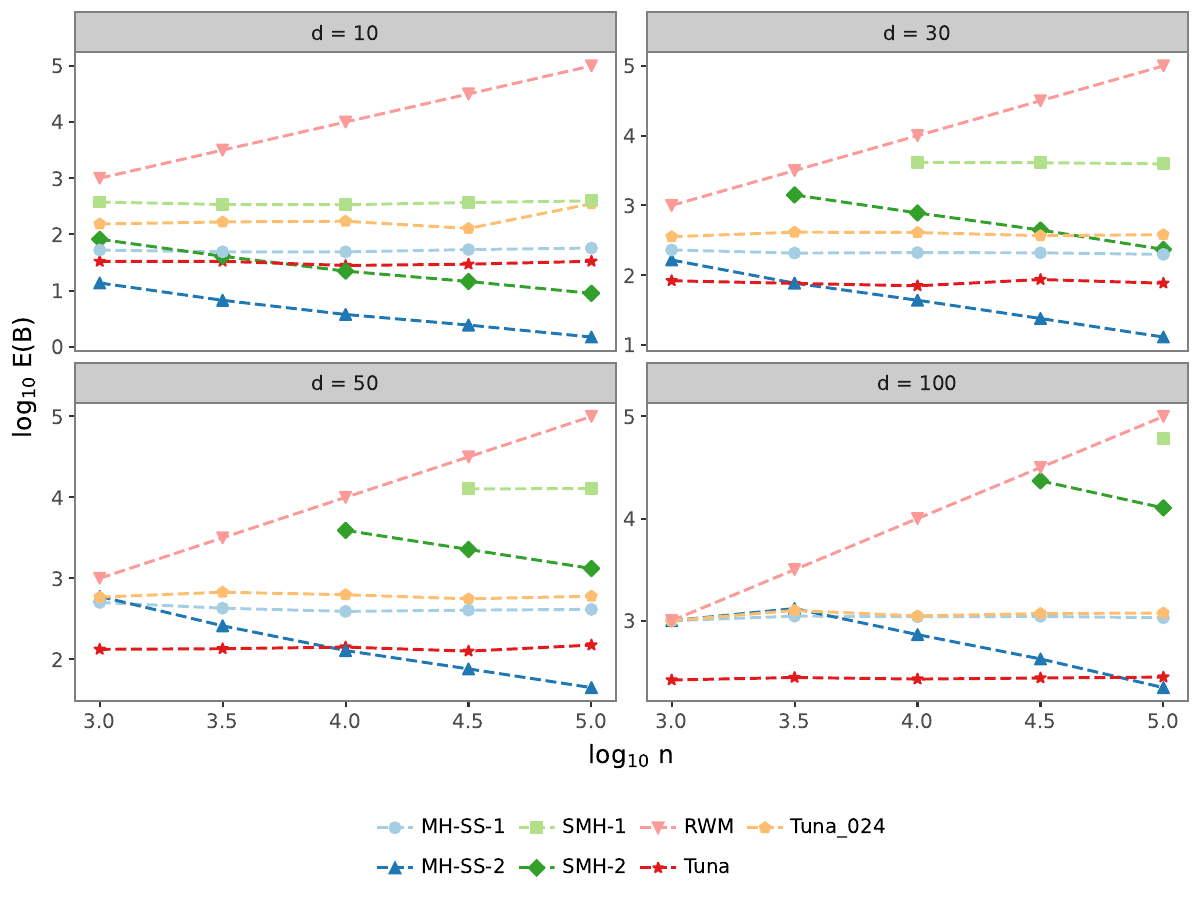}
     
     \captionsetup{justification=centering}
\end{center}
\vspace{-0.5cm}
\caption{Average batch size for Tuna, MH-SS, SMH and RWM for the logistic regression model. For Tuna, we consider two versions with $60\%$ and $24\%$ acceptance rates. Both axes are presented in the logarithm base 10.}
\label{fig:rev1com3_optimal_acc_rate_EB}
\end{figure}

\begin{figure}[H]
\begin{center}
     \includegraphics[scale=0.6]{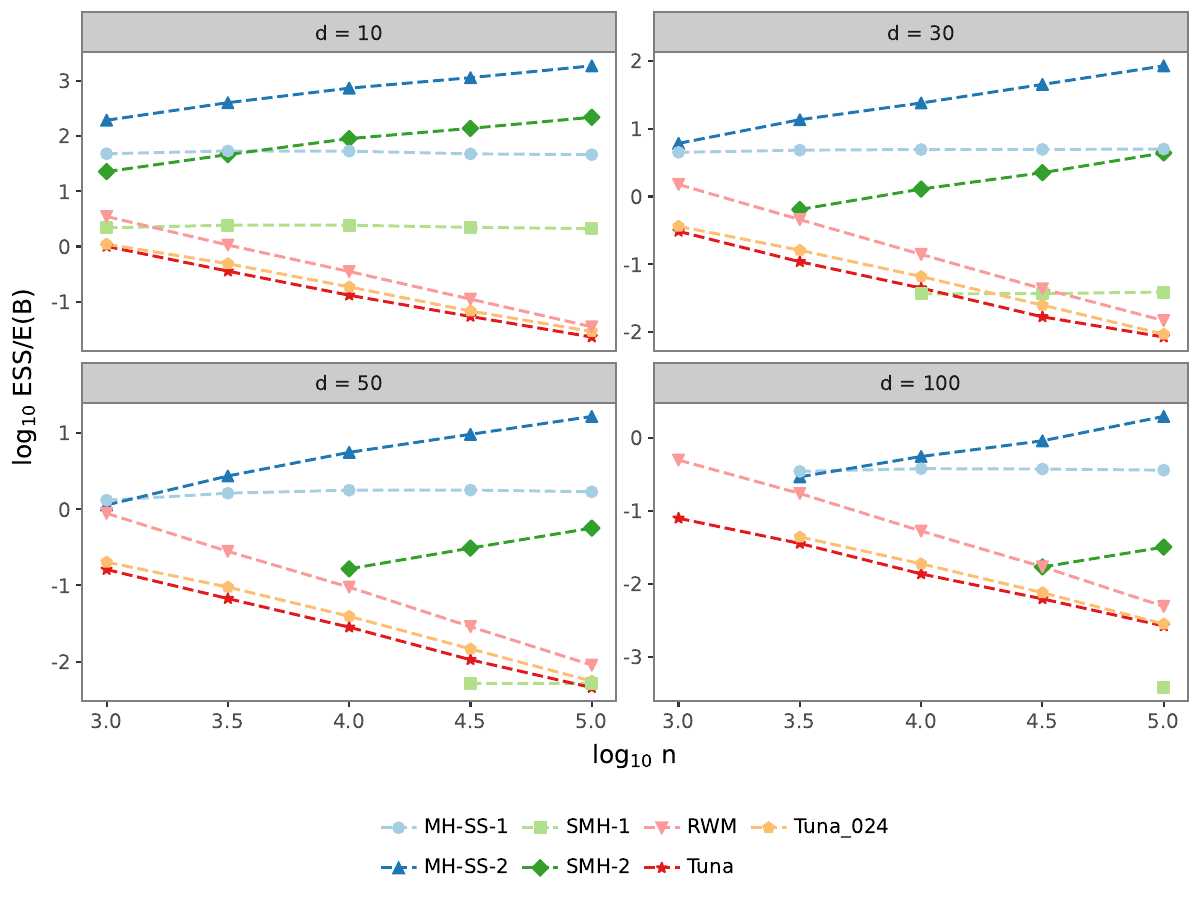} \\ \vspace{-1.2cm}
     \includegraphics[scale=0.6]{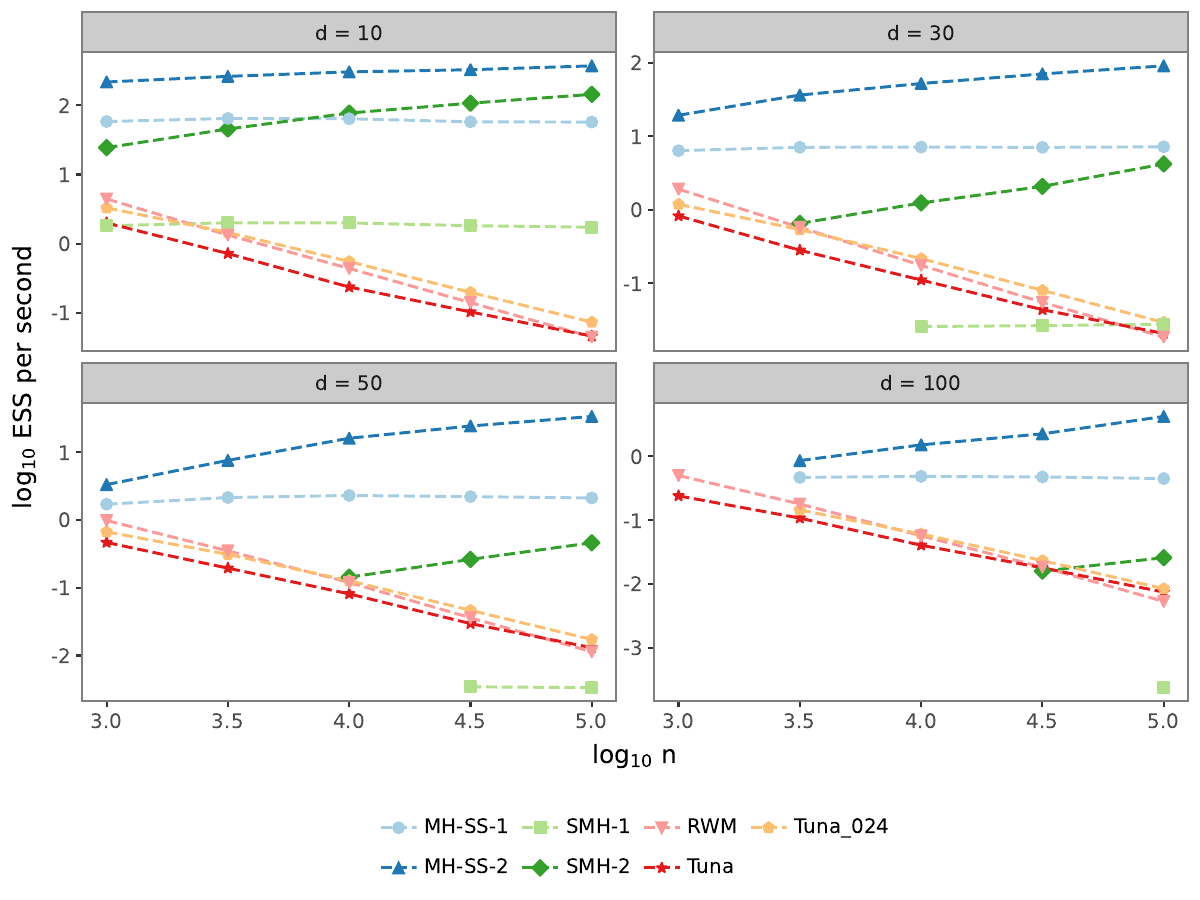}
     \captionsetup{justification=centering}
\end{center}
\vspace{-0.5cm}
\caption{ESS per second of Tuna, MH-SS, SMH and RWM for the logistic regression model. Both axes are presented in the logarithm base 10. Some ESSs are omitted because $\mathbb{E}(B)\geq n$, which implies the use and efficiency of the RWM algorithm.}
\label{fig:rev1com3_optimal_acc_rate_ESS}
\end{figure}

We, therefore, repeated the analysis of the simulated data sets in Section \ref{sec.logisticRegressionSimExpts} but now using Tuna with a larger RWM scaling chosen to obtain $\widehat{\alpha}\approx 0.24$. Figures \ref{fig:rev1com3_optimal_acc_rate_EB} and \ref{fig:rev1com3_optimal_acc_rate_ESS} exhibit the behaviour of expected batch size and efficiency, respectively. They are analogues of Figures \ref{Expected_B_by_N} and \ref{ESS_per_second} but with an extra, yellow line representing the new Tuna results; we also repeat Figure \ref{fig:rev1com3_optimal_acc_rate_ESS} but with efficiency in terms of ESS/$\Expect{B}$.
Figure \ref{fig:rev1com3_optimal_acc_rate_EB} shows that the new, larger scaling is associated with a larger $\Expect{B}$, so the computational cost per iteration is higher; however, Figure \ref{fig:rev1com3_optimal_acc_rate_ESS} makes clear that this is more than compensated for by the improvement in the mixing of the MCMC chain, since the efficiency is higher. Overall, however, in the efficiency plots, whether for ESS/$\Expect{B}$ or ESS/CPU time, the yellow line is only just above the red line that corresponds to Tuna with $\widehat{\alpha}\approx 0.60$, and is never substantially better than the basic RWM.  Except in $d=10$, SMH-1 performs poorly; however, provided there is at least a moderate amount of data ($n\ge 10,000$), MH-SS-1, MH-SS-2 and SMH-2, all perform substantially better than the RWM and than Tuna with either acceptance rate.

\

\end{document}